\documentclass{jfm}
\usepackage{amsmath}
\usepackage{newtxtext}
\usepackage{newtxmath}
\usepackage{mathrsfs}
\usepackage{xcolor}
\usepackage{graphicx}
\usepackage{bm}
\usepackage{natbib}
\usepackage{epstopdf}
\usepackage{lineno}
\usepackage{comment}
\usepackage{tikz}
\usetikzlibrary{decorations.pathreplacing}
\usepackage{hyperref}
\hypersetup{colorlinks=true,
            linkcolor=red,
            anchorcolor=blue,
            citecolor=blue}

\newcommand{\ri}{\mathop{\rm i}\nolimits}
\newcommand{\re}{\mathop{\rm e}\nolimits}

\title{%
Excitation of non-modal perturbations in hypersonic boundary layers by free stream forcing. Part III:  effects of wall temperature and nose bluntness%
}
\author[L. Zhao, M. Sun, Q. Song and M. Dong]%
{ Lei\ns ZHAO$^{1}$,
 Mingze\ns SUN$^{1}$,
 Qinyang\ns SONG$^{1}$
and
Ming\ns DONG$^{2}$%
  \thanks{Email address for correspondence: dongming@imech.ac.cn}}

\affiliation{$^1$Department of Mechanics, Tianjin University, Tianjin, 300072, China\\
$^2$State Key Laboratory of Nonlinear Mechanics, Institute of Mechanics, Chinese Academy of Sciences, Beijing 100190, China
}

\begin{document}
\maketitle

\begin{abstract}
This paper constitutes Part III of a series investigating the excitation of non‑modal perturbations in hypersonic blunt‑nose boundary layers. Based on the numerical shock‑fitting harmonic linearised Navier‑Stokes (SF‑HLNS) approach presented in Part I (Zhao \& Dong, J. Fluid Mech., vol. 1013, 2025, A44) and the asymptotic analysis developed in Part II (Dong et al., J. Fluid Mech., under review), the present work focuses specifically on the influence of nose bluntness (Reynolds number $R$) and wall temperature $T_w$ on receptivity efficiency through seven detailed case studies.  {\color{black}Analysed via the three-step asymptotic model, the downstream perturbation amplitude is governed by two distinct factors: the stagnation {\color{red}vorticity} perturbation, determined through shock-perturbation interaction and the slow-down convection mechanism, and the parabolic evolution of non-modal perturbations driven by the lift-up mechanism. Increasing $R$ strongly amplifies the stagnation vorticity response, with only mild enhancement of the lift-up effect. In contrast, decreasing $T_w$ strengthens the stagnation vorticity perturbation, but slightly weakens the lift-up effect; the net result is an overall intensification of the downstream perturbation amplitude with decreasing  $T_w$. Furthermore, explicit scaling laws for receptivity efficiency with respect to $R$ and $T_w$ are established.}   This study completes the theoretical foundation for non‑modal receptivity in hypersonic blunt boundary layers, and serves as a key step toward a physics-based transition prediction framework.
\end{abstract}



\section{Introduction} 
Hypersonic boundary-layer transition significantly impacts the aerodynamic performance, thermal protection, and stability of high-speed vehicles, primarily due to the sharp increase in skin friction and surface heat flux that occurs during the transition process \citep{Fedorov2011transition,Zhong2012direct}. However, a universally reliable prediction method covering all transition pathways is still lacking. This is largely because existing models do not adequately describe the full sequence of physical processes, ranging from initial receptivity to subsequent nonlinear breakdown. 
A major difficulty in predicting hypersonic boundary-layer transition, particularly for configurations with moderately blunt noses, arises from the prevalence of bypass transition. In such routes, freestream perturbations (such as acoustic, entropy and vortical waves) penetrate the detached bow shock and enter the boundary layer, generating non-modal perturbations that undergo transient growth. If their initial amplitudes are sufficiently high, these perturbations can reach a nonlinear phase before their attenuation, eventually triggering secondary instabilities and the emergence of turbulent spots. Since no exponentially growing modal perturbation emerges during this process, the classical linear stability theory remains inapplicable. 

Although frameworks such as optimal growth theory \citep{andersson1999optimal,paredes2019nonmodal} and input–output (resolvent) analysis \citep{Kamal2023} can describe the downstream evolution of non‑modal perturbations, they do not directly link perturbation formation to freestream forcing. Consequently, a quantitative prediction of bypass transition under realistic disturbance environments remains unfeasible. In contrast, the boundary-region equation (BRE) approach, employed by \cite{Leib1999}, provides a more systematic framework for  accounting for receptivity and the downstream evolution of non‑modal perturbations. In that work, an asymptotic matching technique was applied to quantify the key scaling interactions between freestream turbulence and boundary-layer perturbations near a sharp leading edge of an incompressible boundary layer. The resulting perturbation was then used as the inflow condition for BRE calculations, successfully capturing the formation of streaky structures in the linear regime. Subsequently, \cite{Ricco2007response}  extended this framework to compressible boundary layers, showing that temperature streaks can attain larger amplitudes than velocity streaks. Later, \cite{ricco2011evolution} examined finite-amplitude freestream vortical disturbances, thereby describing the nonlinear evolution of non‑modal streaks and the onset of secondary instabilities. 
However, these studies assume a sharp leading edge, thereby excluding the influences of the bow shock and entropy layer in hypersonic configurations. 

To address these shortcomings, \cite{zhao2025excitation} (hereafter denoted by ZD25 or Part I) recently developed a shock-fitting harmonic linearised Navier-Stokes (SF-HLNS) approach, designed to quantify the receptivity of hypersonic blunt‑nose boundary layers to freestream disturbances. The use of a harmonic {linearised} system enables a systematic and computationally efficient study of receptivity characteristics, while the shock‑fitting technique ensures an accurate description of the interaction between freestream disturbances and the detached bow shock. As a follow-up of this study, \citet{Dong2026} (hereafter denoted by Part II) performed an asymptotic analysis to uncover the intrinsic mechanism that governs the receptivity process. Their analysis revealed that the receptivity is most sensitive to the {\color{red}centreline} region, and a distinct slow‑down convection mechanism was identified. The latter amplifies the {\color{red}centreline} perturbation streamwise vorticity by a factor of  $O(R^{1/2})$ as it convects toward the stagnation region,   where $R$ denotes the Reynolds number based on the nose radius. A detailed literature review of bypass transition can be found in ZD25.

In general, the receptivity system is governed by a high-dimensional set of parameters. For a given oncoming flow  (such as Mach number and temperature) and a prescribed wall shape, the receptivity efficiency depends strongly on the wall temperature, the Reynolds number (or nose radius), and the properties of the freestream forcing. 

In hypersonic wind-tunnel experiments, \cite{stetson1983nosetip} and \cite{borovoy2022laminar} reported the strong influence of nose radius on transition. For small-bluntness configurations, the exponential growth of the Mack modes dominates the perturbation accumulation in laminar phase, which is stabilised as bluntness increases. When the nose radius exceeds a threshold, the non-modal perturbations become  dominant  in the laminar phase, with the bypass transition being promoted as nose bluntness further increases.  To understand the underlying mechanism at moderate bluntness, \cite{paredes2018blunt,paredes2019nose}, used the optimal growth theory (OGT) to  compute the energy gain of  non-modal perturbations in  hypersonic blunt-nose boundary layers, revealing that the transient growth intensifies with increasing Reynolds number or decreasing wall temperature. However,  OGT does not incorporate receptivity; consequently, the predicted initial disturbances may be non‑physical and the maximum energy gain can be overestimated.
Subsequently, \cite{goparaju2022role} employed a random-forcing approach to model  the environmental perturbation, and carried out direct numerical simulations (DNS) on hypersonic boundary-layer transition over a range of nose radii, successfully reproducing the transition reversal phenomenon reported by  \cite{stetson1983nosetip} and \cite{borovoy2022laminar}. By treating the freestream forcing as a series of slow acoustic waves, \cite{guo2025transition} simulated the bypass transition at various moderate nose radii, showing a premature transition as increasing bluntness. 

Wall temperature also plays a key role in boundary‑layer disturbance evolution and subsequent transition. Linear stability analyses indicate that lowering the wall temperature destabili{\color{black}s}es the Mack second mode but stabili{\color{black}s}es the Mack first mode \citep{Mack1987review}. When the wall temperature is sufficiently low,  the second mode near the upper-branch neutral point can radiate acoustic waves to the far field  \citep{Chuvakhov2016}. Using DNS, \cite{Li2025} examined the effect of wall temperature on Görtler instability and transition in hypersonic concave‑wall boundary layers, finding that higher wall temperatures delay streak saturation but increase the saturation amplitude. Moreover, as wall temperature rises, the secondary instability is suppressed and the dominant mode shifts from varicose to sinuous. Furthermore, \cite{Ji2025} proposed a control strategy that delays transition by using surface heating to suppress the fundamental resonance of the Mack second mode. 

Even for a fixed Reynolds number and wall temperature, the type of freestream forcing strongly influences the transition onset. For hypersonic blunt-cone boundary layers, \cite{He2021} investigated the receptivity of Mack modes to various freestream perturbations. They reported that for finite‑duration pulses, the strongest to weakest responses correspond to fast acoustic, entropy, slow acoustic, and vortical pulses, respectively; however, under planar forcing perturbations, the ordering changes to slow acoustic, entropy, vortical, and fast acoustic pulses.  In contrast, for hypersonic blunt‑wedge configurations, ZD25 performed SF-HLNS calculations under  freestream forcing with equal energy norm, and found that acoustic forcing produces the most efficient receptivity of non‑modal perturbations, with fast and slow acoustic waves showing nearly identical effectiveness. The receptivity efficiency for vortical forcing was substantially weaker.

While previous studies have revealed qualitative trends of various factors on receptivity, a systematic investigation that simultaneously accounts for nose bluntness, wall temperature, and freestream forcing is still lacking.  The present work aims to fill this gap by employing the SF‑HLNS framework developed in Part I, together with the asymptotic model established in Part II. 
{\color{black}Specifically, guided by the asymptotic model, the non-modal receptivity process is decomposed into three sequential steps. This study elucidates the impacts of the controlling parameters (including bluntness, wall temperature and properties of the freestream forcing) on the perturbation response at each stage, and establishes the key scaling relations governing this receptivity system. }

The rest of this paper is structured as follows. Section \ref{sec:headings} provides a brief overview of the physical model and numerical methodology. In Section \ref{sec:3}, case studies and preliminary SF‑HLNS results are presented for a range of Reynolds numbers and wall temperatures. Section \ref{sec:4} introduces the asymptotic model and shows the asymptotic predictions of the boundary-layer perturbations for isothermal walls. Building on this analysis, Section \ref{sec:5} explores parametric variations to identify and validate key scaling relations governing the receptivity process. Concluding remarks are given in Section \ref{sec:6}.

\section{Physical model and numerical approaches}
\label{sec:headings}
The physical configuration under investigation is based on Part I (ZD25) of this work series. It involves a hypersonic flow over a blunt wedge boundary layer subjected to infinitesimal freestream disturbances in the form of acoustic, entropy, or vortical waves. These perturbations interact with the detached bow shock, penetrate the boundary layer, and excite non‑modal downstream perturbations. A schematic of the problem is provided in figure 1 of  ZD25.
To quantify the non‑modal receptivity efficiency for various freestream forcings across a broad range of control parameters, a highly efficient and accurate SF-HLNS approach was developed in ZD25. The underlying receptivity mechanism was further elucidated via asymptotic analysis in Part II.
In the present work, we employ both the SF‑HLNS approach and asymptotic theory to systematically examine the influence of wall temperature and nose radius on receptivity efficiency. 

Using the nose radius $r^*$ as the reference length scale, we adopt a dimensionless Cartesian coordinate system $(x,y,z)$ and a dimensionless body-fitted system $(\xi,\eta,z)$, following Part II. The physical velocity field   ${\bf{u}}=(u,v,w)$, density $\rho$, temperature $T$ and viscosity $\mu$ are normalised by their respective freestream values, and the pressure $p$ is normalised by $\rho^*_\infty U_\infty^{*2}$, with the subscript $\infty$ denoting the oncoming quantities. For clarity, the velocity field is also projected into the body-fitted coordinates $(u_\xi,u_\eta,{w})$. This receptivity system is governed by three key parameters, the Mach number $M$, the Reynolds number $R=\rho^*_\infty U^*_\infty r^*/\mu^*_\infty$, and the dimensionless wall temperature $T_w$. Additionally, we choose the Prandtl number $Pr=0.72$ and ratio of specific heats $\gamma=1.4$. The wall geometry is given by Eq. (2.2) in {Part} II, with $\theta$ denoting the half-angle of the wedge.

The governing equations are the dimensionless Navier-Stokes (N-S) equations. In the uniform stream, the infinitesimal perturbations can be expressed as
 \begin{equation}
\frac{\epsilon}2 {\hat \phi_\infty} \re^{\ri (k_1x+k_2y+k_3z-\omega t)}+c.c.,
\label{eq:freestream_perturbation}
 \end{equation}
where $\epsilon\ll 1$ denotes its amplitude, ${\bm k}=(k_1,k_2,k_3)$ is the wavenumber vector, $\omega$ is the frequency, and $c.c.$ denotes the complex conjugate.  The perturbation vector $\hat \phi_\infty=(\hat u_\infty,\hat v_\infty,\hat w_\infty,\hat \rho_\infty,\hat \theta_\infty,\hat p_\infty)$ for freestream fast acoustic, slow acoustic, entropy and vortical forcing can be found in Section 2.2 of Part II. The declination angle is defined as $\vartheta=\tan^{-1}(k_2/k_3)$.
The instantaneous flow field $\phi=({\rho,\bf u},T)$ subject to the freestream forcing is decomposed into a base flow $\bar\phi$ and a harmonic perturbation $\tilde \phi$,
\begin{equation}
\label{eq:decompositioni}\phi(\xi,\eta,z,t)=\bar\phi(\xi,\eta)+\epsilon\tilde \phi(\xi,\eta)\re^{\ri(k_3 z-\omega t)}+c.c..
\end{equation}

In the receptivity study, the selected computational domain spans from the wall $\eta=0$ to the detached bow shock $\eta=\eta_s$; for a symmetric configuration, only the upper-half plane above the {\color{red}centreline} is considered, with $\xi\in[0,\xi_N]$.
 
The base flow $\bar \phi$ is computed using our in-house shock-fitting Navier–Stokes solver, with no-slip  wall boundary conditions applied. The wall is treated as isothermal for the majority of cases. For comparative purposes, one case is computed using an adiabatic wall condition, as will be shown in Table \ref{tab:1}. For the perturbation field, we adopt the  SF-HLNS approach as outlined in   ZD25, with   freestream perturbations given in (\ref{eq:freestream_perturbation}). The upper boundary condition is specified by the linearised Rankine–Hugoniot (R-H) relation ((2.42) of ZD25) and the linearised compatibility condition ((2.39) of ZD25).

\section {Preliminary numerical results}
\label{sec:3}
\subsection{Case studies and base flow}
\begin{figure}
  \begin{center}
  \includegraphics[width = 0.49\textwidth]{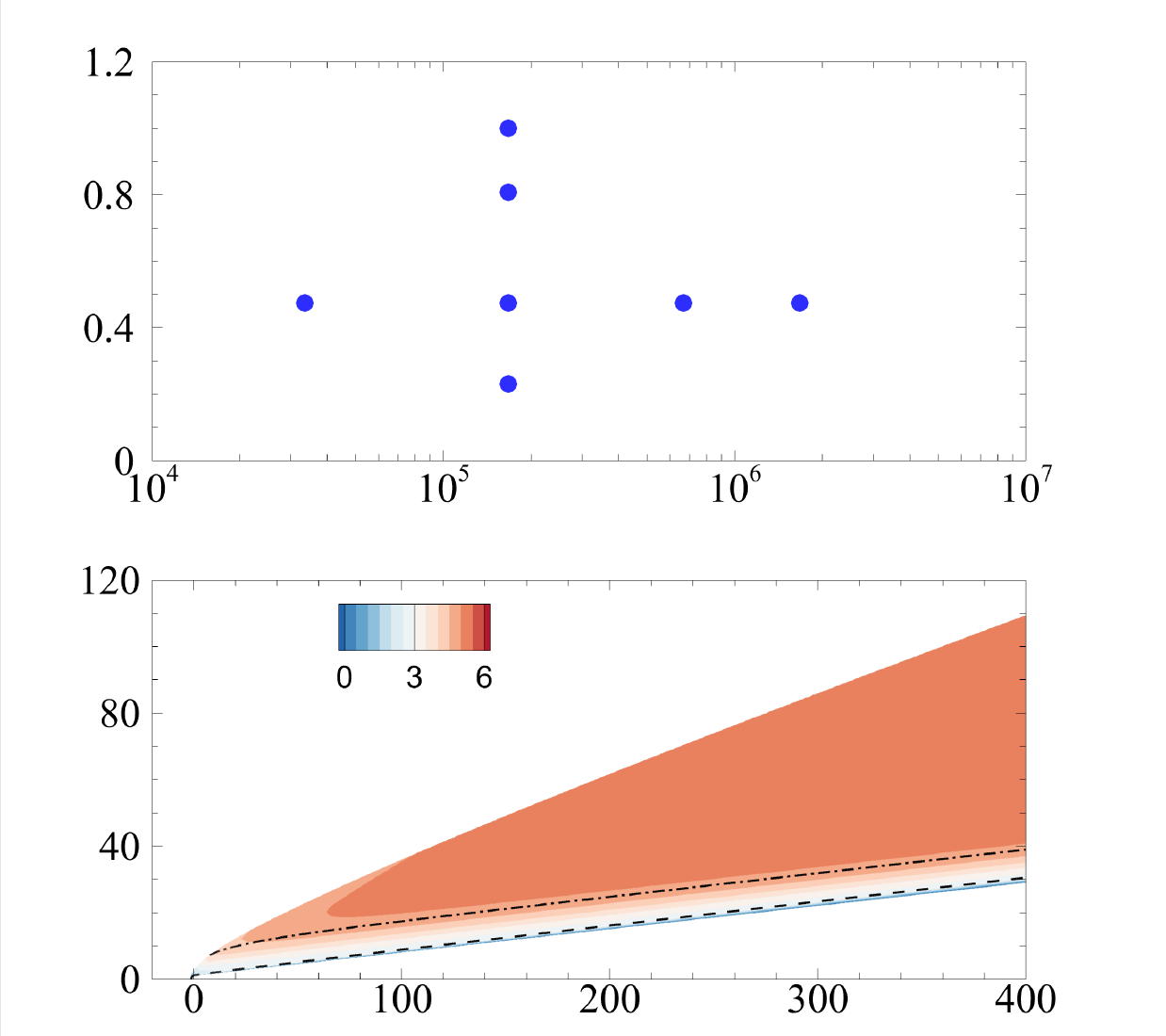}
  \includegraphics[width = 0.49\textwidth]{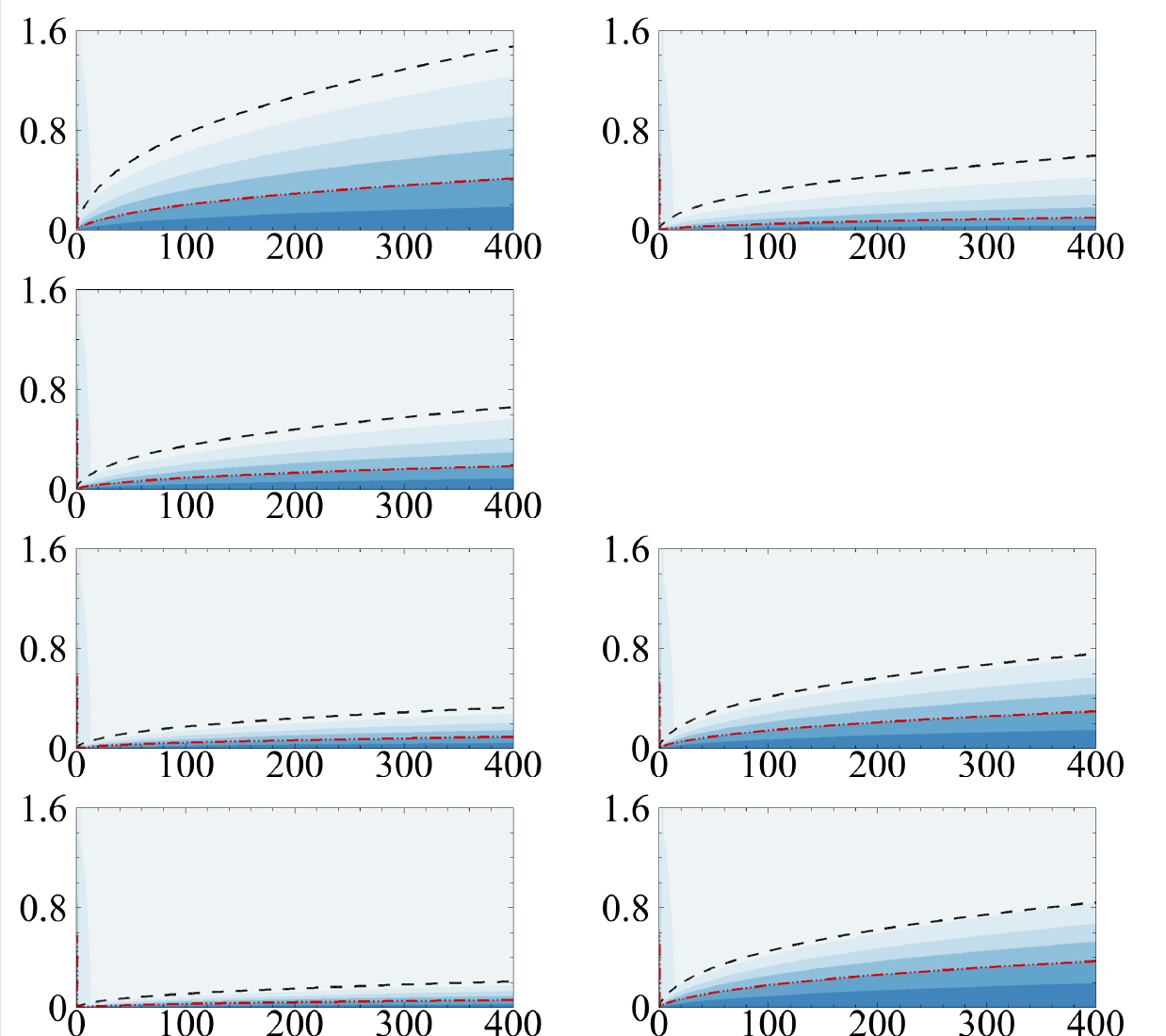}
  \put(-375,115){\rotatebox{90}{$T_w/T_{ad}$}}
  \put(-285,82){$R$}
  \put(-380,160) {$(a)$}
  \put(-380,75) {$(b)$}
  \put(-200,160) {$(c)$}
  \put(-200,115) {$(d)$}
  \put(-200,75) {$(e)$}
  \put(-200,35) {$(f)$}
  \put(-105,160) {$(g)$}
  \put(-105,75) {$(h)$}
  \put(-105,35) {$(i)$}
  \put(-170,150)  {\fontsize{6pt}{6pt}\selectfont$RAT2$}
  \put(-170,110)  {\fontsize{6pt}{6pt}\selectfont$RBT2$}
  \put(-170,70)   {\fontsize{6pt}{6pt}\selectfont$RCT2$}
  \put(-170,30)   {\fontsize{6pt}{6pt}\selectfont$RDT2$}
  \put(-75,150)  {\fontsize{6pt}{6pt}\selectfont$RBT1$}
  \put(-75,70)   {\fontsize{6pt}{6pt}\selectfont$RBT3$}
  \put(-75,30)   {\fontsize{6pt}{6pt}\selectfont$RBT4$}
  \put(-335,60) {$\bar M$}
  \put(-285,-2) {$x$}\put(-375,40) {$y$}
  \put(-140,-6) {$\xi$}
  \put(-50,-6) {$\xi$}
  \put(-195,20) {$\eta$}
  \put(-195,60) {$\eta$}
  \put(-195,100) {$\eta$}
  \put(-195,145) {$\eta$}
  \put(-340,110)  {\fontsize{6pt}{6pt}\selectfont$RAT2$}
  \put(-305,110)  {\fontsize{6pt}{6pt}\selectfont$RBT2$}
  \put(-280,110)   {\fontsize{6pt}{6pt}\selectfont$RCT2$}
  \put(-255,110)   {\fontsize{6pt}{6pt}\selectfont$RDT2$}
  \put(-305,98)  {\fontsize{6pt}{6pt}\selectfont$RBT1$}
  \put(-305,129)   {\fontsize{6pt}{6pt}\selectfont$RBT3$}
  \put(-305,139)   {\fontsize{6pt}{6pt}\selectfont$RBT4 (adiabatic)$}
  \caption{(a): Selection of the parameter studies in the $R$-$T_w$ plane. (b): Contours of the local Mach number $\bar M$ in the $x$-$y$ plane for case RAT2. (c-i): Contours of $\bar M$ in the boundary-layer region for all cases, displayed in the $\xi$-$\eta$ plane .  The dashed, dash-dotted and red dot-dot-dashed lines mark the boundary-layer edge, entropy layer edge and the sonic line (where $\bar M=1$), respectively. Not to scale.}
  \label{fig:baseflow}
  \end{center}
\end{figure}

For non-modal receptivity calculations, seven parameter sets are chosen, as listed in Table \ref{tab:1}. The wall temperature ranges from a low value (1.62) to the adiabatic state ($T_{ad}\approx$7.03), while the Reynolds number varies by a factor of 50. These parameter studies are summarised in figure \ref{fig:baseflow}-(a). For a fixed {\color{black}unit} Reynolds number $R_\infty$, the Reynolds number $R$ is proportional to the nose radius. {\color{cyan}Notably, since  neither the unit Reynolds number nor the dimensional nose radius $r^*$ appears explicitly in the dimensionless governing equations, variations in $R$ also encapsulate  an alternative physical scenario: a change in the unit Reynolds number while holding $r^*$ fixed.}

{\color{black}The base flow is calculated based on direct solution of steady N-S equations, using the in-house DNS code as used in \citep{zhao2025excitation}.}
The computational domain in the streamwise direction spans from $\xi=0$ to $\xi_N=400$, with 1201 grid points employed. In the wall-normal direction, $\eta\in [0,\eta_s]$, with 301 grid points employed. The resolution study has been conducted in ZD25. 
\begin{table}
  \begin{center}
\begin{tabular}{lccccccc}
Case & $M$  & $\theta$  & $R$       & $r^\ast$         & $T_w$    &$T_w/T_{ad}$      &Description \\
RAT2 &      &           & $3.34\times 10^4$ & 1 mm  & 3.33      &  0.47    & isothermal    \\
RBT2 &      &           & $1.67\times 10^5$ & 5 mm  & 3.33      &  0.47     & isothermal   \\
RCT2 &      &           & $6.68\times 10^5$ & 20 mm & 3.33      &  0.47      & isothermal  \\
RDT2 & 5.96 & $4^\circ$ & $1.67\times 10^6$ & 50 mm & 3.33      &  0.47     & isothermal   \\
RBT1 &      &           & $1.67\times 10^5$ & 5 mm  & 1.62      &  0.23     & isothermal   \\
RBT3 &      &           & $1.67\times 10^5$ & 5 mm  & 5.67      &  0.81     & isothermal   \\
RBT4 &      &           & $1.67\times 10^5$ &5 mm   & --        &  --    &adiabatic
\end{tabular}
  \caption{Controlling parameters for case studies. The unit Reynolds number is $R^*_\infty:=R/r^*$={$3.34\times 10^{4}\ {\mathrm{mm}} ^{-1} $}, the  temperature of the oncoming stream is $T^*_\infty=87$ K, and the dimensionless adiabatic-wall temperature is estimated by the recovery temperature $T_{ad}=1+\sqrt{Pr}(\gamma-1)M^2/2\approx7.03$.}  
  \label{tab:1}
  \end{center}
\end{table}

Figure \ref{fig:baseflow}(b) shows the contours of the local Mach number $\bar M=M\sqrt{\bar u^2+\bar v^2} /\sqrt{\bar T}$ for case RAT2 in the $x$-$y$ plane. The Mach number shows an evident gradient in the entropy layer. Panels (c) to (i) display the $\bar M$ contours in the  boundary-layer region for all the cases studied. The flow becomes subsonic where $\bar M<1$ in the near-wall region. Apparently, the boundary-layer thickness decreases with increase of $R$, and decrease of $T_w$.

\begin{figure}
  \begin{center}
  \includegraphics[width = 0.49\textwidth]{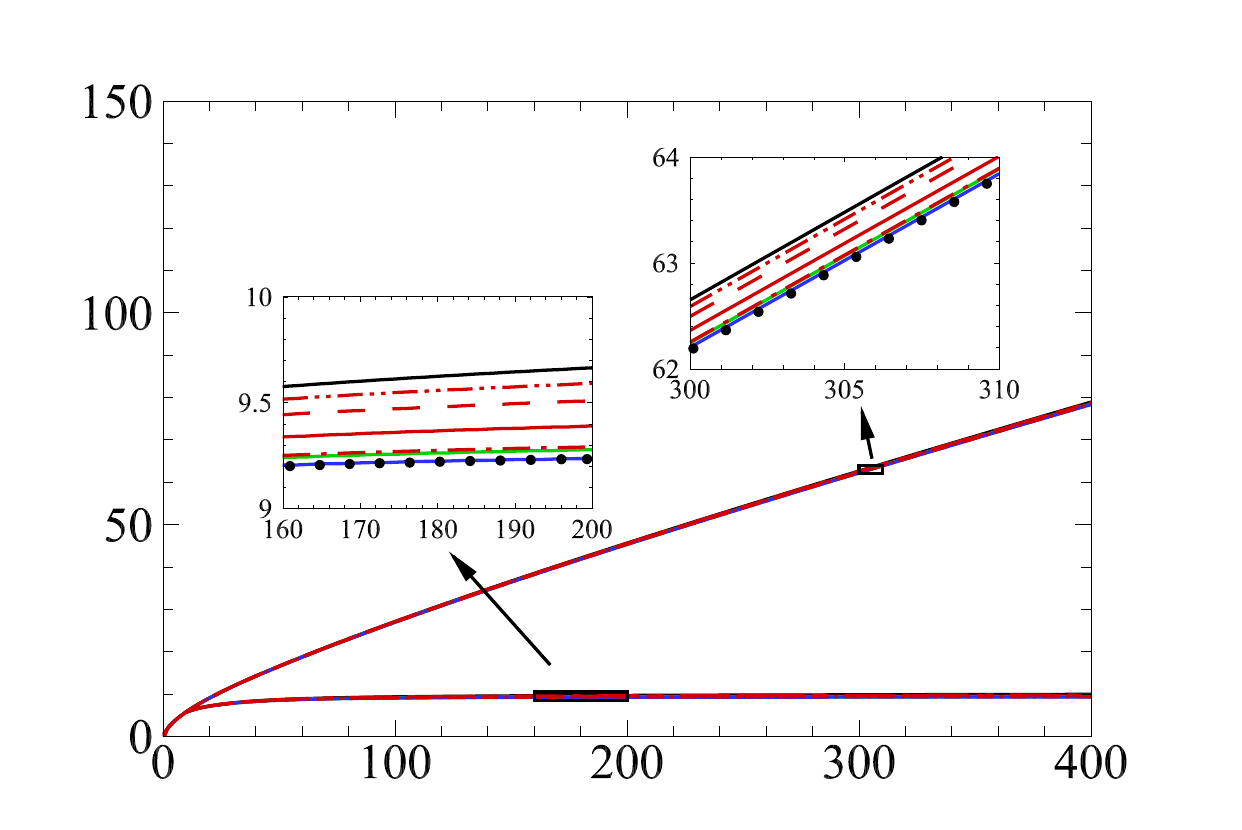}
  \includegraphics[width = 0.49\textwidth]{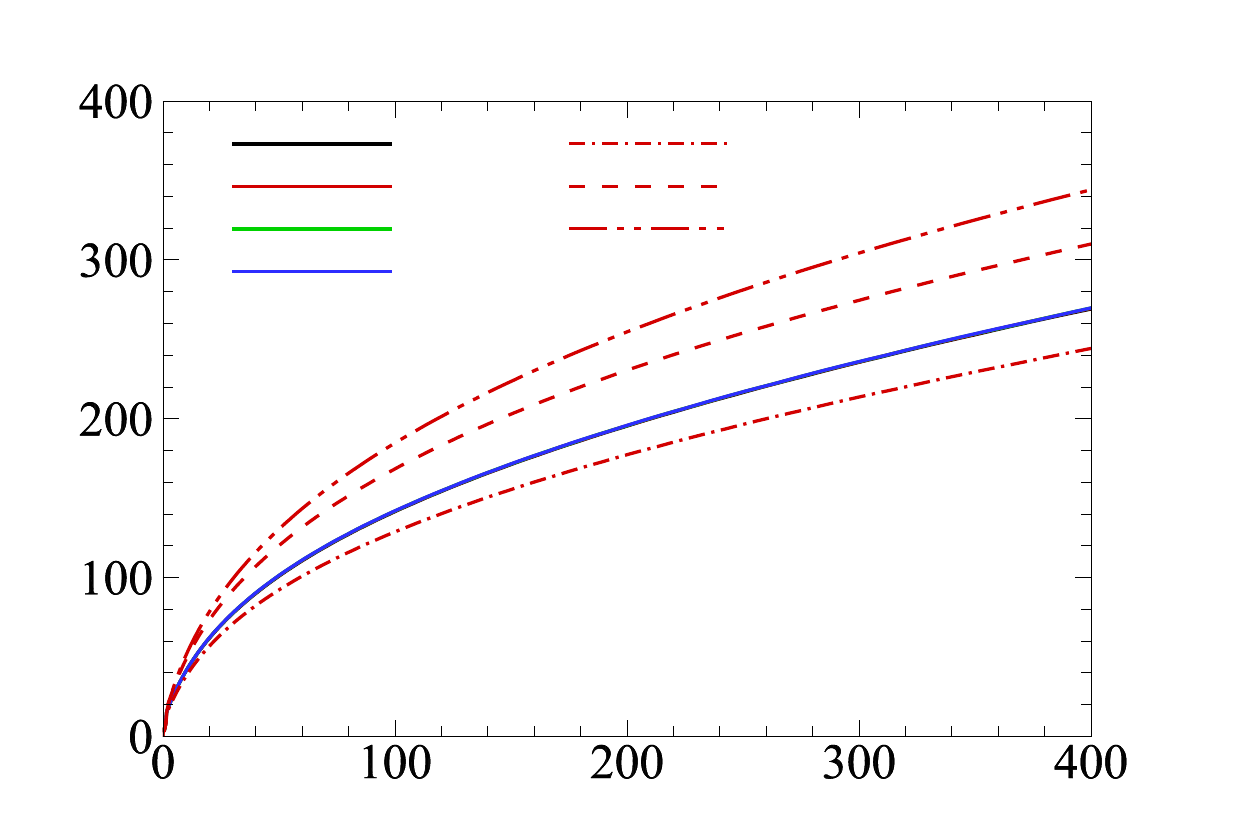}
  \put(-380,120) {$(a)$}
  \put(-190,120) {$(b)$}
  \put(-95,0) {$\xi$}
  \put(-285,0) {$\xi$}
  \put(-190,52){\rotatebox{90} {$\delta_{BL} R^{1/2} $}}
  \put(-375,50) {\rotatebox{90}{$\delta_{SH},\ {\delta_{EL}}$}}
  \put(-245,48) {$shock$}
  \put(-295,25) {$E.L. $}
  \put(-128,102) {\fontsize{6pt}{6pt}\selectfont$RAT2$}
  \put(-128,95)  {\fontsize{6pt}{6pt}\selectfont$RBT2$}
  \put(-128,88)  {\fontsize{6pt}{6pt}\selectfont$RCT2$}
  \put(-128,82)  {\fontsize{6pt}{6pt}\selectfont$RDT2$}
  \put(-78,102)  {\fontsize{6pt}{6pt}\selectfont$RBT1$}
  \put(-78,95)   {\fontsize{6pt}{6pt}\selectfont$RBT3$}
  \put(-78,88)   {\fontsize{6pt}{6pt}\selectfont$RBT4$}
  \\
  \includegraphics[width = 0.49\textwidth]{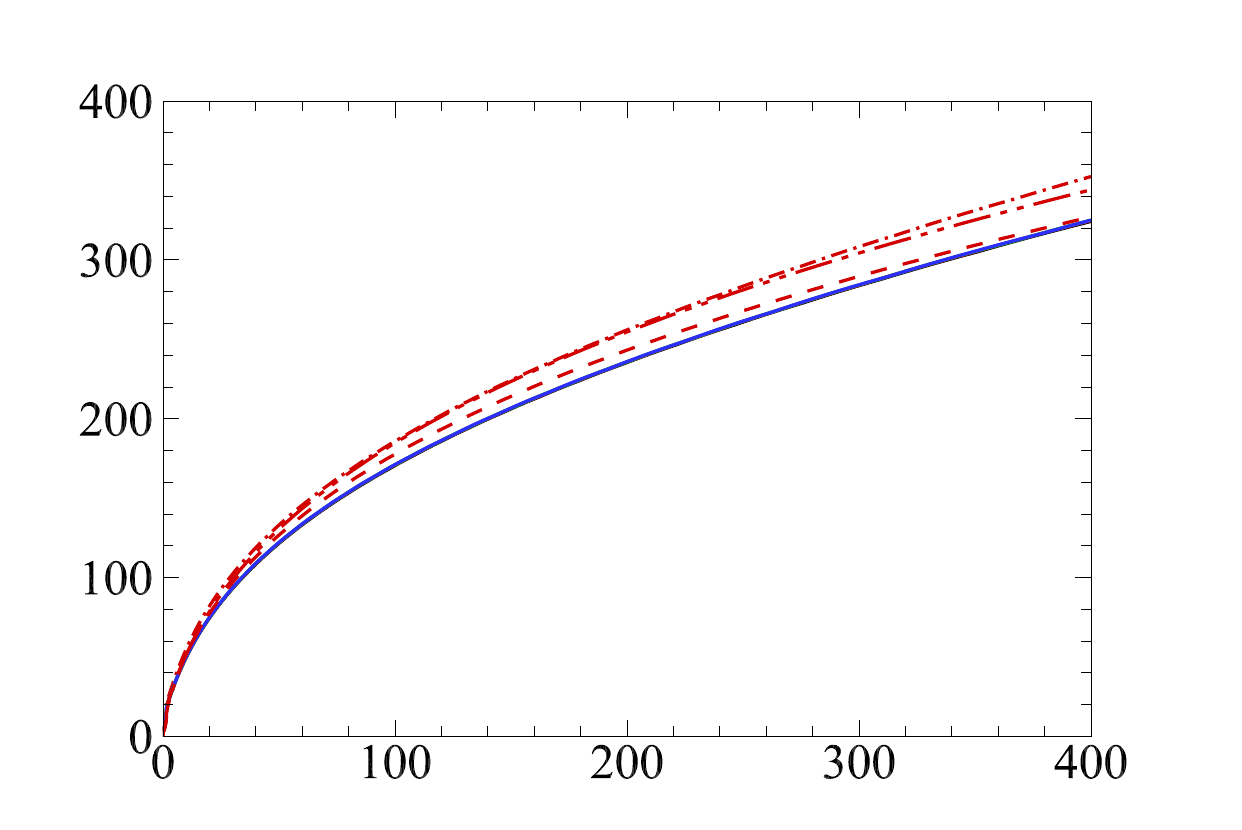}
  \includegraphics[width = 0.49\textwidth]{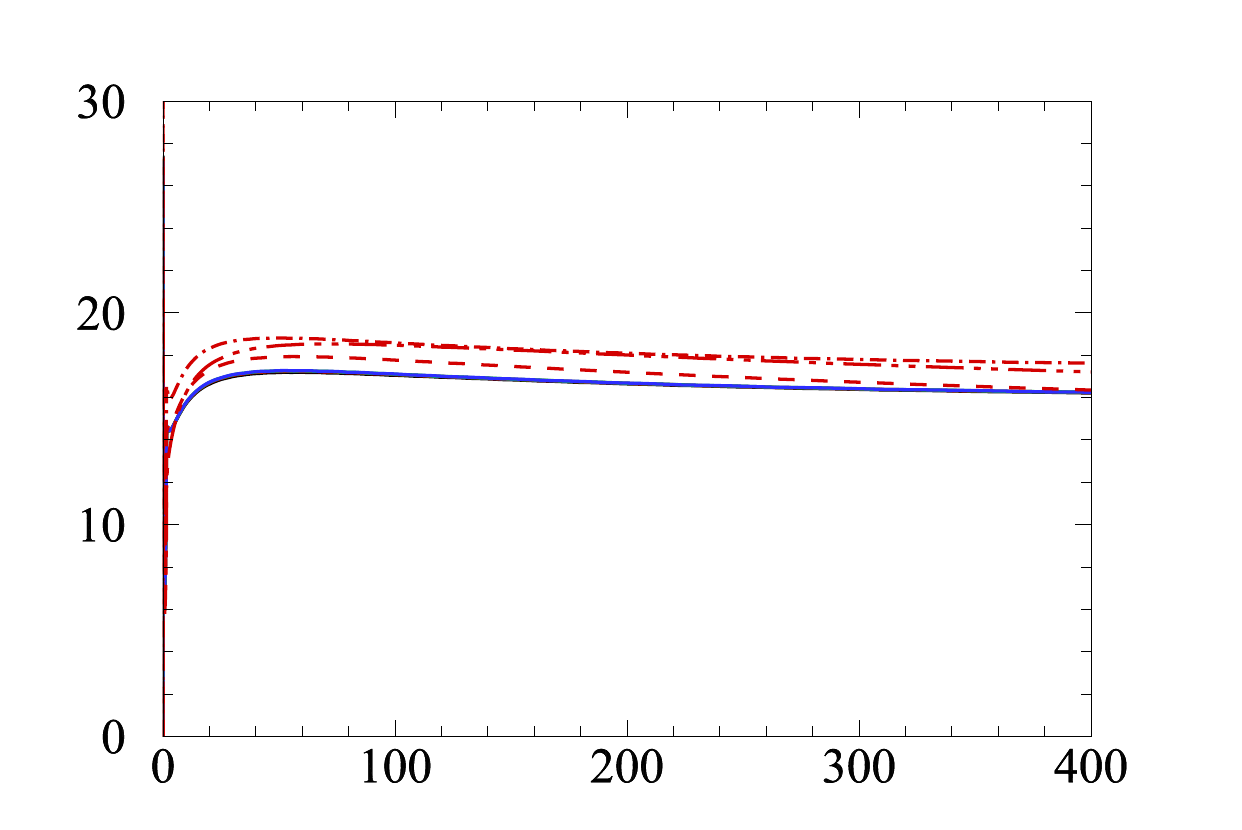}
  \put(-380,120) {$(c)$}
  \put(-190,120) {$(d)$}
  \put(-95,0) {$\xi$}
  \put(-285,0) {$\xi$}
  \put(-195,13){\rotatebox{90} {${\delta_{BL} R^{1/2}(T_w/T_{ad})^{-1/4}}/{\sqrt{\xi } } $}}
  \put(-380,20){\rotatebox{90} {$\delta_{BL}R^{1/2} (T_w/T_{ad})^{-1/4} $}}
  \caption{(a): Streamwise evolution of the thicknesses of the shock layer $\delta_{SH}$ and entropy layer $ {\delta_{EL}}$. (b), (c) and (d) display streamwise evolution of the boundary-layer thickness $\delta_{BL}$ scaled by \textcolor{red}{ $R^{-1/2}$,  $R^{-1/2}(T_w/T_{ad})^{1/4}$, and  $R^{-1/2}(T_w/T_{ad})^{1/4} \sqrt{\xi}$ }, respectively. The circles in (a) depict the Euler solutions.  }
  \label{fig:baseflow_shock_bl}
  \end{center}
\end{figure}

Figure \ref{fig:baseflow_shock_bl}-(a) displays the thicknesses of both the shock layer and entropy layer (defined as in ZD25). In the downstream region, the entropy layer remains  an almost constant thickness, while the shock layer maintains a constant oblique angle of around {10}$^\circ$. Comparisons across  different cases indicate that both thicknesses decrease as $R$ increases and $T_w$ decreases, approaching the solutions of the Euler equations (removing the viscous terms in the N-S equations) shown by the circles in the high-$R$ limit. In figure \ref{fig:baseflow_shock_bl}-(b), the boundary-layer thickness is normalised by $R^{-1/2}$ for all the cases. The curves for the same $T_w$ collapse well across different $R$ values, whereas those for different  $T_w$ remain  largely scattered, with the rescaled $\delta_{BL}$  increasing with $T_w$. Although no theoretical relation for   the dependence of $\delta_{BL}$ on $T_w$ is available,  a curve fit yields an approximate scaling of  $\delta_{BL}\sim (T_w/T_{ad})^{1/4}$, as shown in panel (c). Finally, in figure \ref{fig:baseflow_shock_bl}-(d), the boundary-layer thickness is re-normalised by $\sqrt{\xi}$. The resulting distributions are nearly constant in the downstream region, indicating a self-similar {\color{red}behaviour}.

\begin{figure}
  \begin{center}
  \includegraphics[width = 0.49\textwidth]{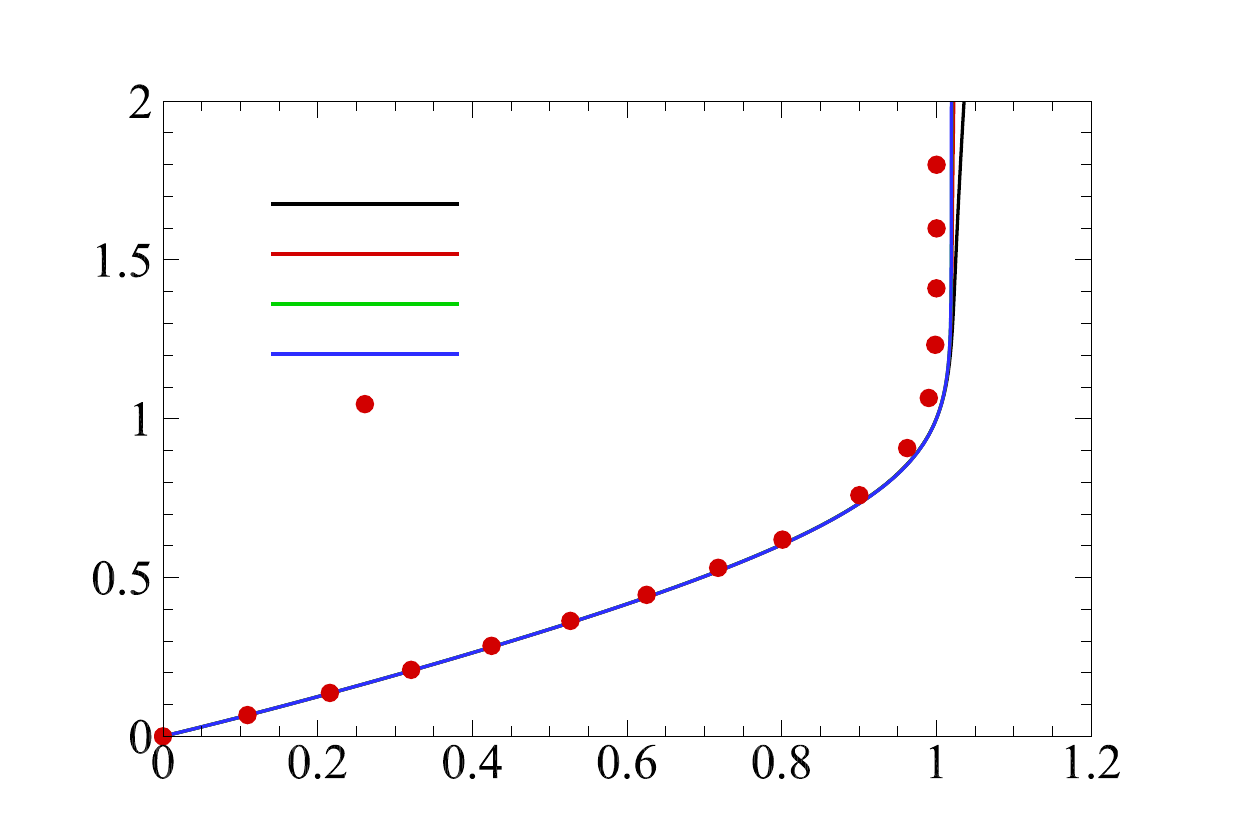}
  \includegraphics[width = 0.49\textwidth]{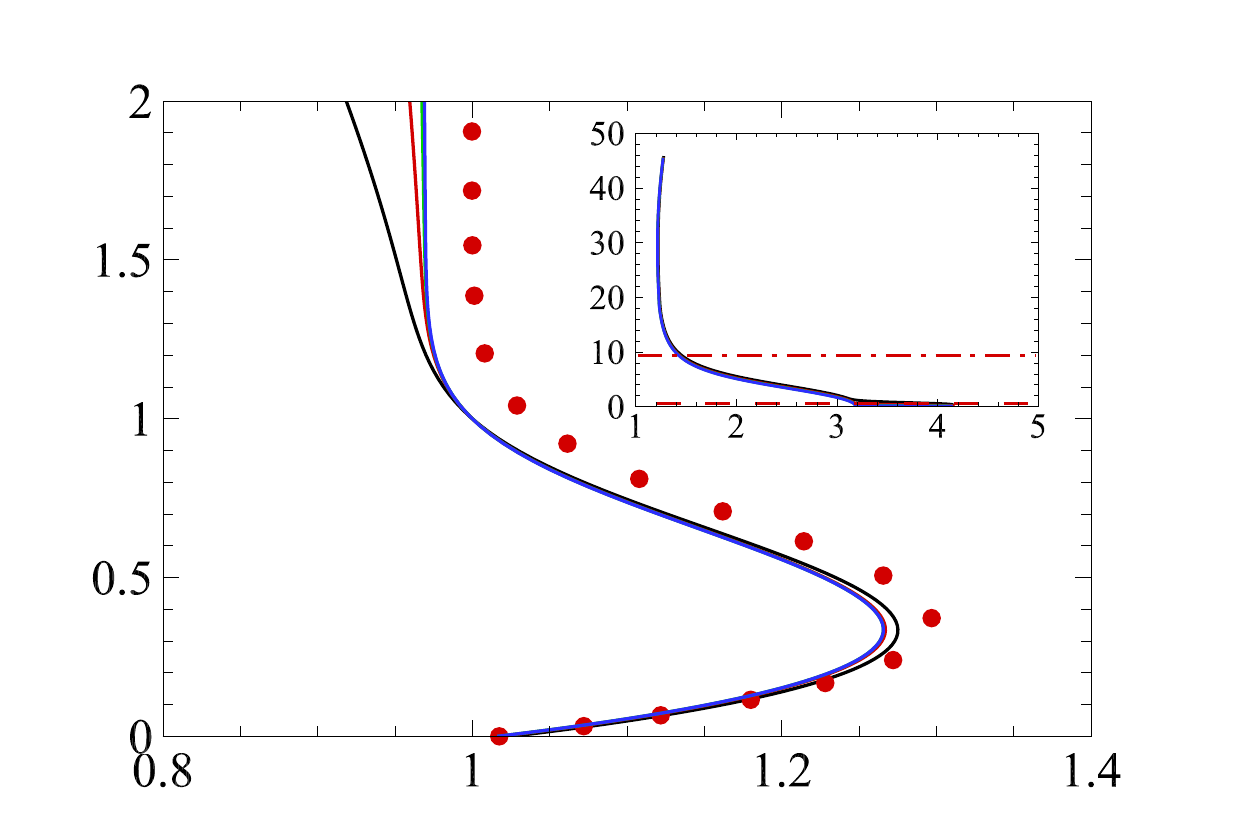}
  \put(-380,120) {$(a)$}
  \put(-190,120) {$(b)$}
  \put(-100,-2) {$\bar{T}/ \bar{T}_e$}
  \put(-290,-2) {$\bar{u}_\xi/ \bar{u}_{e}$}
  \put(-370,50) {\rotatebox{90} {$\eta / \delta_{BL}$}}
  \put(-180,50) {\rotatebox{90} {$\eta / \delta_{BL}$}}
  \put(-65,50) {$\bar{T}$}
  \put(-105,83) {$\eta$}
  \put(-45,65)  {\fontsize{6pt}{6pt}\selectfont$B.L.$}
  \put(-45,73)  {\fontsize{6pt}{6pt}\selectfont$E.L.$}
  \put(-308,93) {$RAT2$}
  \put(-308,85) {$RBT2$}
  \put(-308,77) {$RCT2$}
  \put(-308,69) {$RDT2$}
  \put(-308,61) {$Blasius$}\\
  \includegraphics[width = 0.49\textwidth]{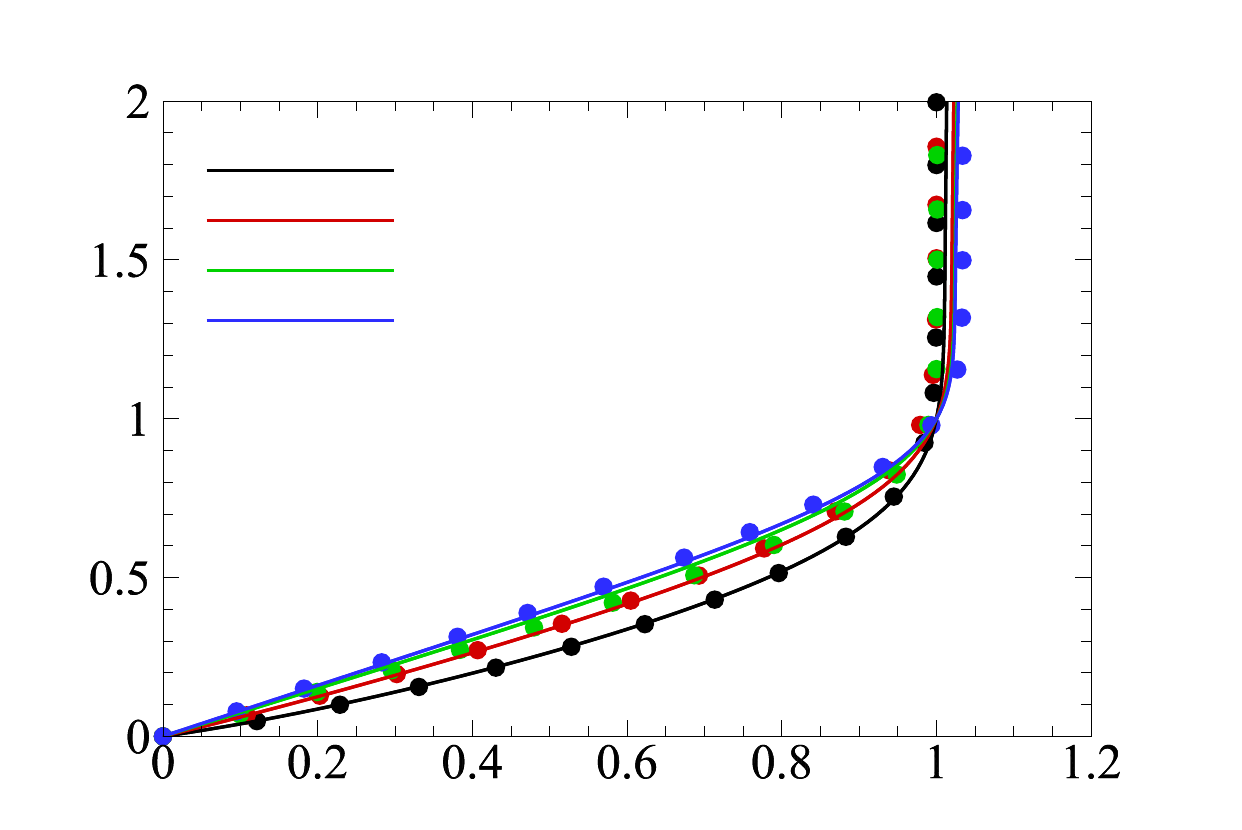}
  \includegraphics[width = 0.49\textwidth]{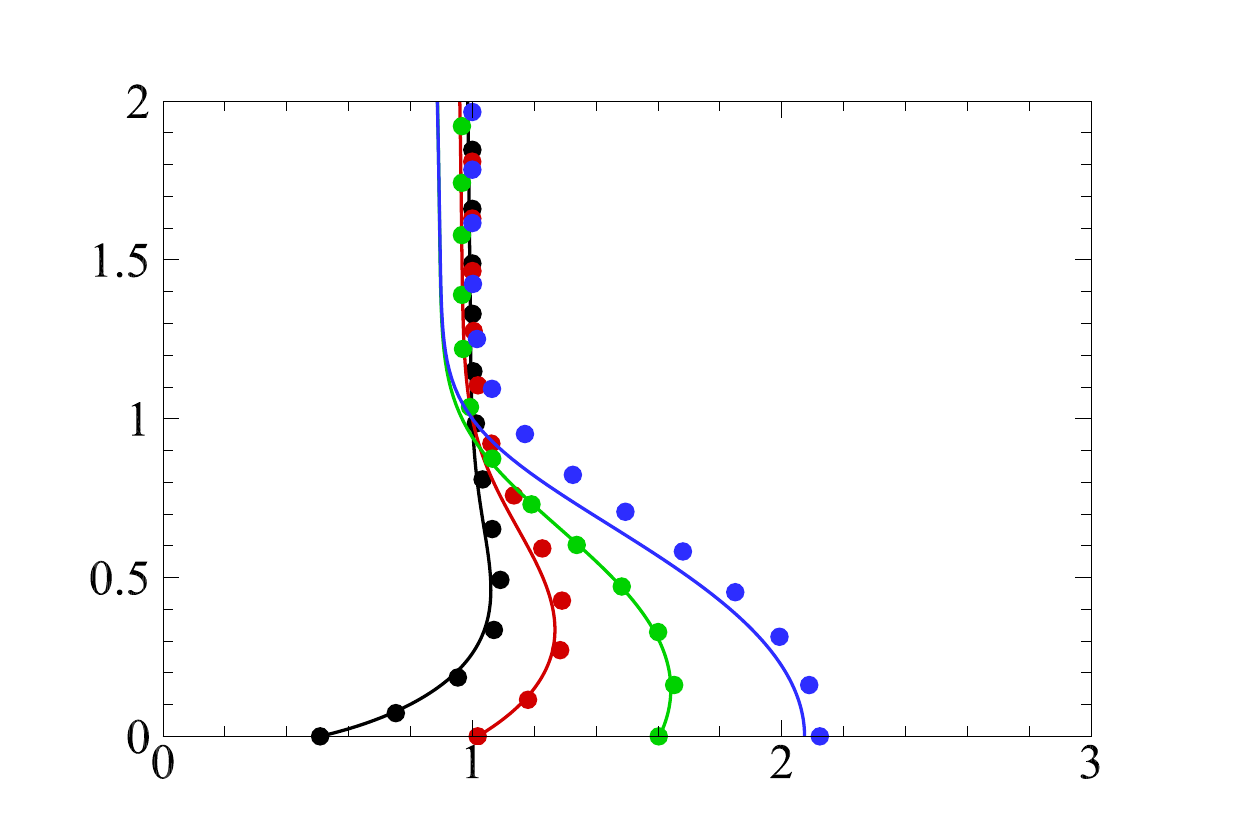}
  \put(-380,120) {$(c)$}
  \put(-190,120) {$(d)$}
  \put(-100,-2) {$\bar{T}/ \bar{T}_e$}
  \put(-290,-2) {$\bar{u}_\xi/ \bar{u}_e$}
  \put(-370,50) {\rotatebox{90} {$\eta / \delta_{BL}$}}
  \put(-180,50) {\rotatebox{90} {$\eta / \delta_{BL}$}}
  \put(-318,98)  {$RBT1$}
  \put(-318,90)  {$RBT2$}
  \put(-318,82)  {$RBT3$}
  \put(-318,74)  {$RBT4$}
  \caption{ Profiles of   tangential velocity $\bar{u}_\xi/ \bar{u}_e$  (left column) and  temperature $\bar{T}/ \bar{T}_e$ (right column) at $\xi = 200$. (a,b): cases RAT2, RBT2, RCT2, and RDT2; (c,d): cases RBT1, RBT2, RBT3 and RBT4. Here, the tangential velocity and temperature at the boundary-layer edge are $\bar u_e\approx 0.82$ and $\bar T_e\approx 3.3$, respectively. The solid lines and circles denote the DNS results and the Blasius solutions, respectively. The in-box of (b) shows the $\bar T$ profiles in the region between the shock position and the wall, with E.L. and B.L. denoting the entropy and boundary layers, respectively.}
  \label{fig:baseflow_profile_x=200_R}
  \end{center}
\end{figure}

It is expected that in the sufficiently downstream region, the blunt-nose effect  becomes negligible, and the rescaled profiles would approach the compressible Blasius similarity solution. To confirm this, we choose  a downstream location at $\xi=200$, and compare the profiles of $\bar u_\xi(\eta/\delta_{BL})/\bar u_e$ and $\bar T(\eta/\delta_{BL})\bar T_e$ for cases that share the same $T_w$ but differ in $R$, as shown in {figures} \ref{fig:baseflow_profile_x=200_R}-(a) and (b). Here, $\bar u_e$ and $\bar T_e$ denote the tangential velocity and  temperature at the boundary-layer edge, respectively. The profiles of $\bar u_\xi/\bar u_e$ almost overlap  within the boundary-layer region,  aligning closely with the  Blasius  solution. In contrast, the temperature profiles display a slight discrepancy from the Blasius solution for $\eta/\delta_{BL}>0.4$. Moreover, the temperature remains a mild wall-normal gradient even in the potential region above the boundary-layer edge, which is due to the entropy-layer effect. The in-box of panel (b) further shows the $\bar T$ profiles in the region between the wall and the shock position, revealing a noticeable increase from the post-shock value $\bar T_s\approx 1.23$ to the boundary-layer-edge value $T_e\approx 3.3$. As case RAT2 corresponds to a thicker boundary layer, its temperature gradient immediately above the boundary layer is greater than other cases.
{Figures} \ref{fig:baseflow_profile_x=200_R}-(c) and (d) further compare the downstream profiles for cases with the same $R$ but different $T_w$. Again, the numerical profiles of $\bar u_\xi/\bar u_e$ match the Blasius solutions across the boundary-layer region, while the $\bar T/\bar T_e$ profiles show {\color{red}slight} deviations from the Blasius solutions for $\eta/\delta_{BL}>$0.4.

\subsection{SF-HLNS calculations for excitation of non-modal perturbations}
\begin{figure}
  \begin{center}
  \includegraphics[width =0.95\textwidth]{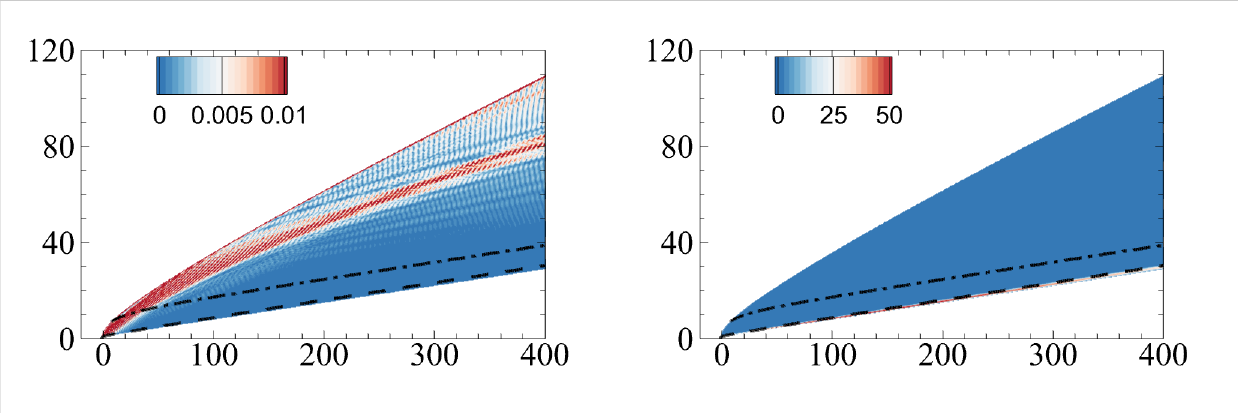}
  \put(-370,110) {$(a)$}
  \put(-190,110) {$(b)$}
  \put(-153,95) {$\left | \tilde{u}_{\xi}\right |  $}
  \put(-332,95) {$\left | \tilde{p} \right |  $}
  \put(-270,5) {$x$}
  \put(-360,60) {$y$}
  \put(-90,5) {$x$}
  \put(-180,60) {$y$}
  \\
  \includegraphics[width =0.225\textwidth]{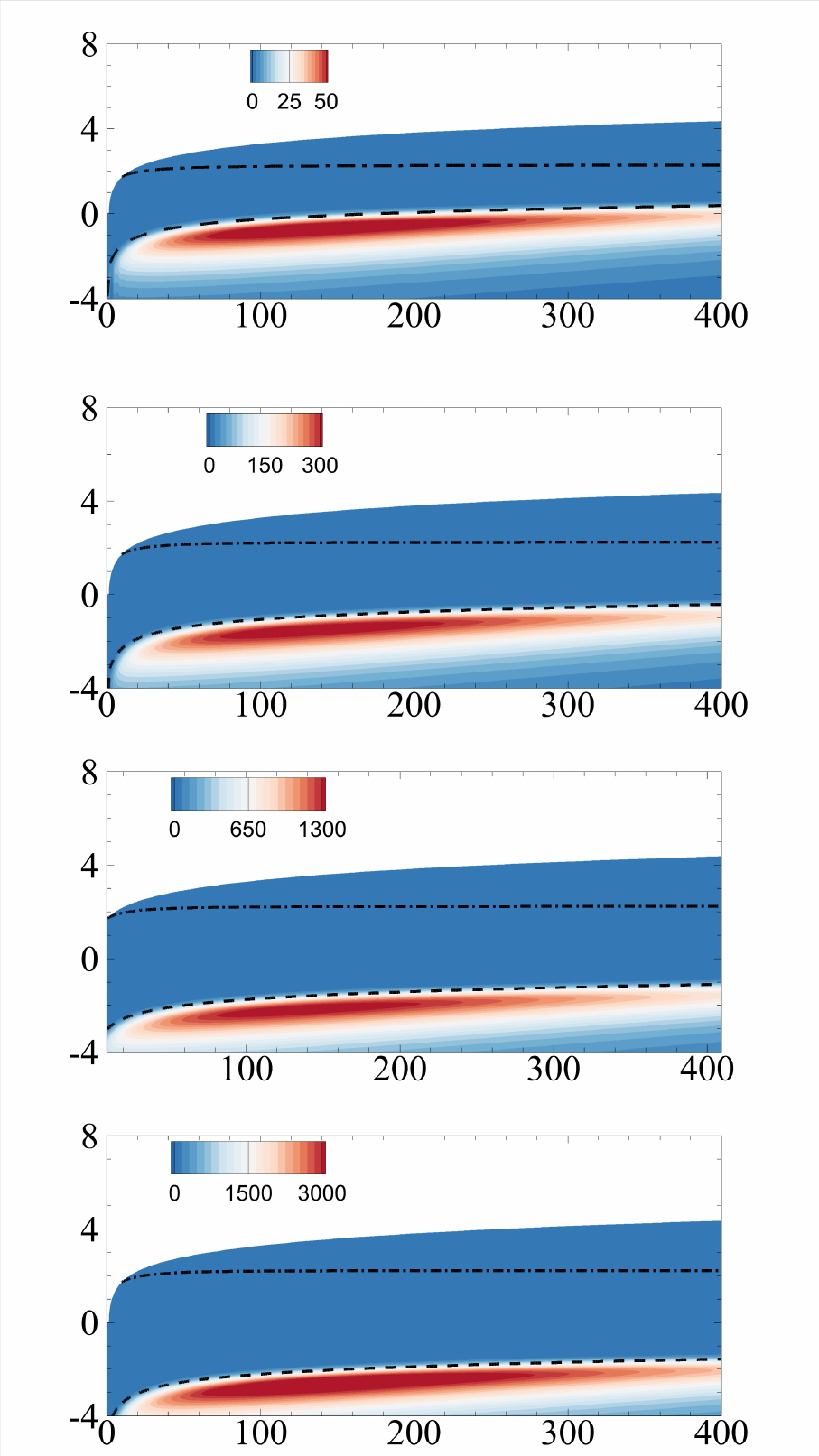}
  \includegraphics[width =0.225\textwidth]{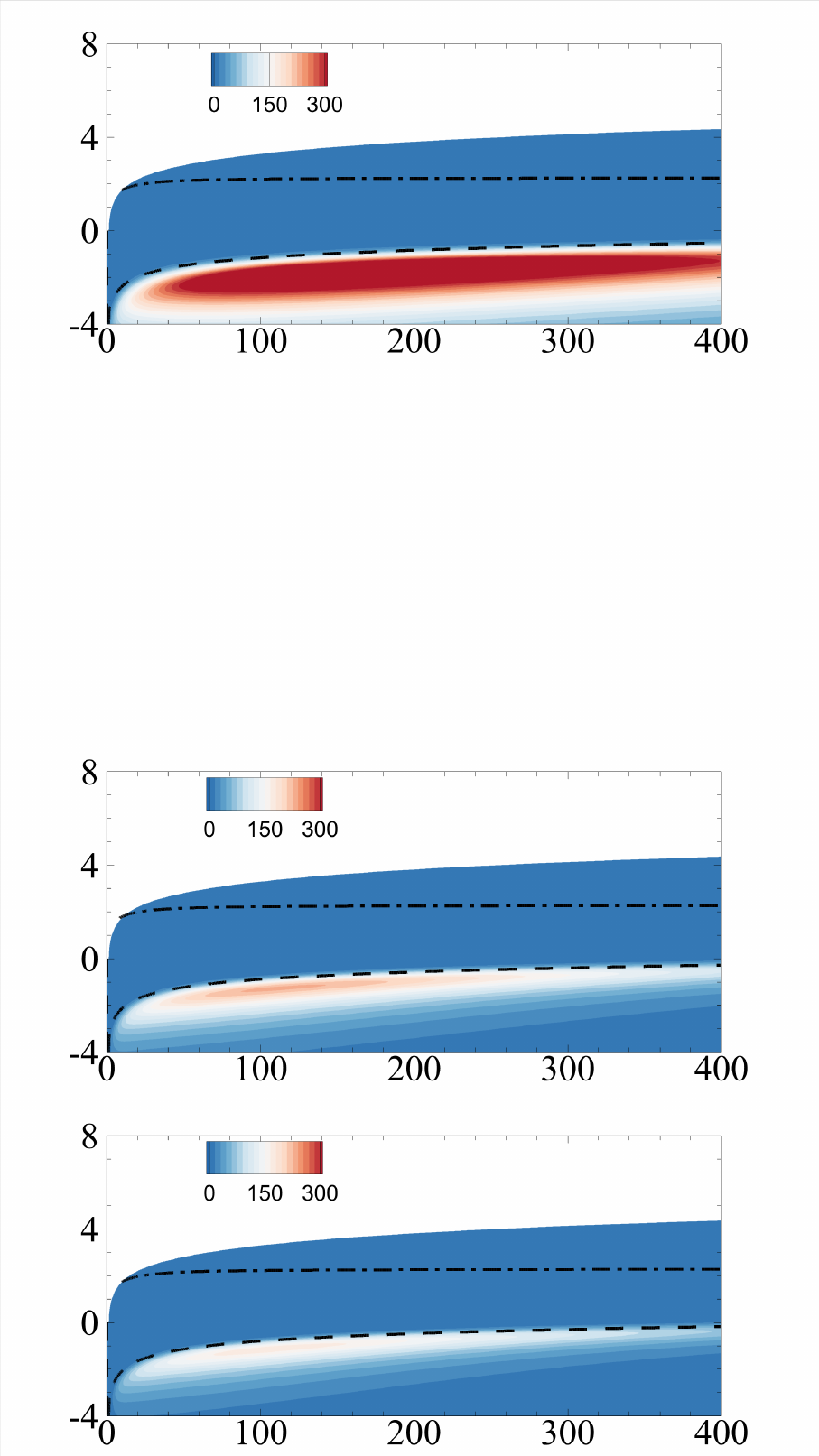}
  \includegraphics[width =0.225\textwidth]{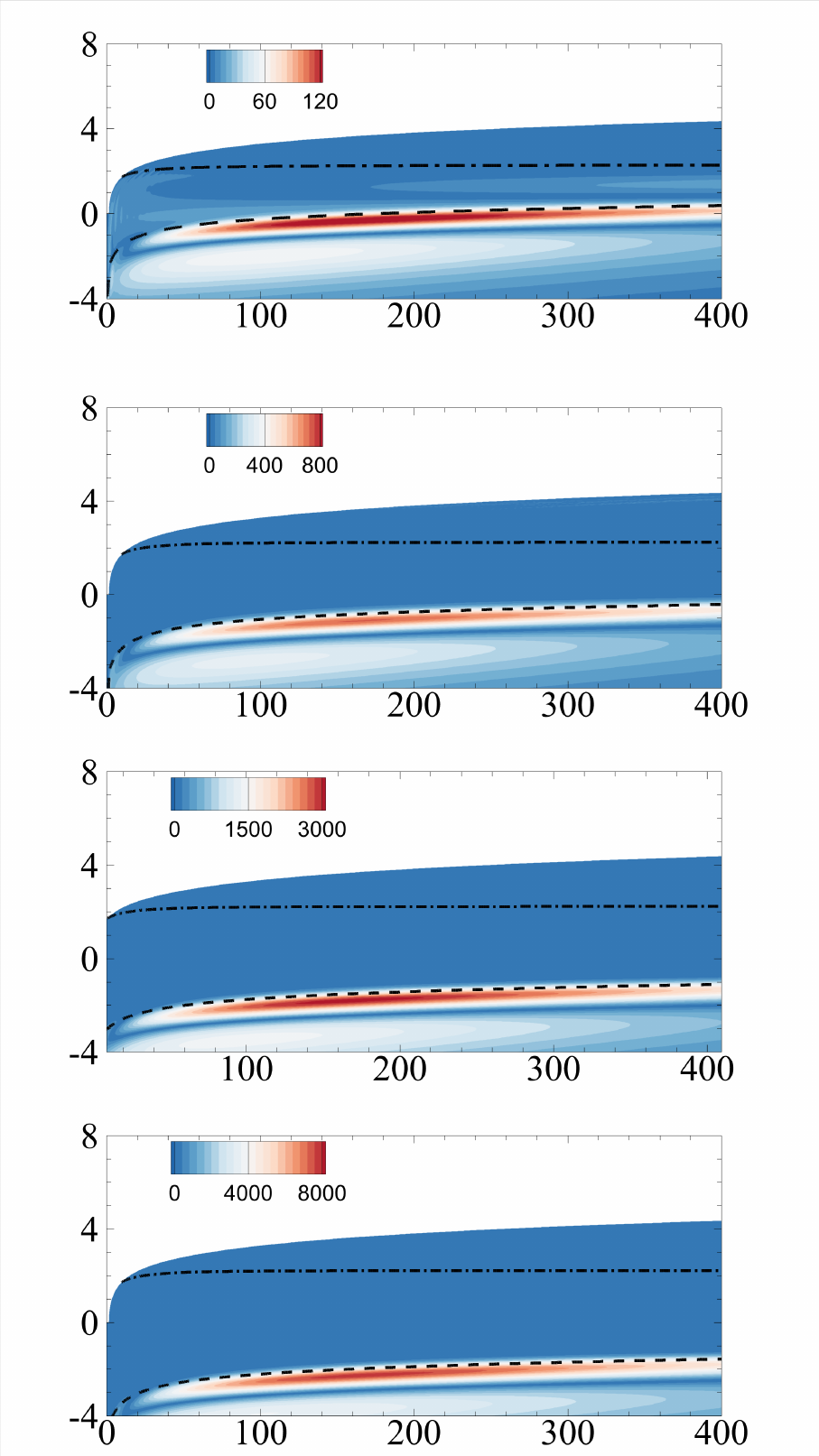}
  \includegraphics[width =0.225\textwidth]{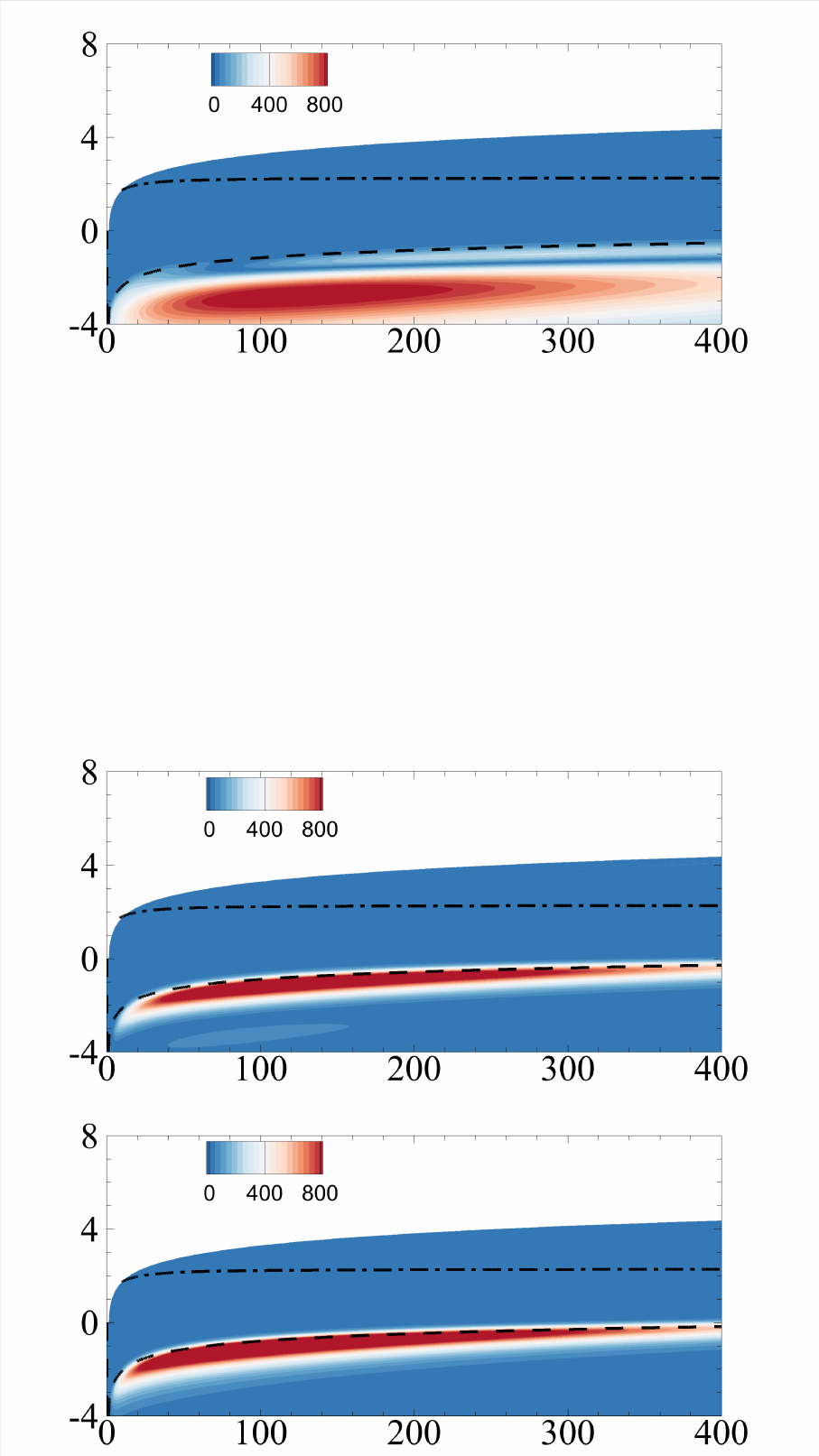}
  \put(-370,150) {$(c)$}
  \put(-180,150) {$(d)$}
  \put(-117,143)  {\fontsize{6pt}{6pt}\selectfont$RAT2$}
  \put(-117,105)  {\fontsize{6pt}{6pt}\selectfont$RBT2$}
  \put(-117,65)   {\fontsize{6pt}{6pt}\selectfont$RCT2$}
  \put(-117,25)   {\fontsize{6pt}{6pt}\selectfont$RDT2$}
  \put(-27,143)  {\fontsize{6pt}{6pt}\selectfont$RBT1$}
  \put(-27,65)   {\fontsize{6pt}{6pt}\selectfont$RBT3$}
  \put(-27,25)   {\fontsize{6pt}{6pt}\selectfont$RBT4$}
  \put(-295,143)  {\fontsize{6pt}{6pt}\selectfont$RAT2$}
  \put(-295,105)  {\fontsize{6pt}{6pt}\selectfont$RBT2$}
  \put(-295,65)   {\fontsize{6pt}{6pt}\selectfont$RCT2$}
  \put(-295,25)   {\fontsize{6pt}{6pt}\selectfont$RDT2$}
  \put(-205,143)  {\fontsize{6pt}{6pt}\selectfont$RBT1$}
  \put(-205,65)   {\fontsize{6pt}{6pt}\selectfont$RBT3$}
  \put(-205,25)   {\fontsize{6pt}{6pt}\selectfont$RBT4$}
  \put(-311,-6) {$\xi$}
  \put(-221,-6) {$\xi$}
  \put(-133,-6) {$\xi$}
  \put(-43,-6) {$\xi$}
  \put(-360,15) {ln$\eta$}
  \put(-360,55) {ln$\eta$}
  \put(-360,95) {ln$\eta$}
  \put(-360,135) {ln$\eta$}
  \put(-138,28) {\fontsize{6pt}{6pt}\selectfont$\left | \tilde{T} \right |  $}
  \put(-315,28) {\fontsize{6pt}{6pt}\selectfont$\left | \tilde{u}_{\xi}  \right |  $}
  \put(-138,65) {\fontsize{6pt}{6pt}\selectfont$\left | \tilde{T} \right |  $}
  \put(-315,65) {\fontsize{6pt}{6pt}\selectfont$\left | \tilde{u}_{\xi}  \right |  $}
  \put(-138,105) {\fontsize{6pt}{6pt}\selectfont$\left | \tilde{T} \right |  $}
  \put(-315,105) {\fontsize{6pt}{6pt}\selectfont$\left | \tilde{u}_{\xi}  \right |  $}
  \put(-138,143) {\fontsize{6pt}{6pt}\selectfont$\left | \tilde{T} \right |  $}
  \put(-315,143) {\fontsize{6pt}{6pt}\selectfont$\left | \tilde{u}_{\xi}  \right |  $}
  \put(-50,28) {\fontsize{6pt}{6pt}\selectfont$\left | \tilde{T} \right |  $}
  \put(-228,28) {\fontsize{6pt}{6pt}\selectfont$\left | \tilde{u}_{\xi}  \right |  $}
  \put(-50,65) {\fontsize{6pt}{6pt}\selectfont$\left | \tilde{T} \right |  $}
  \put(-228,65) {\fontsize{6pt}{6pt}\selectfont$\left | \tilde{u}_{\xi}  \right |  $}
  \put(-50,143) {\fontsize{6pt}{6pt}\selectfont$\left | \tilde{T} \right |  $}
  \put(-228,143) {\fontsize{6pt}{6pt}\selectfont$\left | \tilde{u}_{\xi}  \right |  $}
  \caption{
   {Contours of the (a) $\left | \tilde{p} \right |  $ and  (b) $\left | \tilde{u}_{\xi} \right |  $ in the $x$-$y$ plane for case RAT2. Contours of the (c) $\left | \tilde{u}_{\xi} \right |  $ and (d) $\left | \tilde{T} \right |  $ in the $\xi$-ln$\eta$ plane for all cases.}   The freestream forcing is an entropy wave with $\omega =0,\vartheta= 0^{\circ} ,k_{3}R^{-1/2} =0.02$. The dashed and dash-dotted lines mark the boundary-layer edge and entropy layer edge, respectively.}
  \label{fig:HLNS_Contours}
  \end{center}
\end{figure}

Using SF-HLNS approach, we can calculate the excitation of non-modal perturbations subject to various freestream forcing. The detailed numerical results for case RBT2 forced by all types of freestream perturbations can be found in ZD25 (Part I), whereas those for case RBT4 can be found in Part II. In figure \ref{fig:HLNS_Contours}-(a), we present the pressure perturbation forced by a freestream entropy wave with zero frequency for case RAT2, showing an acoustic beam radiating to the potential region.
The contours of $\tilde u_\xi$, displayed in panel (b), reveal high-level perturbations confined mainly within the boundary-layer region. To visualise the perturbation {\color{red}behaviour} more clearly, panel (c) shows the contours of $\tilde u_\xi$ in the  $\xi-\ln\eta$ plane for all cases.  The excited non-modal perturbation exhibits a well‑defined peak inside the bulk boundary-layer region. This peak initially intensifies with increasing $\xi$, followed by an attenuation after $\xi\approx 140$.
Notably, for the same wall temperature, $T_w=3.33$, the magnitude of $\tilde u_\xi$ increases monotonically with $R$. When $R$ is fixed at a moderate value, lowering the wall temperature leads to a stronger perturbation amplitude.  In contrast, the structure of the perturbation temperature contours, as shown in panel (d), differs markedly.  For $T_w=3.33$ (cases RAT2 to RDT2), $\tilde T$ {\color{black}displays two distinct peaks in the wall-normal direction across all Reynolds numbers: a dominant peak located at the boundary-layer edge (indicated by the dashed lines) and a secondary peak in the near-wall region.}
 The magnitude of $\tilde T$ grows significantly with $R$. For a fixed $R=1.67\times 10^5$, the distribution of the temperature perturbation varies substantially with $T_w$.  At higher wall temperatures (cases RBT3 and RBT4), {\color{black}the dominant peak near the boundary-layer edge intensifies while the secondary peak diminishes. Conversely, at a very low temperature (case RBT1), the near-wall peak becomes dominant.} 

\begin{figure}
  \begin{center}
  \includegraphics[width = 0.49\textwidth]{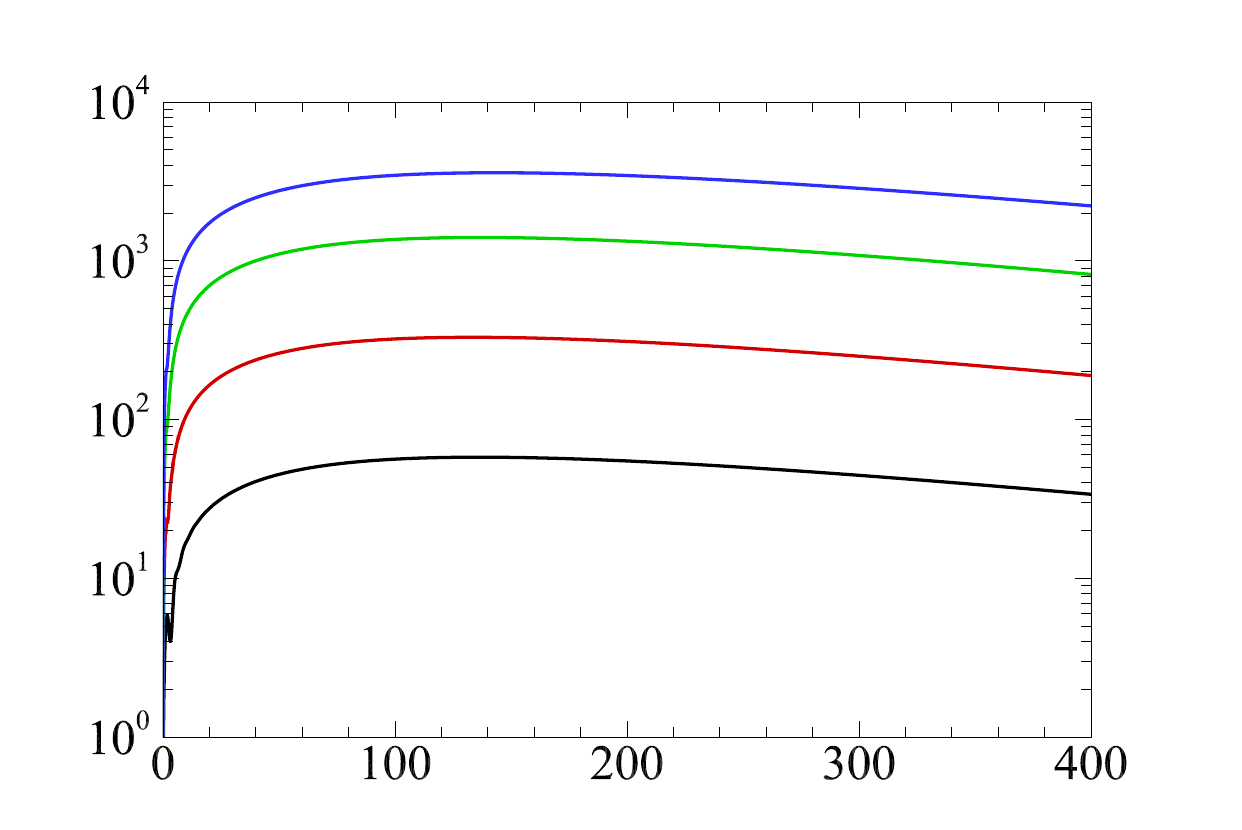}
  \includegraphics[width = 0.49\textwidth]{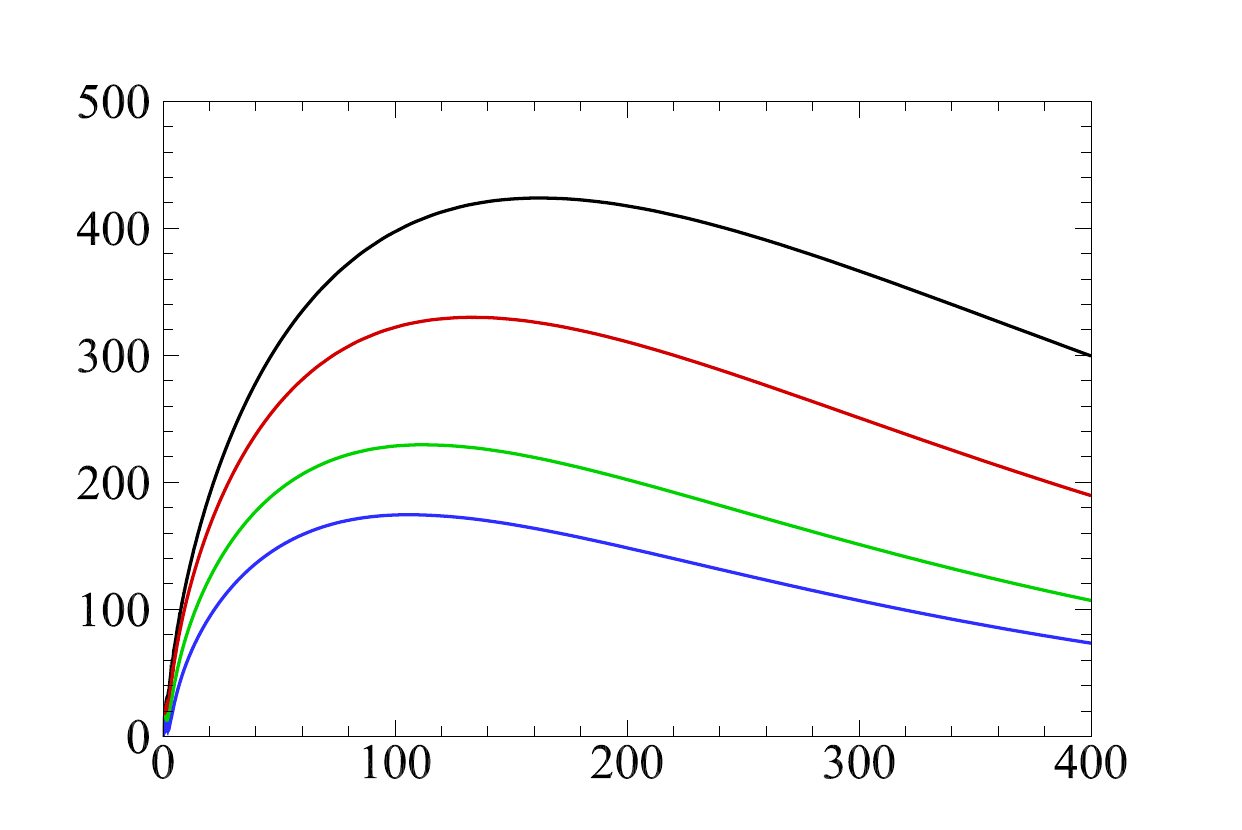}
 \put(-380,120) {$(a)$}
  \put(-190,120) {$(b)$}
  \put(-95,0) {$\xi$}
  \put(-285,0) {$\xi$}
  \put(-380,55) {\rotatebox{90}{$A_u(\xi)$ }}
  \put(-254,36)   {\fontsize{5pt}{5pt}$RAT2$}
  \put(-254,30)   {\fontsize{5pt}{5pt}$RBT2$}
  \put(-254,24)   {\fontsize{5pt}{5pt}$RCT2$}
  \put(-254,18)   {\fontsize{5pt}{5pt}$RDT2$}
  \put(-272,20){\begin{tikzpicture}
    \draw[black,thick] (0,0) -- (0.5,0);
    \draw[red,thick] (0,-0.18) -- (0.5,-0.18);
    \draw[green,thick] (0,-0.36) -- (0.5,-0.36);
    \draw[blue,thick] (0,-0.54) -- (0.5,-0.54);
  \end{tikzpicture}}
  \put(-112,36)  {\fontsize{5pt}{5pt}$RBT1$}
  \put(-112,30)   {\fontsize{5pt}{5pt}$RBT2$}
  \put(-112,24)   {\fontsize{5pt}{5pt}$RBT3$}
  \put(-112,18)   {\fontsize{5pt}{5pt}$RBT4$}
  \put(-130,20){\begin{tikzpicture}
    \draw[black,thick] (0,0) -- (0.5,0);
    \draw[red,thick] (0,-0.18) -- (0.5,-0.18);
    \draw[green,thick] (0,-0.36) -- (0.5,-0.36);
    \draw[blue,thick] (0,-0.54) -- (0.5,-0.54);
  \end{tikzpicture}}
  \\
  \includegraphics[width = 0.49\textwidth]{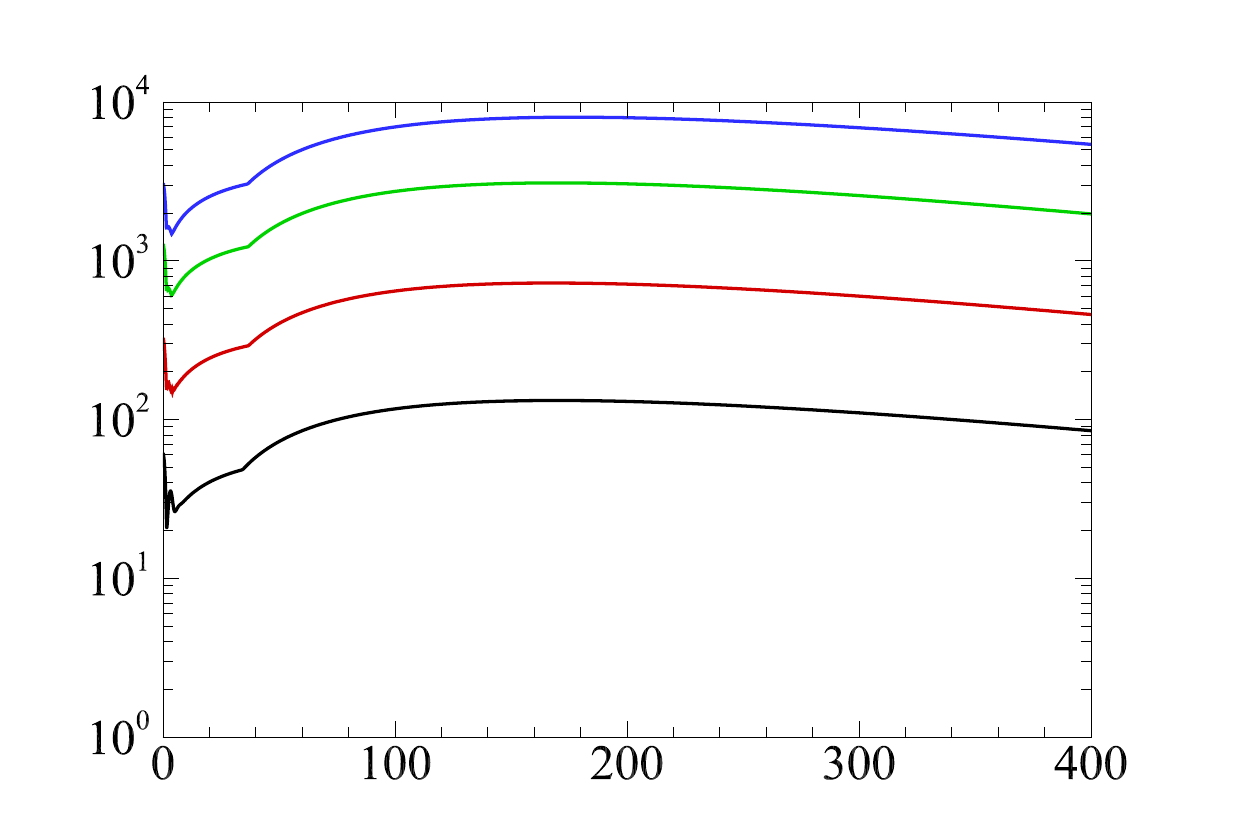}
  \includegraphics[width = 0.49\textwidth]{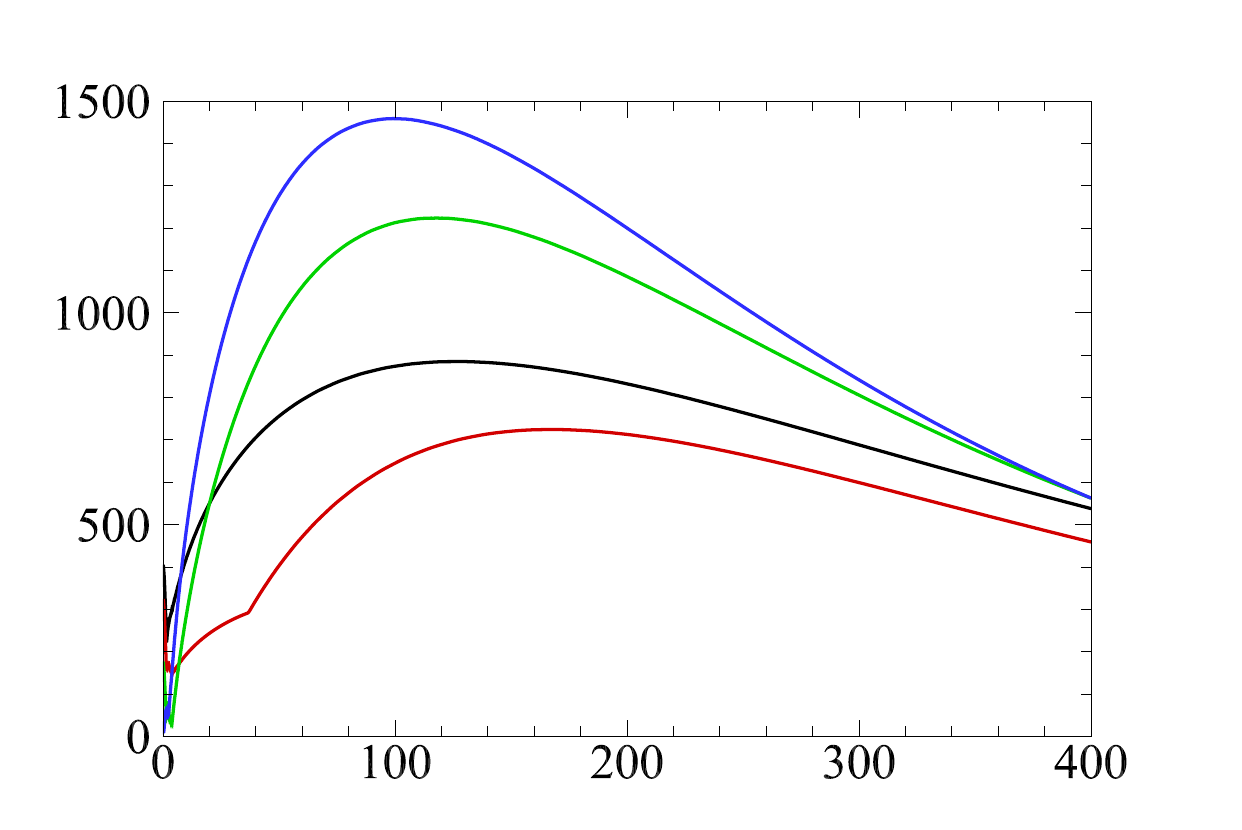}
  \put(-380,120) {$(c)$}
  \put(-190,120) {$(d)$}
  \put(-95,0) {$\xi$}
  \put(-285,0) {$\xi$}
  \put(-380,55) {\rotatebox{90}{$A_T(\xi)$ }}
  \put(-254,36)   {\fontsize{5pt}{5pt}$RAT2$}
  \put(-254,30)   {\fontsize{5pt}{5pt}$RBT2$}
  \put(-254,24)   {\fontsize{5pt}{5pt}$RCT2$}
  \put(-254,18)   {\fontsize{5pt}{5pt}$RDT2$}
  \put(-272,20){\begin{tikzpicture}
    \draw[black,thick] (0,0) -- (0.5,0);
    \draw[red,thick] (0,-0.18) -- (0.5,-0.18);
    \draw[green,thick] (0,-0.36) -- (0.5,-0.36);
    \draw[blue,thick] (0,-0.54) -- (0.5,-0.54);
  \end{tikzpicture}}
  \put(-55,96)  {\fontsize{5pt}{5pt}$RBT1$}
  \put(-55,90)   {\fontsize{5pt}{5pt}$RBT2$}
  \put(-55,84)   {\fontsize{5pt}{5pt}$RBT3$}
  \put(-55,78)   {\fontsize{5pt}{5pt}$RBT4$}
  \put(-72,80){\begin{tikzpicture}
    \draw[black,thick] (0,0) -- (0.5,0);
    \draw[red,thick] (0,-0.18) -- (0.5,-0.18);
    \draw[green,thick] (0,-0.36) -- (0.5,-0.36);
    \draw[blue,thick] (0,-0.54) -- (0.5,-0.54);
  \end{tikzpicture}}
  \\
  \includegraphics[width = 0.49\textwidth]{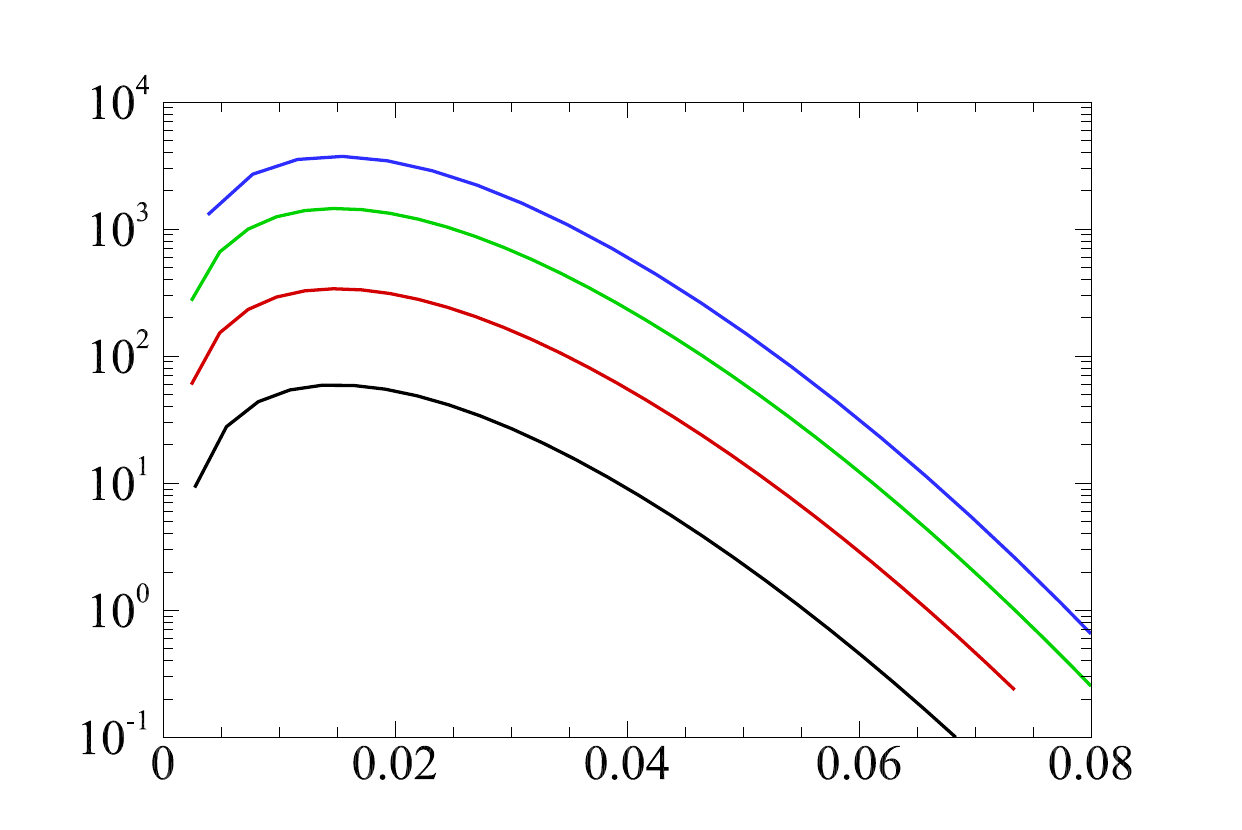}  
  \includegraphics[width = 0.49\textwidth]{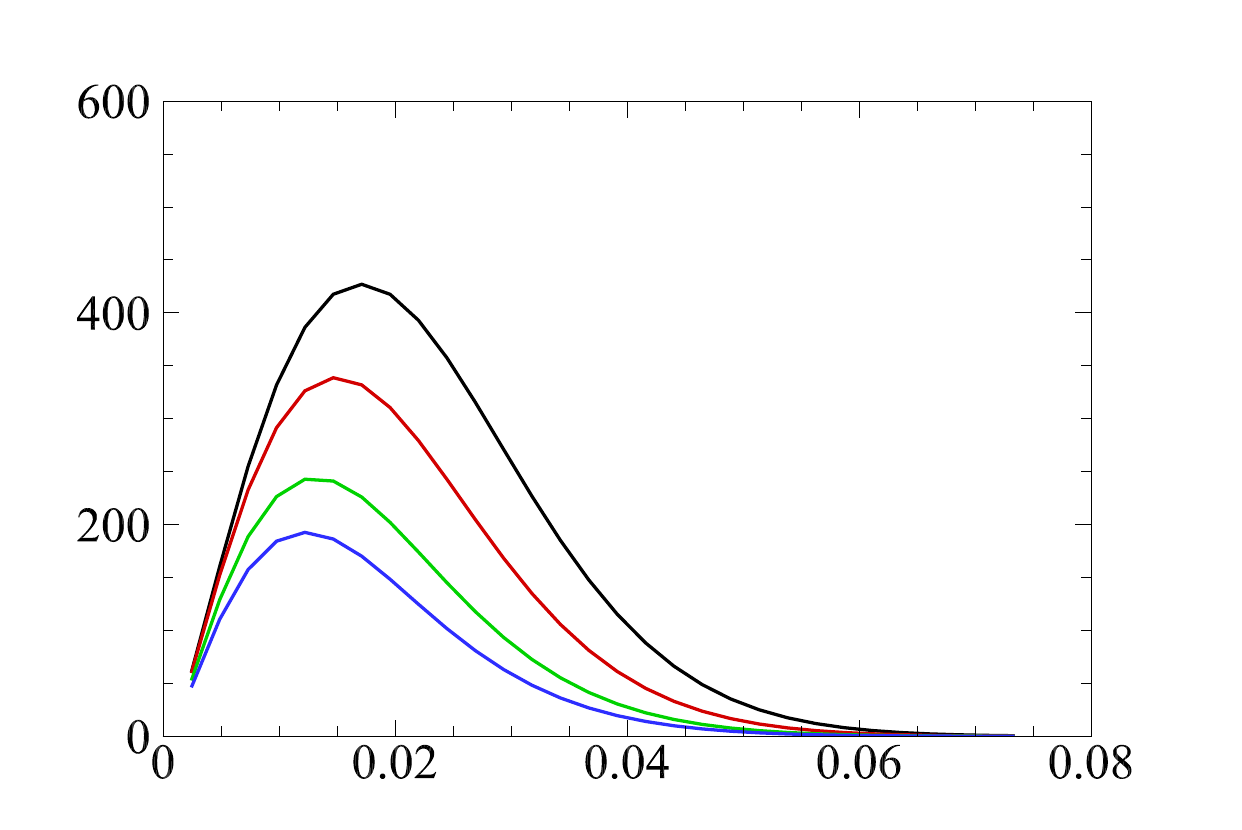}
  \put(-380,120) {$(e)$}
  \put(-190,120) {$(f)$}
  \put(-380,55) {\rotatebox{90}{$A_u(200)$ }}
  \put(-100,-2) {$k_3 R^{-1/2}$}
  \put(-290,-2) {$k_3 R^{-1/2}$}
  \put(-295,36)   {\fontsize{5pt}{5pt}$RAT2$}
  \put(-295,30)   {\fontsize{5pt}{5pt}$RBT2$}
  \put(-295,24)   {\fontsize{5pt}{5pt}$RCT2$}
  \put(-295,18)   {\fontsize{5pt}{5pt}$RDT2$}
  \put(-313,20){\begin{tikzpicture}
    \draw[black,thick] (0,0) -- (0.5,0);
    \draw[red,thick] (0,-0.18) -- (0.5,-0.18);
    \draw[green,thick] (0,-0.36) -- (0.5,-0.36);
    \draw[blue,thick] (0,-0.54) -- (0.5,-0.54);
  \end{tikzpicture}}
  \put(-55,96)  {\fontsize{5pt}{5pt}$RBT1$}
  \put(-55,90)   {\fontsize{5pt}{5pt}$RBT2$}
  \put(-55,84)   {\fontsize{5pt}{5pt}$RBT3$}
  \put(-55,78)   {\fontsize{5pt}{5pt}$RBT4$}
  \put(-72,80){\begin{tikzpicture}
    \draw[black,thick] (0,0) -- (0.5,0);
    \draw[red,thick] (0,-0.18) -- (0.5,-0.18);
    \draw[green,thick] (0,-0.36) -- (0.5,-0.36);
    \draw[blue,thick] (0,-0.54) -- (0.5,-0.54);
  \end{tikzpicture}}
  \caption{{Perturbation evolution forced by  a freestream entropy wave with $\omega =0$ and $\vartheta= 0^{\circ}$. (a,b): Streamwise evolution of $A_u$; (c,d) streamwise evolution of $A_T$; (e,f): Dependence of $A_u(\xi=200)$ on $k_3$. In (a-d), we choose  $k_{3}R^{-1/2} =0.02$.  Left column: Cases with various $R$; right column: cases with various $T_w$.} }
  \label{fig:e_k3Re_A_u}
  \end{center}
\end{figure}
To quantify the  amplitude evolution of non-modal perturbations, we define 
\begin{equation}
   {A_u(\xi)=\max_{\eta}|\tilde u_\xi(\xi,\eta)|,\quad A_T(\xi)=\max_{\eta}|\tilde T_\xi(\xi,\eta)|.} 
\end{equation}
As inferred from figure \ref{fig:HLNS_Contours},  $A_u$  in all cases undergoes transient growth that eventually {\color{black}reaches its peak} at a certain $\xi$. Through a number of numerical {\color{red}tests}, we find that at the {\color{black}peak} location, the local boundary-layer thickness $\delta_{BL}$ becomes comparable in magnitude with the perturbation spanwise wavelength. Given that $\delta_{BL}$ scales as $R^{-1/2}$, the {\color{black}peak} location remains invariant when $k_3R^{-1/2}$ is held constant. In  {figures} \ref{fig:e_k3Re_A_u}-(a,b), we present the $A_u$ evolution for various cases forced by freestream entropy waves with $k_3R^{-1/2}=0.02$. Given the significant variation in the magnitude of $A_u$ with Reynolds number, a logarithmic vertical axis is adopted {\color{black}particularly in panel (a)}.
For the same $T_w$, {\color{black}shown in panel (a)}, increasing $R$ leads to a substantial rise of the magnitude of $A_u$, while the overall distribution of  $A_u$ remains quantitatively similar. In contrast, when $R$ is fixed, {\color{black}shown in panel (b)}, a lower $T_w$ corresponds to a higher {\color{black}maximum} amplitude of $A_u$.

\begin{figure}
  \begin{center}
  \includegraphics[width = 0.49\textwidth]{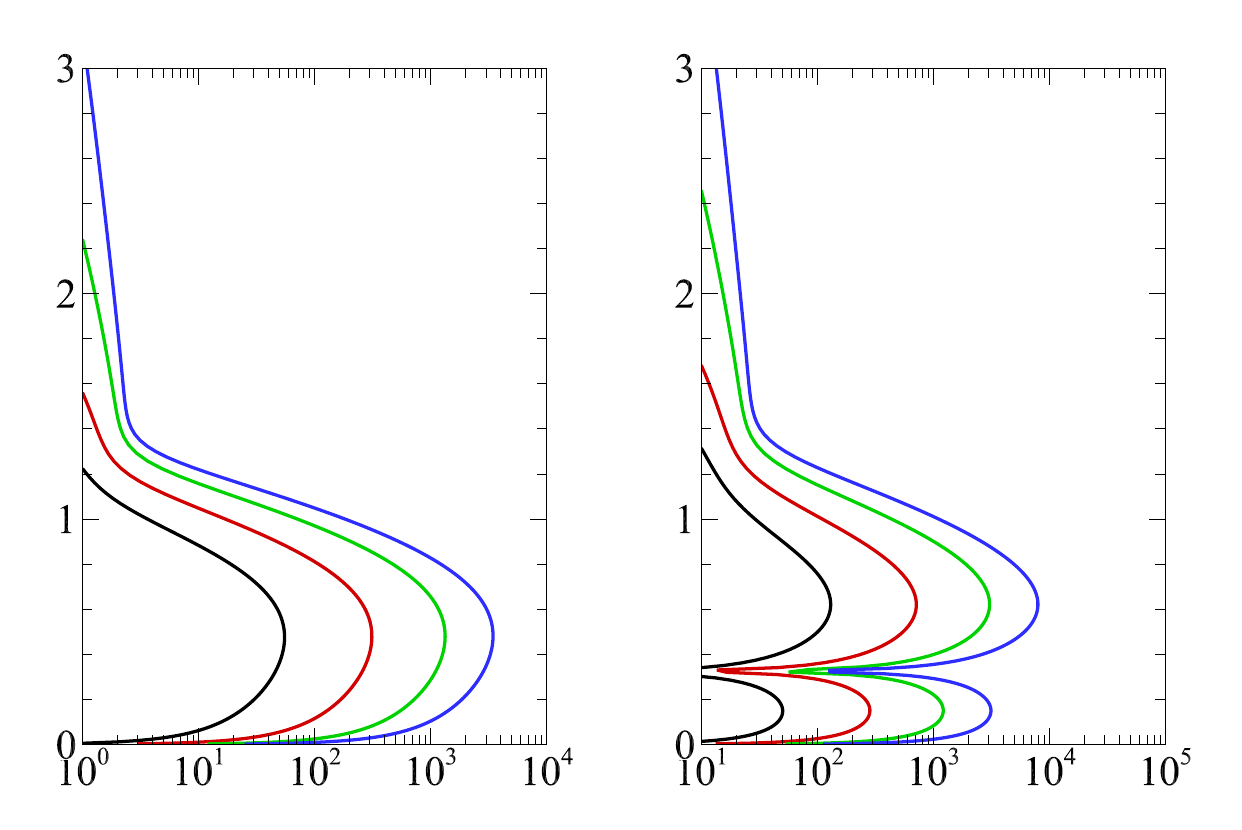} 
  \put(-195,120) {$(a)$}
  \put(-100,120) {$(b)$}
  \put(-145,-5) {$|\tilde{u}_{\xi}|$}
  \put(-50,-5) {$|\tilde{T}|$}
  \put(-190,55) {\rotatebox{90}{$\eta/\delta_{BL}$ }}
  \put(-43,96)  {\fontsize{5pt}{5pt}$RAT2$}
  \put(-43,90)   {\fontsize{5pt}{5pt}$RBT2$}
  \put(-43,84)   {\fontsize{5pt}{5pt}$RCT2$}
  \put(-43,78)   {\fontsize{5pt}{5pt}$RDT2$}
  \put(-60,80){\begin{tikzpicture}
    \draw[black,thick] (0,0) -- (0.5,0);
    \draw[red,thick] (0,-0.18) -- (0.5,-0.18);
    \draw[green,thick] (0,-0.36) -- (0.5,-0.36);
    \draw[blue,thick] (0,-0.54) -- (0.5,-0.54);
  \end{tikzpicture}}
  \put(-138,96)  {\fontsize{5pt}{5pt}$RAT2$}
  \put(-138,90)   {\fontsize{5pt}{5pt}$RBT2$}
  \put(-138,84)   {\fontsize{5pt}{5pt}$RCT2$}
  \put(-138,78)   {\fontsize{5pt}{5pt}$RDT2$}
  \put(-155,80){\begin{tikzpicture}
    \draw[black,thick] (0,0) -- (0.5,0);
    \draw[red,thick] (0,-0.18) -- (0.5,-0.18);
    \draw[green,thick] (0,-0.36) -- (0.5,-0.36);
    \draw[blue,thick] (0,-0.54) -- (0.5,-0.54);
  \end{tikzpicture}}
  \includegraphics[width = 0.49\textwidth]{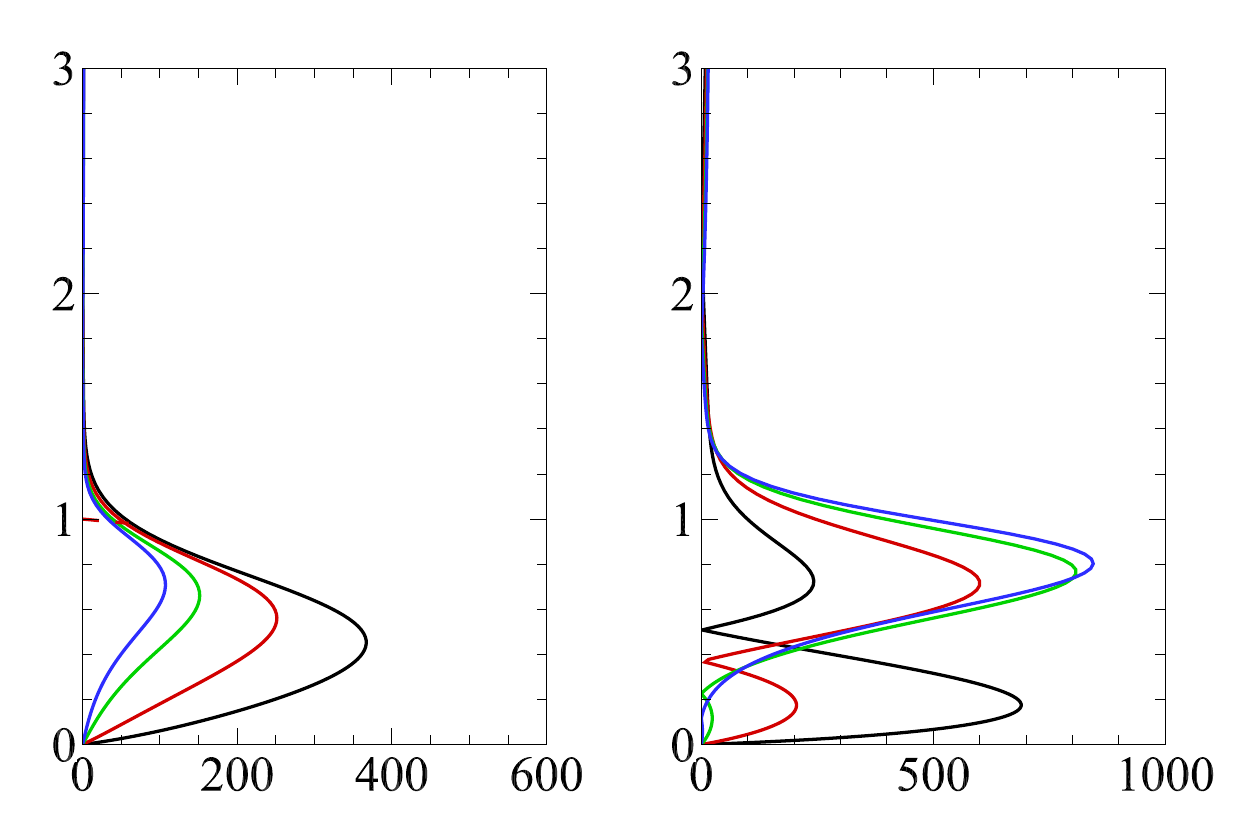} 
  \put(-195,120) {$(c)$}
  \put(-100,120) {$(d)$}
  \put(-145,-5) {$|\tilde{u}_{\xi}|$}
  \put(-50,-5) {$|\tilde{T}|$}
  \put(-190,55) {\rotatebox{90}{$\eta/\delta_{BL}$ }}
  \put(-43,96)  {\fontsize{5pt}{5pt}$RBT1$}
  \put(-43,90)   {\fontsize{5pt}{5pt}$RBT2$}
  \put(-43,84)   {\fontsize{5pt}{5pt}$RBT3$}
  \put(-43,78)   {\fontsize{5pt}{5pt}$RBT4$}
  \put(-60,80){\begin{tikzpicture}
    \draw[black,thick] (0,0) -- (0.5,0);
    \draw[red,thick] (0,-0.18) -- (0.5,-0.18);
    \draw[green,thick] (0,-0.36) -- (0.5,-0.36);
    \draw[blue,thick] (0,-0.54) -- (0.5,-0.54);
  \end{tikzpicture}}
  \put(-138,96)  {\fontsize{5pt}{5pt}$RBT1$}
  \put(-138,90)   {\fontsize{5pt}{5pt}$RBT2$}
  \put(-138,84)   {\fontsize{5pt}{5pt}$RBT3$}
  \put(-138,78)   {\fontsize{5pt}{5pt}$RBT4$}
  \put(-155,80){\begin{tikzpicture}
    \draw[black,thick] (0,0) -- (0.5,0);
    \draw[red,thick] (0,-0.18) -- (0.5,-0.18);
    \draw[green,thick] (0,-0.36) -- (0.5,-0.36);
    \draw[blue,thick] (0,-0.54) -- (0.5,-0.54);
  \end{tikzpicture}}
  \caption{Profiles of $|\tilde{u}_{\xi}|$ (a,c) and $|\tilde{T}|$ (b,d) at $\xi=300$  under entropy forcing with $\omega =0$, $\vartheta= 0^{\circ}$ and $k_{3}R^{-1/2} =0.02$ for all cases. (a,b) Cases with $T_w=3.33$ for various  $R$; (c,d) Cases with $R=1.67\times 10^5$ for  various $T_w$.}
  \label{fig:e_xi=300_k3Re=0.02_uT}
  \end{center}
\end{figure}

{\color{black}The perturbation velocity profiles $\tilde u_\xi$ at a downstream position $\xi=300$ across all cases are compared in figure \ref{fig:e_xi=300_k3Re=0.02_uT}-(a) and (c), where the wall-normal coordinate is normalised by the local boundary-layer thickness. For the same $T_w$ with various $R$ values (cases RAT2 to RDT2), shown in panel (a), the profiles of  $\tilde u_\xi$ show similar shapes despite differences in magnitude. Notably, the normalised peaking locations are identical for these cases. In contrast, for a fixed $R$ with varying $T_w$, the profile shapes differ significantly, with the normalised peak locations shifting toward the wall as $T_w$ reduces.}

Figures \ref{fig:e_k3Re_A_u}-(c) and (d) display the streamwise evolution of $A_T$ for all the cases considered. For a fixed wall temperature, the amplitude $A_T$ follows a trend similar to that of $A_u$, but with a notably higher magnitude. However, its variation with $T_w$ at a constant Reynolds number, {\color{black}shown in panel (d),} reveals a different pattern. Increasing $T_w$ {\color{black}from cases RBT2 to RBT4} not only enhances the {\color{black}maximum} amplitude of $A_T$, but also accelerates its decay in the downstream region. {\color{black}Notably, reducing $T_w$ from cases RBT2 to RBT1 also leads to an increase of $A_T$, indicating a non-monotonic dependence of the maximum $A_T$ on $T_w$. This non-monotonicity is elucidated by the perturbation temperature profiles. A demonstration of such profiles at $\xi=300$ is shown in figures \ref{fig:e_xi=300_k3Re=0.02_uT}-(b) and (d). For a fixed  $T_w$, the $\tilde T$ profiles maintain a consistent shape across different $R$, displaying two distinct peaks   with the   boundary-layer-edge peak being dominant. In contrast, the wall temperature substantially {\color{red}affects} the $\tilde T$-profile shape. As $T_w$ increases from cases RBT2 to RBT4, the edge peak intensifies while the near-wall peak diminishes. Conversely,   as $T_w$ decreases from cases RBT2 to RBT1, the near-wall peak intensifies and becomes dominant. Although the edge peak weakens with decreasing $T_w$, the near-wall peak for   $T_w=1.62$ overwhelms the edge peak. Consequently, the overall maximum $A_T$ decreases from $T_w=1.62$ to  $T_w=3.33$, accounting for the non-monotonic variation.}

In panels {(e) and (f)}, we plot the dependence of $A_u$ at $\xi=200$  on the rescaled wavenumber $k_3R^{-1/2}$. Each curve clearly exhibits a distinct peak near $k_3R^{-1/2}\approx {0.015}$. This position corresponds to a ratio of the spanwise wavelength $\lambda_z=2\pi/k_3$ to the boundary layer thickness of $\lambda_z/\delta_{BL}\approx${2.14}, indicating that the condition for preferred amplification occurs when the spanwise wavelength matches the {\color{black}magnitude of the} characteristic boundary-layer thickness.

Overall, this receptivity system is governed by a high‑dimensional set of parameters. The first set comprises the oncoming flow parameters ($M$, $T_e$) and the parameters that define the wall shape, both of which are fixed in the present series of studies. The second parameter set includes $T_w$, $R$ (or the nose radius), whose impacts on the receptivity efficiency {are} the central focus of this study. The third parameter set includes the type of the freestream forcing (fast acoustic, slow acoustic, entropy and vortical waves) and their associated  parameters (such as frequency $\omega$, spanwise wavenumber $k_3$, declination angle $\vartheta$). Even with the efficient SF‑HLNS approach, it remains impractical to fully characterise the receptivity response across the entire parameter space, with the numerical results shown in this section insufficient to capture this receptivity characteristics. Hence, an asymptotic analysis is needed to reduce the complexity of the system, as will be presented in the following section.

\section{Asymptotic model for non-modal receptivity with isothermal walls}
\label{sec:4}
In Part II of this work series \citep{Dong2026}, an asymptotic theory was developed to quantify non‑modal receptivity in hypersonic blunt boundary layers, revealing the key mechanisms governing the excitation of non-modal perturbations by various free-stream forcing.  {\color{black}While that study focused on a single validation case ($M=5.96$, $R=167000$ and adiabatic wall), the present work investigates the receptivity characteristics across a broad parameter space defined by $R$ and $T_w$. To this end, we employ the asymptotic model established previously to help illustrate the key scaling relations.}

\begin{figure}  
\begin{center}
\includegraphics[width=0.95\textwidth]{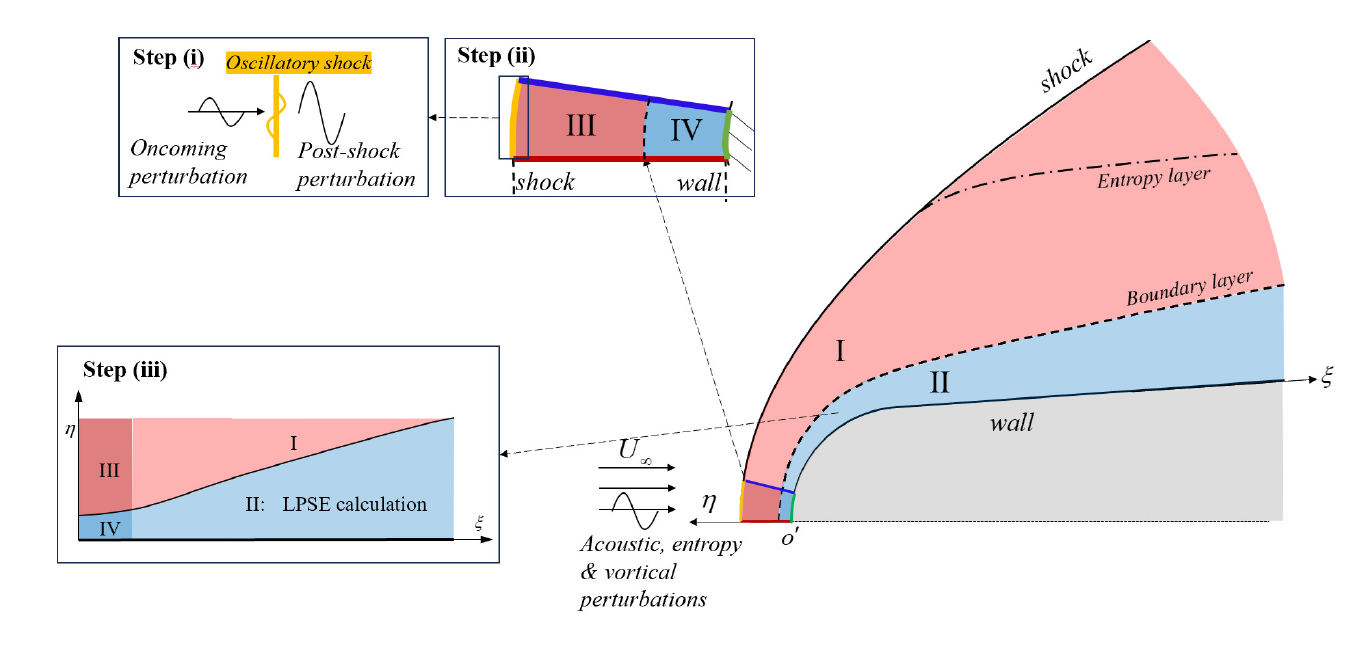}
  \caption{Schematic of the physical configuration and the three key steps of the asymptotic model. Regions I and II represent the bulk inviscid flow and the viscous boundary layer, respectively, while Regions III and IV denote the inviscid and boundary-layer regions along the {\color{red}centreline}. Step (i) addresses the interaction between freestream perturbations and the bow shock; Step (ii) describes the perturbation evolution from the post-shock region to the stagnation point along the {\color{red}centreline}; Step (iii) covers the downstream development of non-modal perturbations and the subsequent formation of streaks}\label{fig:sketch}
  \end{center}
\end{figure}
{\color{black} Figure \ref{fig:sketch} presents a schematic of the physical configuration for non-modal receptivity, outlining the three key steps of the asymptotic model for predicting the receptivity efficiency.} Step (i): freestream perturbations, whether acoustic, entropy, or vortical waves, interact with the bow shock around the {\color{red}centreline}, exciting post-shock perturbations.  This interaction, governed by the linearised R-H relation, involves shock oscillation and renders freestream {\color{cyan}acoustic} and entropy forcing more effective. Step (ii): these post-shock perturbations propagate toward the stagnation point, with the streamwise vorticity being amplified until the boundary-layer region due to the slow-down convection mechanism. Step (iii): the excited streamwise vortices propagate downstream within the boundary layer,  generating longitudinal streaks (non-modal perturbations) via the lift-up mechanism.

\begin{figure}
  \begin{center}
  \includegraphics[width = \textwidth]{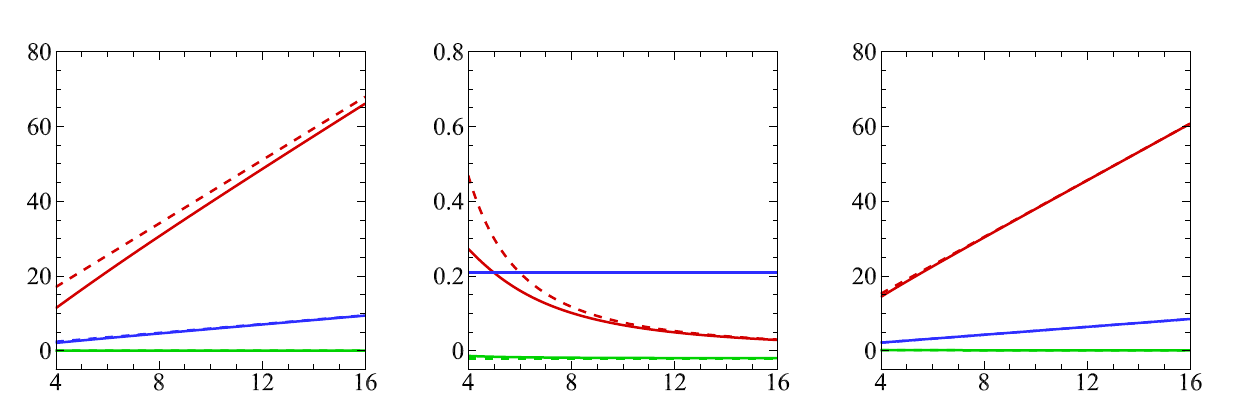}  
  \put(-390,120) {$(a)$}
  \put(-265,120) {$(b)$}
  \put(-135,120) {$(c)$}
  \put(-390,40) {\rotatebox{90}{$\hat\rho_{ps}, \hat u_{ps}, \hat p_{ps}$}}
  \put(-325,0) {$M$}
  \put(-195,0) {$M$}
  \put(-70,0) {$M$}
  \put(-211,100)  {\fontsize{5pt}{5pt}(\ref{eq:Perturbation_ps})}
  \put(-184,100)  {\fontsize{5pt}{5pt}(\ref{eq:Post_high_M})}
  \put(-230,90)  {$\hat \rho_{ps}$}
  \put(-230,80)  {$\hat u_{ps}$}
  \put(-230,70)  {$\hat p_{ps}$}
  \put(-210,70){\begin{tikzpicture}
    \draw[red,thick] (0,0) -- (0.5,0);
    \draw[red,dashed,thick] (1,0) -- (1.5,0);
    \draw[green,thick] (0,-0.4) -- (0.5,-0.4);
    \draw[green,dashed,thick] (1,-0.4) -- (1.5,-0.4);
    \draw[blue,thick] (0,-0.8) -- (0.5,-0.8);
    \draw[blue,dashed,thick] (1,-0.8) -- (1.5,-0.8);
  \end{tikzpicture}}
  \caption{ Dependence of the post-shock quantities $\hat\rho_{ps}$, $\hat u_{ps}$ and $\hat p_{ps}$ on $M$ forced by freestream acoustic $(a)$, vortical $(b)$ and entropy $(c)$ perturbations. Solid lines: numerical results for (\ref{eq:Perturbation_ps}); dashed lines: high-$M$ asymptotic regime (\ref{eq:Post_high_M}). For the vortical {\color{cyan}forcing}, panel (b), we have chosen $\hat \Omega_2=1$ and $k_3=8$ for demonstration.}  \label{fig:line_rup_shock}
  \end{center}
\end{figure}

{\color{black}
\subsection{Prediction of post-shock perturbations around the {\color{red}centreline}}
For step (i), the post-shock perturbations depend on two factors: the characteristics of the incident forcing and the strength of the bow shock. These factors are independent of the Reynolds number and wall temperature. Since the interaction occurs near the {\color{red}centreline}, the bow shock is approximated as a normal shock, the strength of which is determined solely by the oncoming Mach number. Following \cite{Dong2026}, we can evaluate the low-frequency freestream perturbation  density   $\hat \rho_\infty$, velocity $\hat u_\infty$ and pressure $\hat p_\infty$  for various forcing types  under a unit energy norm:
 \begin{equation}
\Big({\hat \rho_\infty},{\hat u_\infty},{\hat p_\infty}\Big)\sim \left\{
\begin{array}{ll}
\vspace{.2cm} \displaystyle
   \Big(\frac{M}{\sqrt{2}},-\frac{1}{\sqrt{2}M},\frac{1}{\sqrt{2}M}\Big) & \mbox{for fast or slow acoustic forcing},\\
   \vspace{.2cm} \displaystyle \Big(0,\frac{\hat\Omega_2}{k_3},0\Big) & \mbox{for vortical forcing},
   \\
   \vspace{.2cm} \displaystyle  \Big(\sqrt{\gamma-1}M,0,0\Big) & \mbox{for entropy forcing},\end{array}
 \right.
\label{eq:estimate_FS}
 \end{equation}
where {$\hat \Omega_2$} represents the  vertical vorticity perturbation in the free stream. Notably, these quantities are independent of the declination angle $\vartheta$, but the frequency $\omega$ is assumed to be much smaller than unity. It is evident that the freestream density perturbation  $\hat \rho_\infty$ dominates for acoustic and entropy forcing, exceeding the $\hat u_\infty$ amplitude of vortical forcing by   $O(M)$.  Furthermore, the amplitude of vortical forcing scales linearly with $\hat \Omega_2$, but is inversely proportional to $k_3$.

The post-shock perturbation is governed by the linearised R-H relation. Combining (4.4), (4.5) and (3.8) in \cite{Dong2026}, we obtain the post-shock  perturbation density $\hat\rho_{ps}$, velocity {$\hat u_{ps}$} and pressure $\hat p_{ps}$:
\refstepcounter{equation}
$$
 \hat\rho_{ps}= \frac{(\gamma+1)M^2[4+(\gamma-1)M^2]\hat \rho_\infty}{[2+(\gamma-1)M^2]^2}+\frac{4(\gamma+1)M^2\hat u_\infty}{[2+(\gamma-1)M^2]^2}-\frac{2\gamma(\gamma+1)M^4\hat p_\infty}{[2+(\gamma-1)M^2]^2},\eqno{(\theequation a)}\label{eq:Perturbation_ps}
$$
$$
{\hat u_{ps}}=\frac{2\hat\rho_\infty}{(\gamma+1)M^2}+\frac{[2-(\gamma-1)M^2]\hat u_\infty}{(1+\gamma)M^2}-\frac{2\gamma\hat p_\infty}{1+\gamma},\eqno{(\theequation b)}
$$
$$
\hat p_{ps}=\frac{2\hat\rho_\infty}{\gamma+1}+\frac{4\hat u_\infty}{\gamma+1}-\frac{(\gamma-1)\hat p_\infty}{\gamma+1}.\eqno{(\theequation c)}
$$
Notably, they only depend on the oncoming Mach number $M$. 
The solid lines in figure \ref{fig:line_rup_shock} {\color{red}display} the variation in post-shock quantities as functions of $M$ under various types of freestream forcing. Under freestream acoustic and entropy forcing,  the post-shock   density perturbation $\hat \rho_{ps}$ dominates and increases monotonically with $M$. Conversely, under   vortical forcing, $\hat \rho_{ps}$ is dominant   only for $M<5$. It decreases with $M$ monotonically, while the post-shock pressure perturbation   $\hat p_{ps}$ remains invariant.  Consequently, $\hat p_{ps}$ becomes the dominant post-shock perturbation for $M>5$. Although the case studies considered in this paper focus on a single Mach number, $M=5.96$, figure \ref{fig:line_rup_shock} clearly establishes the parametric dependence required for the upper boundary conditions for the asymptotic model for non-modal receptivity.}

{\color{black}In the high-$M$ limit, the post-shock response to each type of freestream forcing can be summarised as:
 \begin{equation}
\Big({\hat \rho_{ps}},{\hat u_{ps}},{\hat p_{ps}}\Big)_{M\gg 1}\sim \left\{
\begin{array}{ll}
\vspace{.2cm} \displaystyle
   \Big(\frac{(\gamma+1)M}{\sqrt{2}(\gamma-1)},{-\frac{\gamma - 1}{\sqrt{2}(\gamma + 1)M}},\frac{\sqrt{2}M}{\gamma+1}\Big) & \mbox{for acoustic forcing},\\
   \vspace{.2cm} \displaystyle \frac{\hat\Omega_2}{k_3}\Big(\frac{4(\gamma+1)}{(\gamma-1)^2M^2},-\frac{\gamma-1}{\gamma+1},\frac{4}{\gamma+1}\Big) & \mbox{for vortical forcing},
   \\
   \vspace{.2cm} \displaystyle  \Big(\frac{(\gamma+1)M}{\sqrt{\gamma-1}},\frac{2\sqrt{\gamma-1}}{(\gamma+1)M},\frac{2\sqrt{\gamma-1}M}{\gamma+1}\Big) & \mbox{for entropy forcing},\end{array}
 \right.
\label{eq:Post_high_M}
 \end{equation}
These asymptotic predictions are plotted as dashed lines in Fig.~\ref{fig:line_rup_shock}. While evident deviations are observed at low Mach numbers, particularly for $\hat \rho_{ps}$ in panels (a) and (b), the data collapse onto the asymptotic scalings as $M$ increases, thereby validating the high-$M$ analysis. Notably, the post-shock perturbations under acoustic and entropy forcing appear to be comparable, whereas those to vortical forcing are significantly weaker, indicating a rather weak vortical receptivity efficiency. 
}

\subsection{Perturbation evolution from  shock to  stagnation point}
\label{sec:asyptotic_theory}
{\color{black}For step (ii) of the asymptotic model as sketched in figure \ref{fig:sketch}, we analyse the perturbation evolution along the {\color{red}centreline} from the bow shock to the stagnation point. As established in \cite{Dong2026}, this process involves two distinct layers: an inviscid bulk region ($\xi\sim R^{-1/2}$ and $\eta\sim 1$) and a viscous boundary layer ($\xi\sim R^{-1/2}$ and $\eta\sim R^{-1/2}$).
The inviscid solution is independent of the Reynolds number and wall temperature. Consequently, it remains identical to that in \cite{Dong2026}, as both studies consider the same Mach number. Therefore, we restrict our analysis in this subsection to the boundary-layer region. For convenience, we introduce a local coordinate system
\begin{equation}
  X=R^{1/2}\xi=O(1),\quad  Y=R^{1/2}\eta=O(1).
\end{equation}
}

\subsubsection{Base flow region IV}\label{sec:base_flow}
{\color{black}The base flow in the boundary layer around the stagnation region can be approximated by Taylor expansion in terms of $\xi$,
\refstepcounter{equation}
$$
\bar u_\xi=\bar U_0(Y)+\xi\bar  U_1(Y)+O(\xi^2)=\bar U_0(Y)+R^{-1/2}X\bar U_1(Y)+O(R^{-1}),\eqno{(\theequation a)}
$$
$$
\bar u_\eta=R^{-1/2}\Big(\bar V_0(Y)+\xi\bar V_1(Y)+O(\xi^2)\Big)=R^{-1/2}\Big(\bar V_0(Y)+R^{-1/2}X\bar V_1(Y)+O(R^{-1})\Big),\eqno{(\theequation b)}
$$
$$
\bar w=0,\eqno{(\theequation c)}
$$
$$
\bar\rho=\bar R_0(Y)+\xi\bar R_1(Y)+O(\xi^2)=\bar R_0(Y)+R^{-1/2}X\bar R_1(Y)+O(R^{-1}),\eqno{(\theequation d)}
$$
$$\bar T=
M^2\Big(\bar T_0(Y)+\xi\bar T_1(Y)+O(\xi^2)\Big)=M^2\Big(\bar T_0(Y)+R^{-1/2}X\bar T_1(Y)+O(R^{-1})\Big),\eqno{(\theequation e)}$$
$$
\bar P=\bar P_{00}+\xi\bar P_{10}+\frac{\xi^2}{2}P_{20}+O(R^{-3/2})=\bar P_{00}+R^{-1/2}X\bar P_{10}+\frac{R^{-1}X^2}2\bar P_{20}+O(R^{-3/2}),\eqno{(\theequation f)}
$$
where the subscripts 0 and 1 denote the zeroth- and first-order derivatives with respect to $\xi$, respectively, and $\bar P_{00}$, $\bar P_{10}$ and $\bar P_{20}$ represent the zeroth-, first- and second-order derivatives with respect {\color{red}to} $\xi$ of $\bar P$ in the boundary-layer region (and are independent of $Y$), respectively. Note that the wall-normal velocity {\color{red}$\bar u_\eta$} in the boundary layer scales as $O(R^{-1/2})$ because the region-III solution dictates a linear {\color{red}behaviour} near the wall, $\tilde u_\eta\sim \eta$  as $\eta\to 0$.
Exploiting the symmetry about  the {\color{red}centreline} ($\xi=0$), we obtain
\begin{equation}
    \bar U_0=\bar V_1=\bar R_1=\bar T_1=\bar P_{10}=0.
\end{equation}
Consequently, the expansion is reduced to
}
\begin{equation}
(\bar u_\xi,\bar u_\eta,\bar w,\bar \rho,\bar T,\bar P)=\Big(\frac{X\bar U_1(Y)}{R^{1/2}},\frac{\bar V_0(Y)}{R^{1/2}},0,\bar R_{0}(Y),M^2\bar T_{0}(Y),P_{00}+\frac{X^2P_{20}}{2R}\Big)+\cdots.\label{eq:Base_flow_quantity}
\end{equation}
{\color{black}Although the region-III base flow was detailed in Part II, we briefly revisit the governing equations and boundary conditions here to facilitate interpretation of the numerical results across varying wall temperatures.} 
Substituting (\ref{eq:Base_flow_quantity}) into the N-S equations and {\color{red}collecting} the leading-order terms, we derive the following governing equations:
\refstepcounter{equation}$$
\bar R_0(\bar U_1+\bar V_0')+\bar V_0\bar R_0'=0,\quad \bar R_0(\bar U_1^2+\bar V_0\bar U_1')+{P_{20}}-({\bar \mu_{0}}\bar U_1')'=0,\eqno{(\theequation a,b)}
\label{eq:EQ_BL}
$$
$$
\bar R_0\bar V_0\bar T_0'-\frac{1}{Pr}(\bar \mu_0\bar T_0')'=0,\quad \bar R_0\bar T_0=\gamma P_{00},\eqno{(\theequation c,d)}
$$
where $Pr=0.72$ is the Prandtl number, and $\bar \mu_0=\bar \mu_0(\bar T_0)=(1+\tilde C)(M^2\bar T_{0})^{3/2}/(M^2\bar T_{0}+\tilde C)$ with $\tilde C=110.4 \mathrm{K} /T_\infty$. 
The boundary conditions for isothermal walls are 
\begin{equation}
    \bar U_1(0)=\bar V_0(0)=0,\quad \bar T_0(0)={\color{cyan}M^{-2}}T_w, \quad (\bar V_0',\bar T_0)\to (C_v,T_{00})\quad \mbox{as } Y\to \infty,\label{eq:BC_base}
\end{equation}
where $T_{00}$ and $C_v$ are the region-III temperature and velocity gradient at the wall. 

\begin{figure}
  \begin{center}
  \includegraphics[width = 0.49\textwidth]{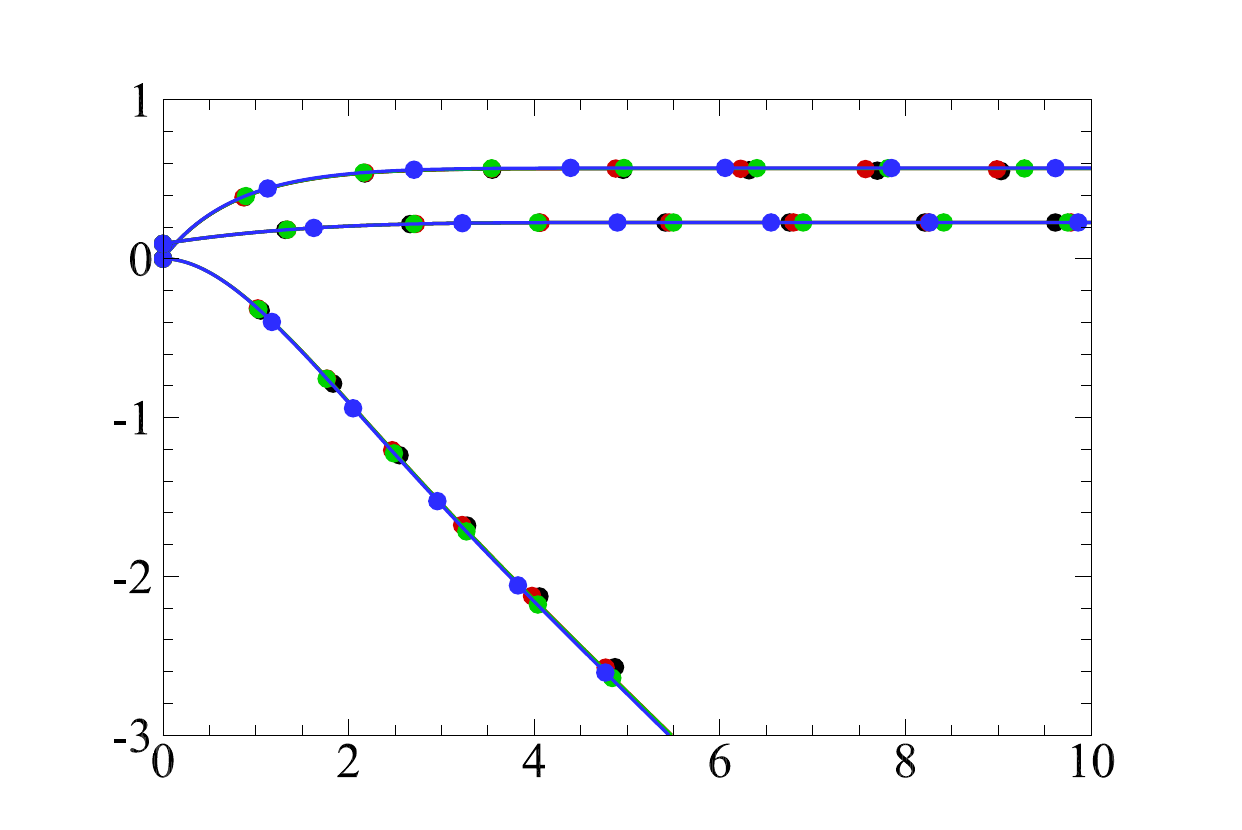}
  \put(-190,110){$(a)$}
  \put(-95,0) {$Y$}
  \includegraphics[width = 0.49\textwidth]{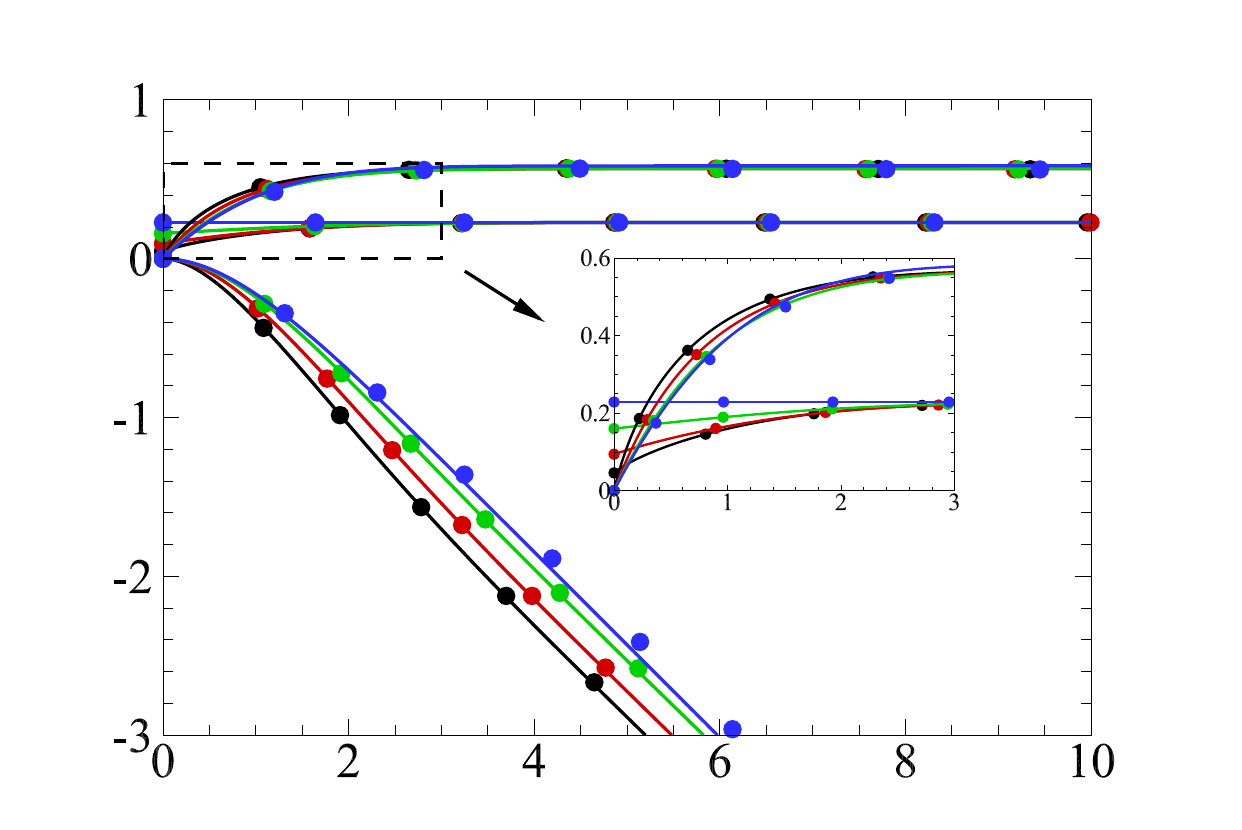}
  \put(-95,0) {$Y$}
  \put(-190,110){$(b)$}
  \put(-270,33) {\fontsize{5pt}{5pt}\selectfont$RAT2$}
  \put(-270,28) {\fontsize{5pt}{5pt}\selectfont$RBT2$}
  \put(-270,23) {\fontsize{5pt}{5pt}\selectfont$RCT2$}
  \put(-270,18) {\fontsize{5pt}{5pt}\selectfont$RDT2$}
  \put(-300,25) {\fontsize{6pt}{6pt}\selectfont$\bar{V}_0$}
  \put(-300,83) {\fontsize{6pt}{6pt}\selectfont$\bar{T}_0$}
  \put(-300,102) {\fontsize{6pt}{6pt}\selectfont$\bar{U}_1$}
  \put(-250,19){\begin{tikzpicture}
    \draw[black,thick] (0,0) -- (0.5,0);
    \draw[red,thick] (0,-0.18) -- (0.5,-0.18);
    \draw[green,thick] (0,-0.36) -- (0.5,-0.36);
    \draw[blue,thick] (0,-0.54) -- (0.5,-0.54);
    \node[draw,black,circle,fill,inner sep=0.8pt] at (1.0,0) {};
    \node[draw,red,circle,fill,inner sep=0.8pt] at (1.0,-0.18) {};
    \node[draw,green,circle,fill,inner sep=0.8pt] at (1.0,-0.36) {};
    \node[draw,blue,circle,fill,inner sep=0.8pt] at (1.0,-0.54) {};
    \end{tikzpicture}}
  \put(-250,40) {\fontsize{4pt}{4pt}\selectfont $Asym.$}
  \put(-230,40) {\fontsize{4pt}{4pt}\selectfont SF-DNS}
  \put(-80,33) {\fontsize{5pt}{5pt}\selectfont$RBT1$}
  \put(-80,28) {\fontsize{5pt}{5pt}\selectfont$RBT2$}
  \put(-80,23) {\fontsize{5pt}{5pt}\selectfont$RBT3$}
  \put(-80,18) {\fontsize{5pt}{5pt}\selectfont$RBT4$}
  \put(-60,19){\begin{tikzpicture}
    \draw[black,thick] (0,0) -- (0.5,0);
    \draw[red,thick] (0,-0.18) -- (0.5,-0.18);
    \draw[green,thick] (0,-0.36) -- (0.5,-0.36);
    \draw[blue,thick] (0,-0.54) -- (0.5,-0.54);
    \node[draw,black,circle,fill,inner sep=0.8pt] at (1.0,0) {};
    \node[draw,red,circle,fill,inner sep=0.8pt] at (1.0,-0.18) {};
    \node[draw,green,circle,fill,inner sep=0.8pt] at (1.0,-0.36) {};
    \node[draw,blue,circle,fill,inner sep=0.8pt] at (1.0,-0.54) {};
    \end{tikzpicture}}
  \put(-60,40) {\fontsize{4pt}{4pt}\selectfont $Asym.$}
  \put(-40,40) {\fontsize{4pt}{4pt}\selectfont SF-DNS}
  \put(-110,23) {\fontsize{6pt}{6pt}\selectfont$\bar{V}_0$}
  \put(-110,84) {\fontsize{6pt}{6pt}\selectfont$\bar{T}_0$}
  \put(-110,103) {\fontsize{6pt}{6pt}\selectfont$\bar{U}_1$}
  \put(-80,53) {\fontsize{6pt}{6pt}\selectfont$\bar{T}_0$}
  \put(-90,76) {\fontsize{6pt}{6pt}\selectfont$\bar{U}_1$}
  \caption{Comparison of base-flow solutions in region IV obtained by asymptotic predictions (solid lines) and N-S solutions (circles).   (a): Cases for $T_w=3.33$ with various $R$; (b): cases for $R=1.67\times 10^5$ with various $T_w$.} \label{fig:base_Bl}
  \end{center}
\end{figure}

{\color{black}In contrast to the adiabatic cases examined in Part II, the cold wall boundary conditions give rise to thermal boundary layers with $O(1)$ temperature and density gradients. This not only complicates the derivation of the boundary-layer solution, but also enhances the perturbation response near the stagnation point, as will be demonstrated in figure \ref{fig:Per_Bl}-(c,d) in the subsequent subsection. Figure \ref{fig:base_Bl} displays base-flow profiles in the boundary-layer region for all considered cases. The profiles of $\bar U_1$ and $\bar T_0$ approach constant values as $Y\to \infty$, while $\bar V_0$ exhibits linear growth proportional to $Y$, consistent with the matching conditions in (\ref{eq:BC_base}). For a fixed  $T_w$, shown in panel (a), the profiles for various Reynolds numbers collapse onto a single  curve. However, as indicated in panel (b), reducing the wall temperatures steepens the near-wall gradients of $\bar U_1$ and $\bar T_0$, and elevates the magnitude of $|\bar V_0|$ by a constant shift. The asymptotic predictions exhibit excellent agreement with the N-S solutions across all cases, thereby confirming the predictive fidelity of the asymptotic model.}

\subsubsection{Perturbations in  region IV}
Similar to the base-flow expansion in \ref{sec:base_flow}, we expand the boundary-layer perturbation based on the Taylor expansion and the symmetric nature about the {\color{red}centreline}:
\refstepcounter{equation}
$$( \tilde u_\xi,\tilde u_\eta,\tilde w,\tilde \rho,\tilde T,\tilde p)=\Big(\frac{X\bar u_1({Y})}{R^{1/2}},\frac{\bar v_0({Y})}{R^{1/2}},\bar w_0({Y}),\bar \rho_0({Y}),M^2 \bar\theta_0({Y}), \hat p_{00}\Big)\re^{\ri k_3 z}+c.c.+\cdots.\eqno{(\theequation)}\label{eq:perturbation_BL_sta}
$$
{\color{black}Consistent with  equations (3.30) and (3.31) of Part II, we reformulate the governing equations and boundary conditions for the region‑IV perturbations:}
\refstepcounter{equation}
$$
(\bar U_1+\bar V_0')\bar\rho_0+\bar R_0'\bar v_0+\bar V_0\bar \rho_0'+\bar R_0(\bar u_1+\bar v_0'+\ri k_3\bar w_0)=0,\eqno{(\theequation a)}$$
$$ \bar R_{0}(2\bar U_1\bar u_1+\bar U_1'\bar v_0+\bar V_0\bar u_1')+(\bar U_1^2+\bar U_1'\bar V_0)\bar\rho_0=(\bar \mu_{0}\bar u_1')'+(\mu_T\bar U_1'\bar \theta_0)', \eqno{(\theequation b)}\label{eq:perturbation_BL}
$$
$$\bar R_{0}\bar V_0\bar w_0'+\ri k_3 \hat p_{00}=(\bar\mu_{0}\bar w_0')',\eqno{(\theequation c)}$$
$$
 \bar R_0(\bar V_0\bar\theta_0'+\bar T_0'\bar v_0)+\bar V_0\bar T_0'\bar\rho_0=\frac{1}{Pr}(\bar\mu_0\bar\theta_0'+\mu_T\bar T_0'\bar\theta_0)',\quad \frac{\hat p_{00}}{P_{00}}=\frac{\bar\rho_0}{\bar R_0}+\frac{\bar \theta_0}{\bar T_0},\eqno{(\theequation d,e)}
$$
\begin{equation}
    \bar u_1(0)=\bar v_0(0)=\bar \theta_0(0)=0,\quad (\bar u_1',\bar w_0,\bar\theta_0)\to (0,-\frac{\ri k_3 \hat p_{00}}{R_{00}C_v}\ln Y+C_1,\tilde \theta_0(0))\quad\mbox{as }Y\to \infty,
    \label{eq:asym_BC}
\end{equation}
where {$\mu_T$=d$\bar \mu_0/$d$\bar T_0$}, and 
\begin{equation}
  C_1=B_1+\frac{\ri k_3 \hat p_{00}}{2R_{00}C_v}\ln R,  \label{eq:C1}
\end{equation}
with $R_{00}$ being the region-III density solution at the wall and $B_1$ being a constant determined by the inviscid solution.

The linear system (\ref{eq:perturbation_BL}) with (\ref{eq:asym_BC}) can be solved numerically using the same approach as in Part II. {\color{black}Particularly, (\ref{eq:perturbation_BL}c) is self-contained, which can be solved analytically:
\begin{equation}
\bar w_0={\cal W}_p+{\cal C}{\cal W}_c,\label{eq:bar_w_0}
\end{equation}
where $\cal C$ is a constant to be determined, and the particular and complementary solutions are expressed as
\begin{equation}
{\cal W}_p={\ri k_3 \hat p_{00}}\int_0^Y\frac{\re^{\int_0^{\bar Y}A_0d\tilde Y}}{\mu_{00}}\Big(\int_0^{\bar Y}\re^{-\int_0^{\breve Y}A_0d\tilde Y}d\breve Y\Big)d\bar Y,
\quad
{\cal W}_c=\int_0^Y\re^{\int_0^{\bar Y}A_0d\tilde Y}d\bar Y,
\end{equation}
where $A_0(Y)=\frac{\bar R_0\bar V_0-\bar \mu_0'}{\bar\mu_0}$. If $\bar \mu_0$ and $\bar R_0$ are constant, then the solution recovers to that of the adiabatic configuration, as given by (3.36) of Part II. However, the gradients of these quantities within the boundary layer render the problem  more intricate. 
The upper-limit behaviours of these solutions are
\begin{equation}
{\cal W}_p\to -\frac{\ri k_3 \hat p_{00}}{R_{00}C_v}\ln Y+D_{1p},\quad {\cal W}_c\to D_{1c}\quad\mbox{as }Y\to \infty,
\end{equation}
where $D_{1p}$ and $D_{1c}$ are constants that can be obtained numerically.
Comparing with the matching condition (\ref{eq:asym_BC}), we can obtain
\begin{equation}
{\cal C}=(C_1-D_{1p})/D_{1c}.
\label{eq:C}
\end{equation}
}

\begin{figure}
  \begin{center}
  \includegraphics[width = 0.49\textwidth]{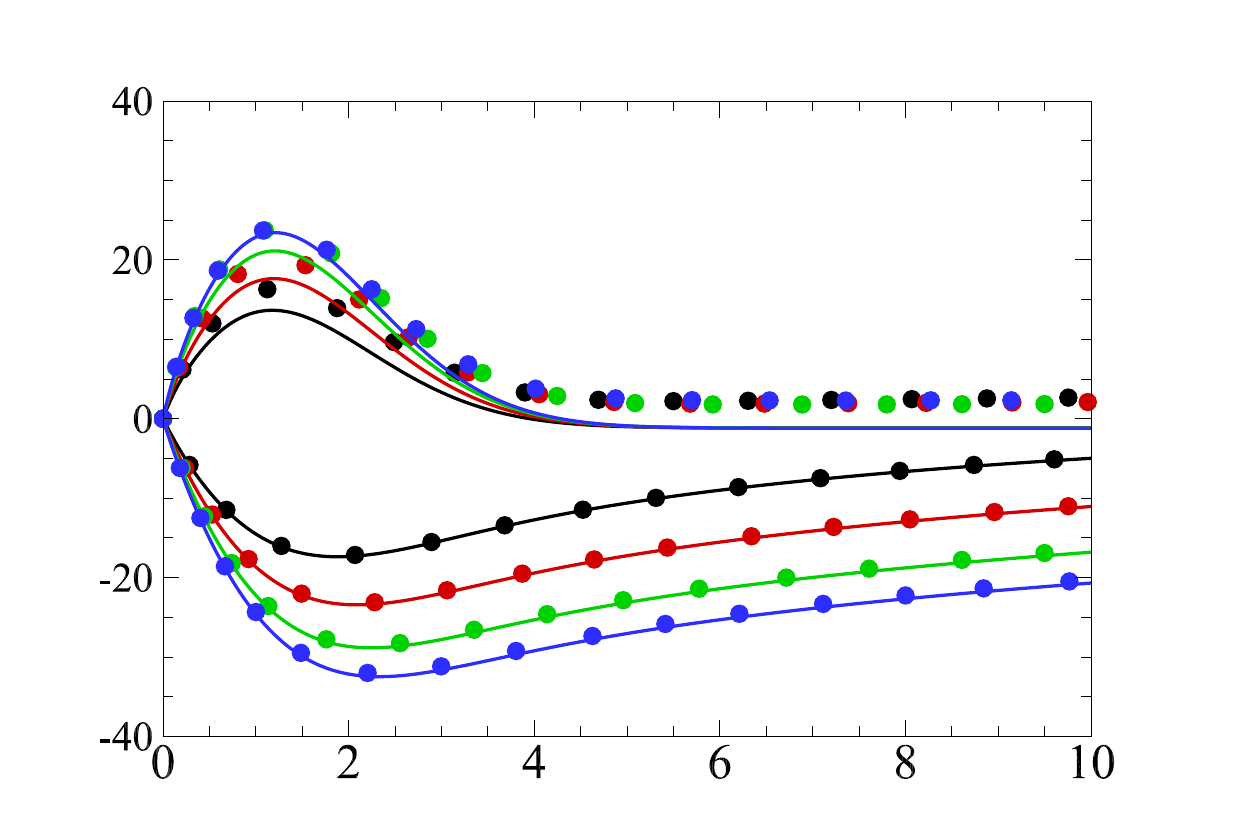}
  \put(-190,110){$(a)$}
  \put(-95,0) {$Y$}
  \includegraphics[width = 0.49\textwidth]{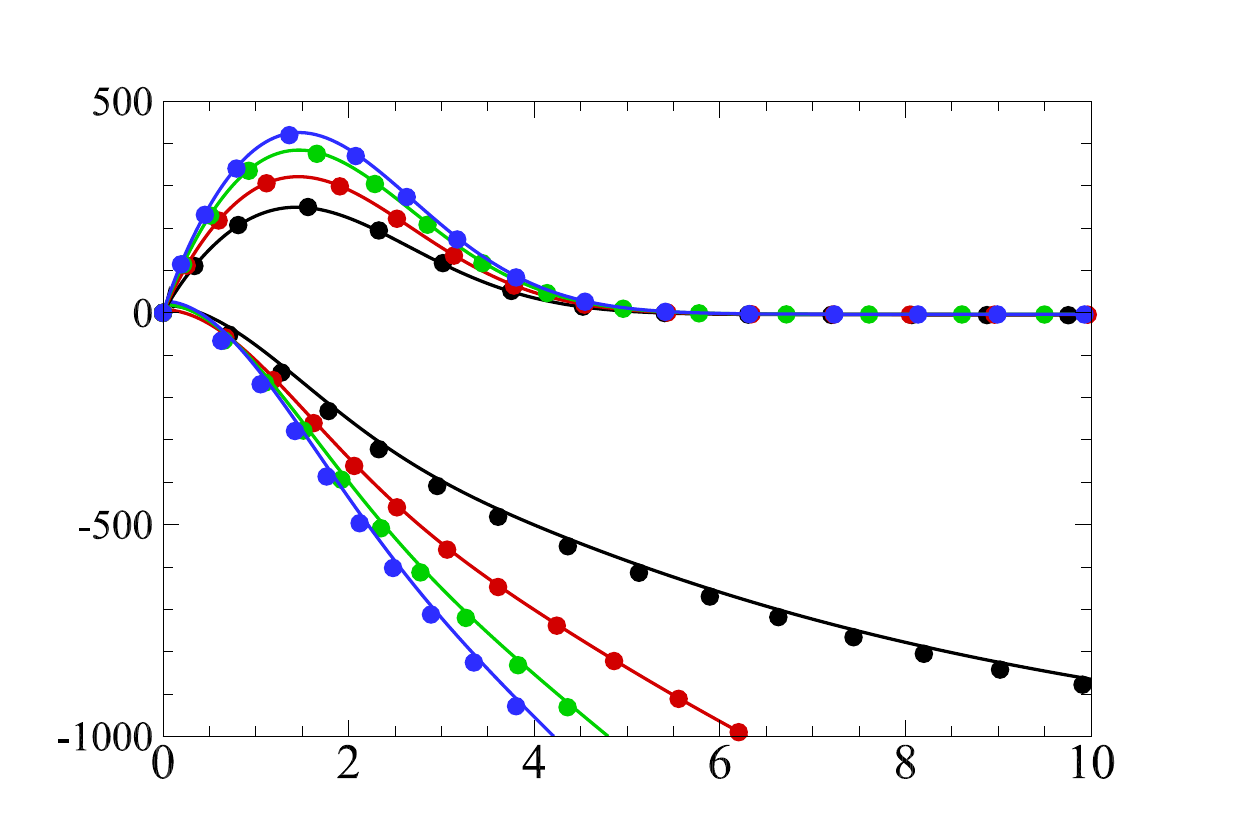}
  \put(-95,0) {$Y$}
  \put(-190,110){$(b)$}
  \put(-270,93) {\fontsize{5pt}{5pt}\selectfont$RAT2$}
  \put(-270,88) {\fontsize{5pt}{5pt}\selectfont$RBT2$}
  \put(-270,83) {\fontsize{5pt}{5pt}\selectfont$RCT2$}
  \put(-270,78) {\fontsize{5pt}{5pt}\selectfont$RDT2$}
  \put(-300,20) {$-\ri\tilde w$}
  \put(-300,72) {$\tilde{u}_1$}
  \put(-250,79){\begin{tikzpicture}
    \draw[black,thick] (0,0) -- (0.5,0);
    \draw[red,thick] (0,-0.18) -- (0.5,-0.18);
    \draw[green,thick] (0,-0.36) -- (0.5,-0.36);
    \draw[blue,thick] (0,-0.54) -- (0.5,-0.54);
    \node[draw,black,circle,fill,inner sep=0.8pt] at (1.0,0) {};
    \node[draw,red,circle,fill,inner sep=0.8pt] at (1.0,-0.18) {};
    \node[draw,green,circle,fill,inner sep=0.8pt] at (1.0,-0.36) {};
    \node[draw,blue,circle,fill,inner sep=0.8pt] at (1.0,-0.54) {};
    \end{tikzpicture}}
  \put(-250,100) {\fontsize{4pt}{4pt}\selectfont $Asym.$}
  \put(-230,100) {\fontsize{4pt}{4pt}\selectfont SF-HLNS}
  \put(-110,50) {$R^{1/2}\tilde u_\eta$}
  \put(-105,84) {$M^2\tilde{\theta}$}
  \\
  \includegraphics[width = 0.49\textwidth]{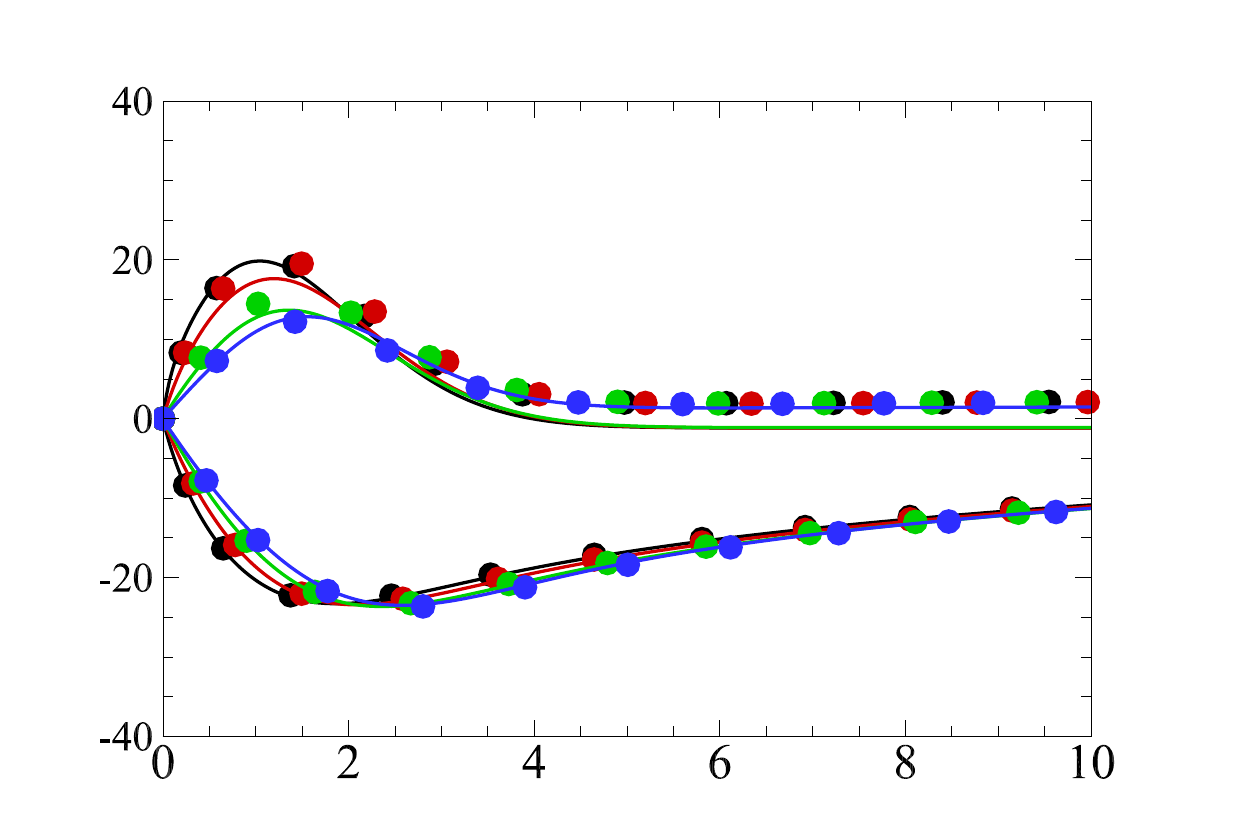}
  \put(-190,110){$(c)$}
  \put(-95,0) {$Y$}
  \includegraphics[width = 0.49\textwidth]{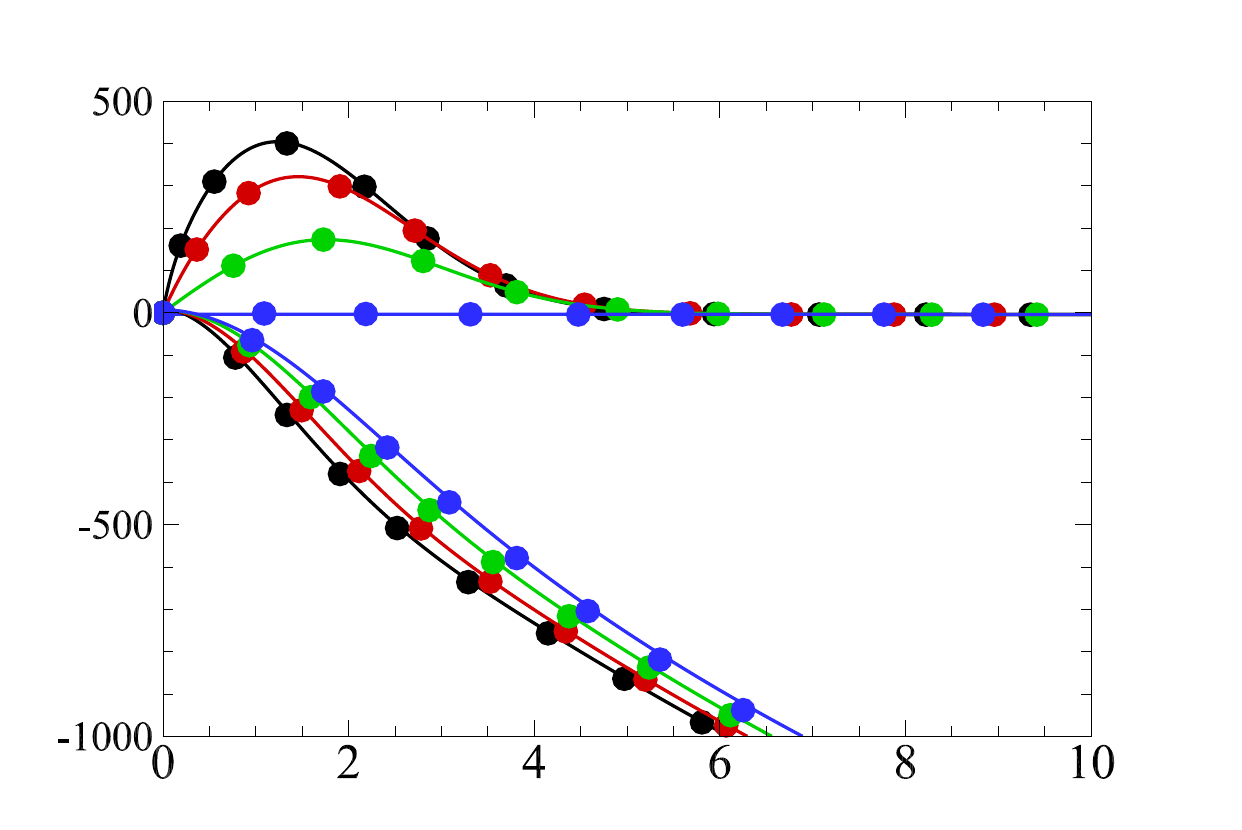}
  \put(-95,0) {$Y$}
  \put(-190,110){$(d)$}
  \put(-270,93) {\fontsize{5pt}{5pt}\selectfont$RBT1$}
  \put(-270,88) {\fontsize{5pt}{5pt}\selectfont$RBT2$}
  \put(-270,83) {\fontsize{5pt}{5pt}\selectfont$RBT3$}
  \put(-270,78) {\fontsize{5pt}{5pt}\selectfont$RBT4$}
  \put(-300,25) {$-\ri\tilde w$}
  \put(-300,72) {$\tilde{u}_1$}
  \put(-250,79){\begin{tikzpicture}
    \draw[black,thick] (0,0) -- (0.5,0);
    \draw[red,thick] (0,-0.18) -- (0.5,-0.18);
    \draw[green,thick] (0,-0.36) -- (0.5,-0.36);
    \draw[blue,thick] (0,-0.54) -- (0.5,-0.54);
    \node[draw,black,circle,fill,inner sep=0.8pt] at (1.0,0) {};
    \node[draw,red,circle,fill,inner sep=0.8pt] at (1.0,-0.18) {};
    \node[draw,green,circle,fill,inner sep=0.8pt] at (1.0,-0.36) {};
    \node[draw,blue,circle,fill,inner sep=0.8pt] at (1.0,-0.54) {};
    \end{tikzpicture}}
  \put(-250,100) {\fontsize{4pt}{4pt}\selectfont $Asym.$}
  \put(-230,100) {\fontsize{4pt}{4pt}\selectfont SF-HLNS}
  \put(-110,43) {$R^{1/2}\tilde u_\eta$}
  \put(-105,84) {$M^2\tilde{\theta}$}
  \caption{{\color{black}Perturbation profiles of $(\tilde u_1,\tilde w)$ (left column) and $(\tilde\theta,\tilde u_\eta)$ (right column)  in region IV obtained by asymptotic predictions (solid lines) and SF-HLNS calculations (circles). The freestream forcing is entropy wave with $\omega =0$, $\vartheta= 0^{\circ}$ and $k_{3}=8$. Top row: cases with $T_w=3.33$ for various  $R$; bottom row: cases with $R=1.67\times 10^5$ for  various $T_w$.  Here, $\tilde{u}_1$, $\tilde u_\eta$ and $\tilde{\theta}$ {\color{red}are} real, while $\tilde w$ is pure imaginary.}} \label{fig:Per_Bl}
  \end{center}
\end{figure}

{\color{black}
Figure \ref{fig:Per_Bl}-(a) and (b) present the perturbation profiles for cases with $T_w=3.33$ across various $R$ values for a specific freestream entropy forcing. $\tilde u_1$ and $\tilde \theta$ approach constant values in the limit of $Y\to \infty$, and  increasing $R$  intensifies these perturbations in the near-wall region. The {\color{red}centreline} perturbation $\tilde w$ is pure imaginary and, according to the matching conditions in (\ref{eq:asym_BC}), behaves as  $-\ri k_3\hat p_{00}/(R_{00}C_v)\ln Y+C_1$ as $Y\to \infty$. Consistent with this logarithmic scaling, panel (a) reveals that  $-\ri \tilde w$ for different $R$ values exhibits parallel   {\color{red}behaviour} for $Y\gg 1$. However, increasing $R$ uniformly amplifies the magnitude of  $\tilde w$  across all $Y$. Furthermore, the matching conditions  (\ref{eq:asym_BC}) predict that $\tilde u_\eta \to k_3^2\hat p_{00}/(R_{00}C_v)Y\ln Y$ as $Y\to \infty$. This divergent {\color{red}behaviour} is clearly observed in panel (b), where increasing $R$ correspondingly elevates the magnitude of  $\tilde u_\eta$ at any given $Y$. 

Figure \ref{fig:Per_Bl}-(c) and (d) further examine the perturbation profiles for a fixed $R$ with various $T_w$ values. Overall, in the far-field region, $Y\gg 1$, the perturbation $\tilde u_1$, $\tilde w$ and $\tilde \theta$ for all $T_w$ values almost collapse. Conversely,   $\tilde u_\eta$ profiles exhibit a constant vertical  shift,   attributed to a high-order correction to the leading-order $O(Y\ln Y)$ term.  In contrast, within the near-wall region,  the  gradients of $\tilde u_1$, $\tilde w$, $\tilde \theta$ and $\tilde u_\eta$  all steepen as  $T_w$ reduces.

In summary, the  magnitudes and near-wall gradients of these perturbations   increase with  $R$ and {\color{cyan}decrease} with $T_w$. Additionally, the asymptotic predictions (circles) exhibit excellent agreement with the SF-HLNS calculations for all cases, thereby confirming the accuracy of our asymptotic theory.
}

\begin{figure}
  \begin{center}
  \includegraphics[width = 0.49\textwidth]{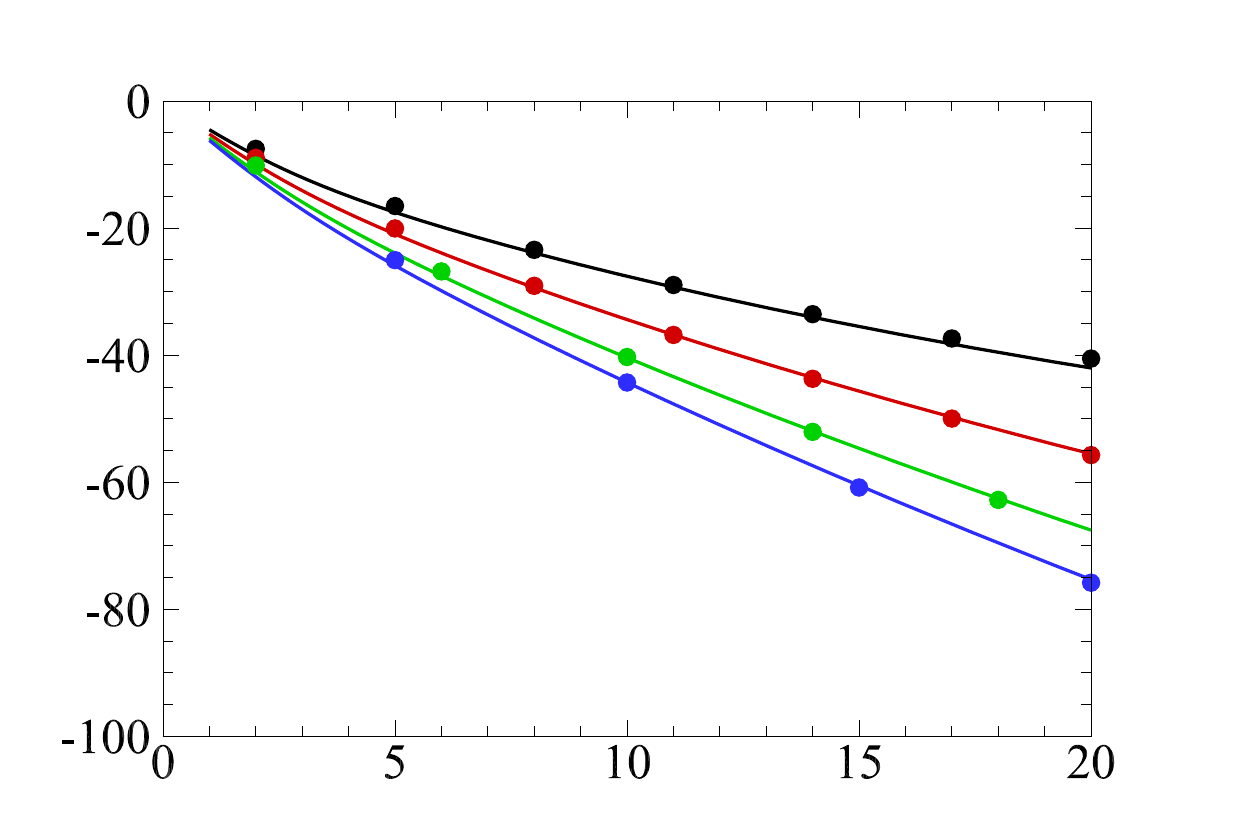}
  \put(-190,110){$(a) $}
  \put(-95,0) {$k_3$}
  \put(-190,15) {\rotatebox{90}{$-\ri\bar\Omega_\xi(0)$, $-\ri R^{-\frac12}\tilde\Omega_{\xi 0}$}}
  \includegraphics[width = 0.49\textwidth]{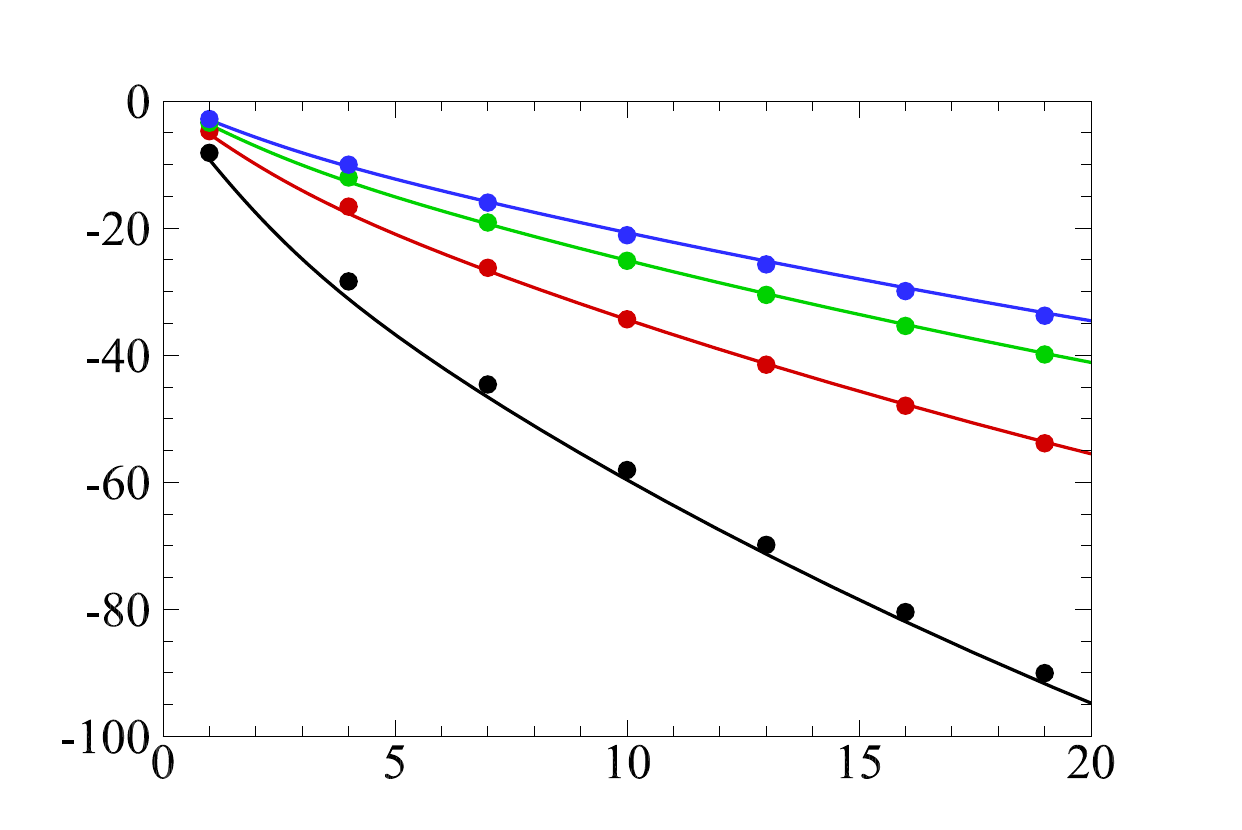}
  \put(-95,0) {$k_3$}
  \put(-190,110){$(b)$}
  \put(-340,33) {\fontsize{5pt}{5pt}\selectfont$RAT2$}
  \put(-340,28) {\fontsize{5pt}{5pt}\selectfont$RBT2$}
  \put(-340,23) {\fontsize{5pt}{5pt}\selectfont$RCT2$}
  \put(-340,18) {\fontsize{5pt}{5pt}\selectfont$RDT2$}
  \put(-320,19){\begin{tikzpicture}
    \draw[black,thick] (0,0) -- (0.5,0);
    \draw[red,thick] (0,-0.18) -- (0.5,-0.18);
    \draw[green,thick] (0,-0.36) -- (0.5,-0.36);
    \draw[blue,thick] (0,-0.54) -- (0.5,-0.54);
    \node[draw,black,circle,fill,inner sep=0.8pt] at (1.0,0) {};
    \node[draw,red,circle,fill,inner sep=0.8pt] at (1.0,-0.18) {};
    \node[draw,green,circle,fill,inner sep=0.8pt] at (1.0,-0.36) {};
    \node[draw,blue,circle,fill,inner sep=0.8pt] at (1.0,-0.54) {};
    \end{tikzpicture}}
  \put(-320,40) {\fontsize{4pt}{4pt}\selectfont $Asym.$}
  \put(-304,40) {\fontsize{4pt}{4pt}\selectfont $SF-HLNS$}
  \put(-150,33) {\fontsize{5pt}{5pt}\selectfont$RBT1$}
  \put(-150,28) {\fontsize{5pt}{5pt}\selectfont$RBT2$}
  \put(-150,23) {\fontsize{5pt}{5pt}\selectfont$RBT3$}
  \put(-150,18) {\fontsize{5pt}{5pt}\selectfont$RBT4$}
  \put(-130,19){\begin{tikzpicture}
    \draw[black,thick] (0,0) -- (0.5,0);
    \draw[red,thick] (0,-0.18) -- (0.5,-0.18);
    \draw[green,thick] (0,-0.36) -- (0.5,-0.36);
    \draw[blue,thick] (0,-0.54) -- (0.5,-0.54);
    \node[draw,black,circle,fill,inner sep=0.8pt] at (1.0,0) {};
    \node[draw,red,circle,fill,inner sep=0.8pt] at (1.0,-0.18) {};
    \node[draw,green,circle,fill,inner sep=0.8pt] at (1.0,-0.36) {};
    \node[draw,blue,circle,fill,inner sep=0.8pt] at (1.0,-0.54) {};
    \end{tikzpicture}}
  \put(-130,40) {\fontsize{4pt}{4pt}\selectfont $Asym.$}
  \put(-114,40) {\fontsize{4pt}{4pt}\selectfont $SF-HLNS$}
  \caption{ Rescaled perturbation streamwise vorticity at the stagnation point for all cases forced by freestream entropy waves ({with $\omega =0$ and $\vartheta= 0^{\circ}$}) vs. $k_3$. Solid lines: asymptotic predictions; circles: {SF-HLNS} calculations. {\color{black} $- \ri \bar{\Omega}_{\xi} (0)$ is for the asymptotic predictions and $- \ri R^{1/2} \bar{\Omega}_{\xi0}$ is for the SF-HLNS calculations.}} \label{fig:e_omega_TW}
  \end{center}
\end{figure}

Of particular interest is the perturbation streamwise vorticity {$\tilde \Omega_{\xi}:=\partial_\eta\tilde w-\partial_z\tilde u_\eta$}, which, to leading order, scales as $R^{1/2}\bar \Omega_\xi=R^{1/2}\bar w_0'$, and serves as the seed for the subsequent lift-up mechanism. {\color{black}Based on the solution (\ref{eq:bar_w_0}), we can predict the streamwise vorticity at the stagnation region 
\begin{equation}
\bar \Omega_\xi(0)=\bar w_0'(0)={\cal C}.
\end{equation}
}
By evaluating the wall-normal gradient of $\tilde w$ at the wall under various freestream forcing, we determine the dependence of the stagnation streamwise vorticity perturbation $\bar \Omega_\xi(0)$ on the key controlling parameters.
Figure \ref{fig:e_omega_TW} displays the asymptotic predictions of $\bar \Omega_\xi(0)$ as a function of $k_3$ across all cases. Since $\bar \Omega_\xi$ is purely imaginary,  its imaginary part is displayed by multiplying by  $-\ri$. Note that  results for case RBT4 were already reported in Part II. Under freestream entropy forcing, the response of $\bar \Omega_\xi(0)$  increases with {$k_3$} for all cases.  At a fixed $k_3$, its amplitude  increases with  $R$, but decreases with $T_w$. 
The SF-HLNS calculations of the stagnation vorticity $\tilde \Omega_{\xi}(0,0)$, denoted by $\tilde\Omega_{\xi 0}$, are  plotted alongside the asymptotic predictions. The two sets of results align closely across the full parameter space, validating the asymptotic model.

\subsubsection{Composite solution}
Combining the solutions in regions III and IV, we can construct a composite solution by combining the boundary-layer perturbation $\bar \phi_0$ with the bulk-layer perturbation $\hat \phi_0$, expressed as
\begin{equation}
\left\{\begin{array}{ll}
    \vspace{.2cm}    \tilde u_\xi&=\Big[\hat u_1(\eta)+\bar u_1(R^{1/2}\eta)-\hat u_1(0)\Big]\xi, \\
       \vspace{.2cm}   \tilde u_\eta&\displaystyle=\hat v_0(\eta)+R^{-1/2}\bar v_0(R^{1/2}\eta)+\frac{k_3^2\hat p_{0}(0)}{R_{00}C_v}\eta\ln \eta-b_0\eta, \\ \vspace{.2cm}
         \tilde w&\displaystyle=\hat w_0(\eta) +\bar w_0(R^{1/2}\eta)+\frac{\ri k_3\hat p_{0}(0)}{R_{00}C_v}{\ln\eta}-C_1,\\ \vspace{.2cm}
(\tilde \rho,\tilde \theta)&=\Big(\hat \rho_0(\eta),\hat \theta_0(\eta)\Big)+\Big(\bar \rho_0(R^{1/2}\eta),\bar\theta_0(R^{1/2}\eta)\Big)-\Big(\hat \rho_0(0),\hat \theta_0(0)\Big),\\
\tilde p&=\hat p_0(\eta),
       \end{array}
\right.\label{eq:composite}
\end{equation}
where $(\hat u_1,\hat v_0,\hat w_0,\hat \rho_0,\hat \theta_0,\hat p_0)$ are leading-order solutions in region III (illustrated in Section 3.2.1 of Part II), and the constants $b_0$ and $C_1$ can be obtained by fitting the asymptotic behaviours of $\bar v_0$ and $\bar w_0$ in the large-$Y$ limit. 

\begin{figure}
  \begin{center}
  \includegraphics[width = 0.49\textwidth]{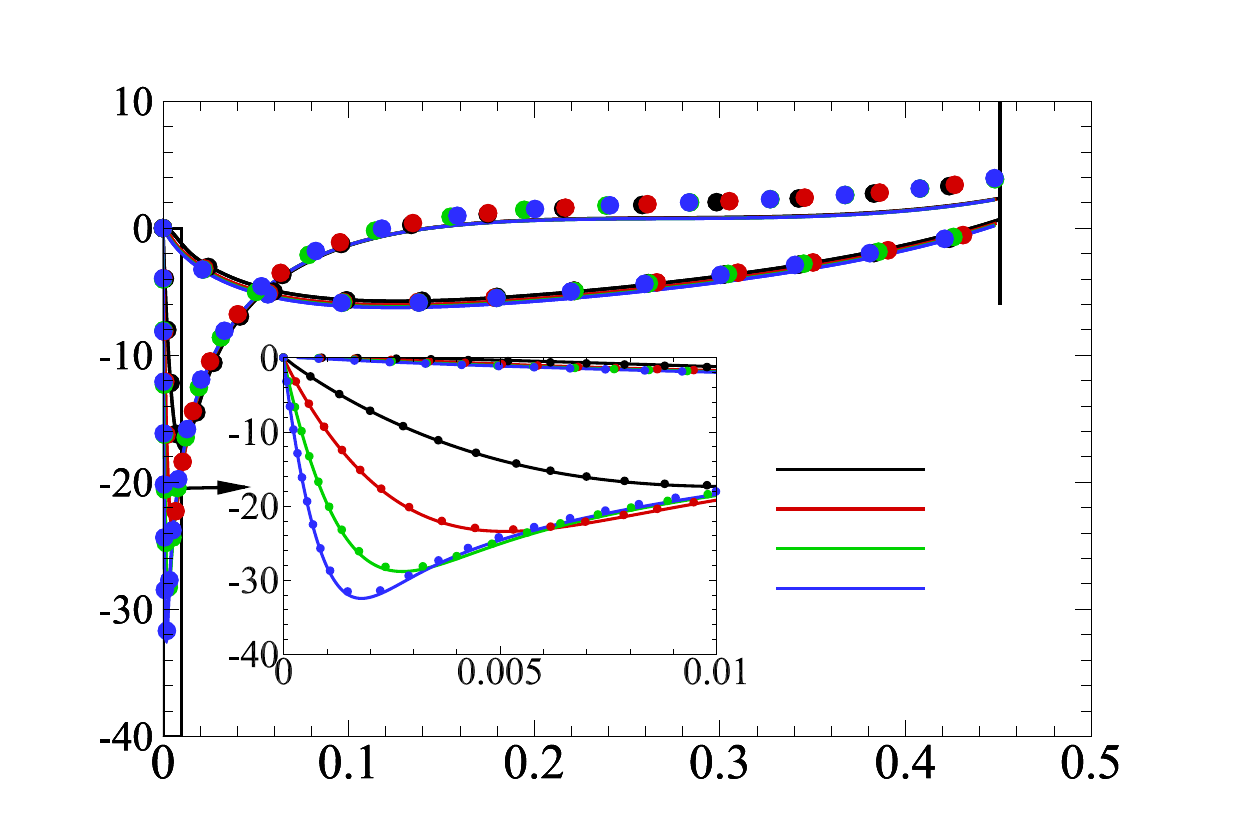}  
  \includegraphics[width = 0.49\textwidth]{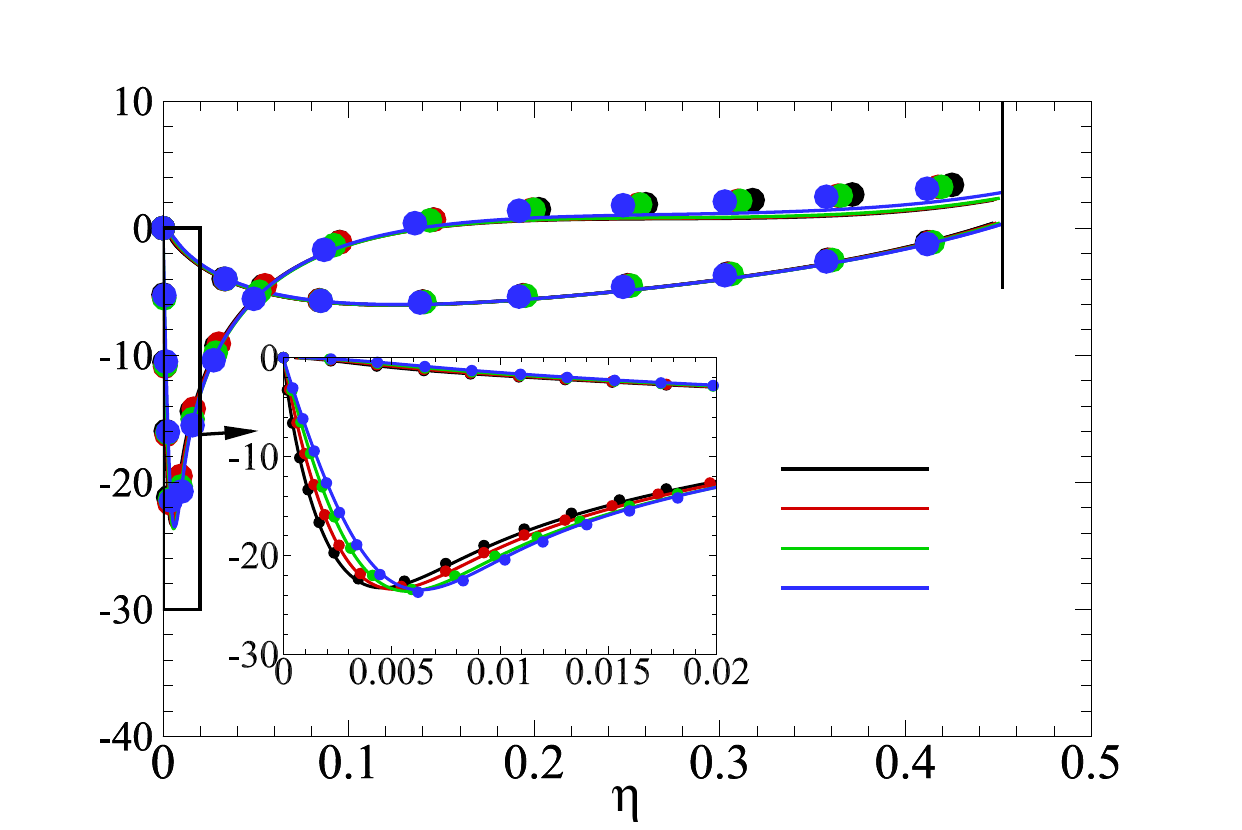}
  \put(-380,120) {$(a)$}
  \put(-190,120) {$(b)$}
  \put(-92,5){$\eta$}
  \put(-285,5){$\eta$}
  \put(-185,50){\rotatebox{90}{$\tilde u_\eta,-\ri\tilde w$}} 
  \put(-70,77){$\tilde u_\eta$}
  \put(-90,97){$-\ri\tilde w$}
  \put(-45,52){\fontsize{5pt}{5pt}\selectfont$RBT1$}
  \put(-45,46){\fontsize{5pt}{5pt}\selectfont$RBT2$}
  \put(-45,40){\fontsize{5pt}{5pt}\selectfont$RBT3$}
  \put(-45,34){\fontsize{5pt}{5pt}\selectfont$RBT4$}
  \put(-100,40){$-\ri\tilde w$}
  \put(-100,60){$\tilde u_\eta$}
  \put(-375,50){\rotatebox{90}{$\tilde u_\eta,-\ri\tilde w$}} 
  \put(-260,77){$\tilde u_\eta$}
  \put(-280,97){$-\ri\tilde w$}
  \put(-238,52){\fontsize{5pt}{5pt}\selectfont$RAT2$}
  \put(-238,46){\fontsize{5pt}{5pt}\selectfont$RBT2$}
  \put(-238,40){\fontsize{5pt}{5pt}\selectfont$RCT2$}
  \put(-238,34){\fontsize{5pt}{5pt}\selectfont$RDT2$}
  \put(-290,40){$-\ri\tilde w$}
  \put(-290,62){$\tilde u_\eta$}
  \put(-240,75){Shock}
  \put(-50,75){Shock}
  \caption{Profiles of $\tilde u_\eta$ and $\tilde w$ at the {\color{red}centreline}  under entropy forcing with $\omega =0$, $\vartheta= 0^{\circ}$ and $k_{3}=8$. (a): Cases for $T_w=3.33$ with various $R$; (b): cases for $R=1.67\times 10^5$ with various $T_w$. In the present study, $\tilde u_\eta$ is real, while $\tilde w$ is pure imaginary.}
  \label{fig:com_vw_h_ek3=8_xi=0}
  \end{center}
\end{figure}

Under the same freestream entropy wave forcing, the profiles of the composite solutions $\tilde u_\eta$ and $\tilde w$ are displayed in figure \ref{fig:com_vw_h_ek3=8_xi=0}. The asymptotic predictions are in good agreement with the SF-HLNS calculations. In the near‑wall region, $\tilde w$ increases {\color{red}remarkablely} and peaks at a position where $\eta\sim R^{-1/2}$. As seen in panel (a), the peak amplitude of $\tilde w$ is greater for a higher Reynolds number. When $R$ is fixed, as illustrated in panel (b), variations in wall temperature do not alter the peak value of the  $\tilde w$ profile. However,  the peak locations shift toward the wall as $T_w$ reduces, indicating a strengthening of the streamwise vorticity  $\tilde \Omega_\xi$.

\subsection{Evolution of non-modal perturbation in  downstream boundary layers}
We now turn our attention to step (iii). Upon obtaining the composite solution (\ref{eq:composite}), we employ the LPSE approach to calculate the downstream evolution of non-modal perturbations, using the composite solution as the initial condition. The scaling relations and LPSE approach are identical to those in section 5 of Part II. 

\begin{figure}
  \begin{center}
  \includegraphics[width = 0.49\textwidth]{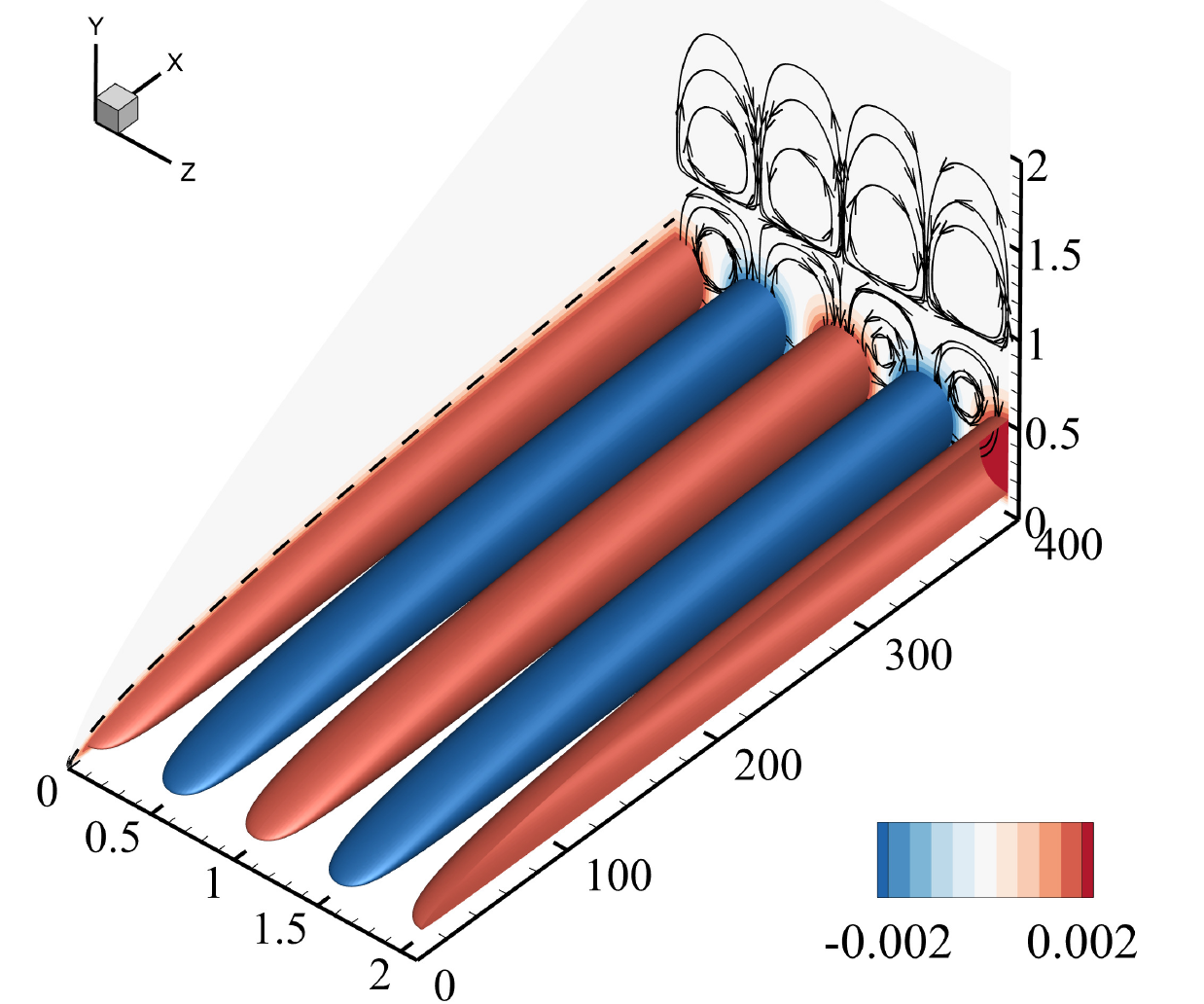} 
  \includegraphics[width = 0.49\textwidth]{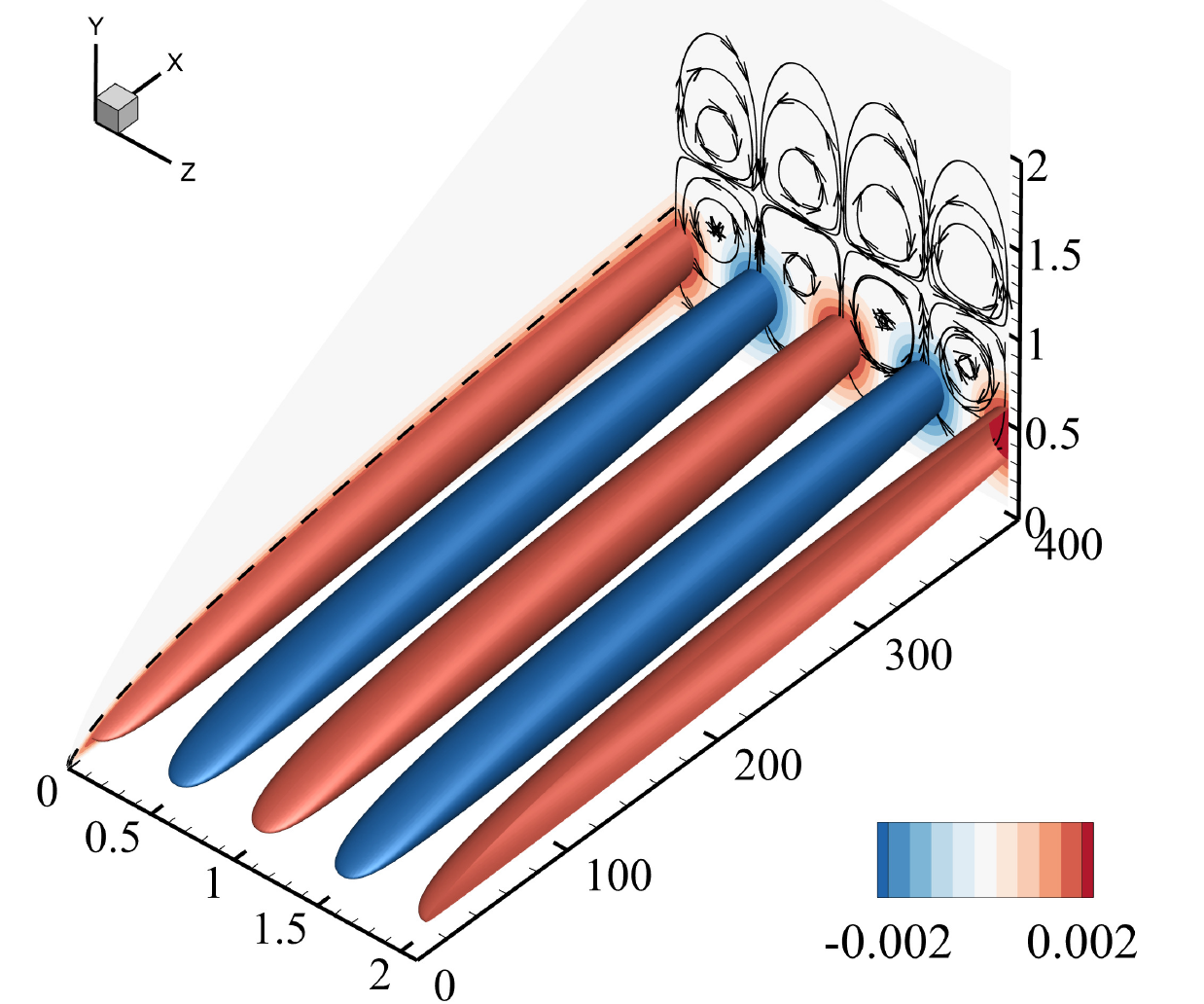}
  \put(-380,150) {$(a)$}
  \put(-205,100) {$\eta $}
  \put(-360,5) {$z/z_l$}
  \put(-265,25) {$\xi$}
  \put(-236,33) {$R^{-1} {\bar u_\xi}$}
  \put(-320,105) {B.L.} 
  \put(-190,150) {$(b)$}
  \put(-15,100) {$\eta $}
  \put(-170,5) {$z/z_l$}
  \put(-75,25) {$\xi$}
  \put(-46,33) {$R^{-1} {\bar u_\xi}$}
  \put(-130,105) {B.L.}
  \\
  \includegraphics[width = 0.49\textwidth]{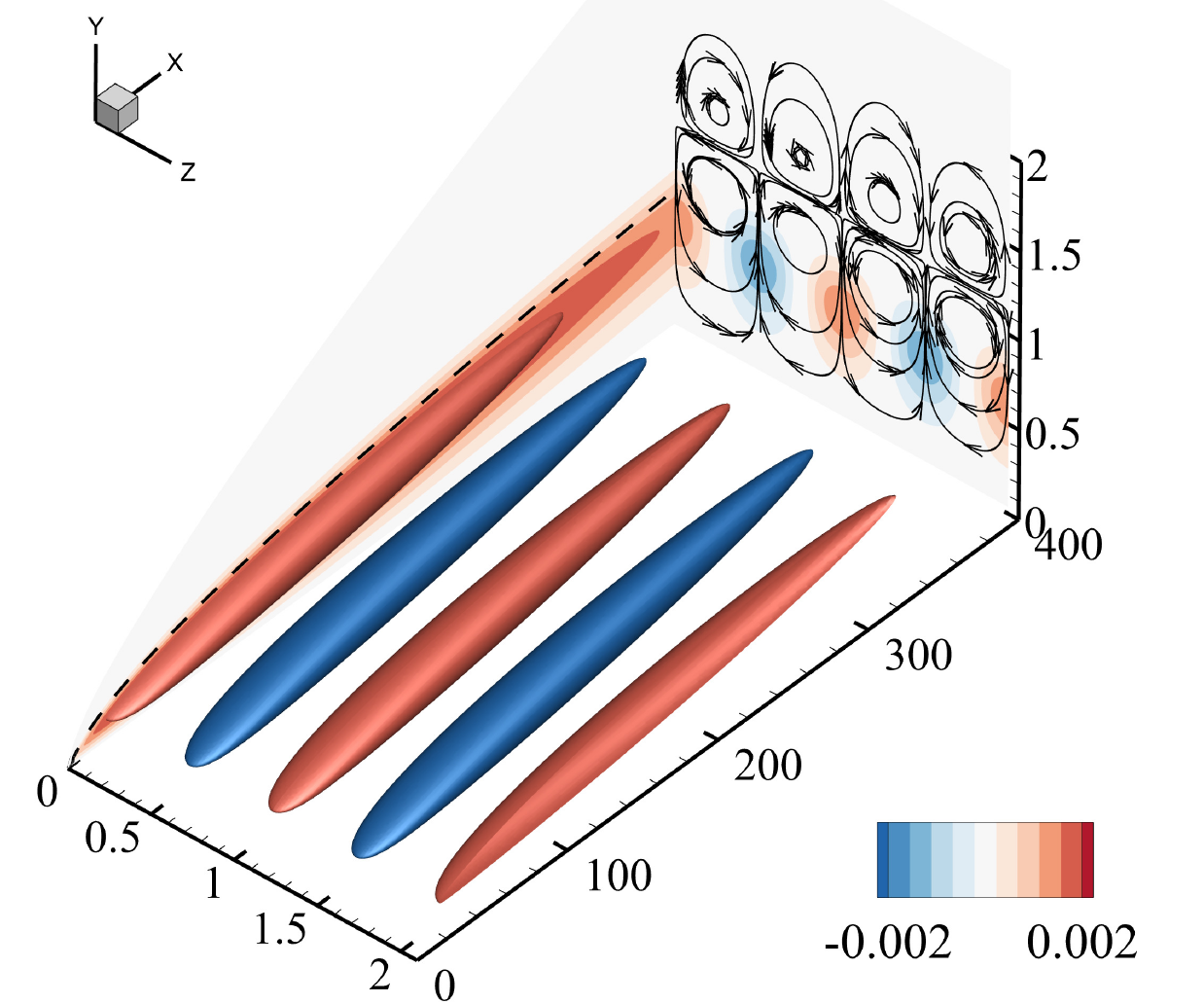}  
  \includegraphics[width = 0.49\textwidth]{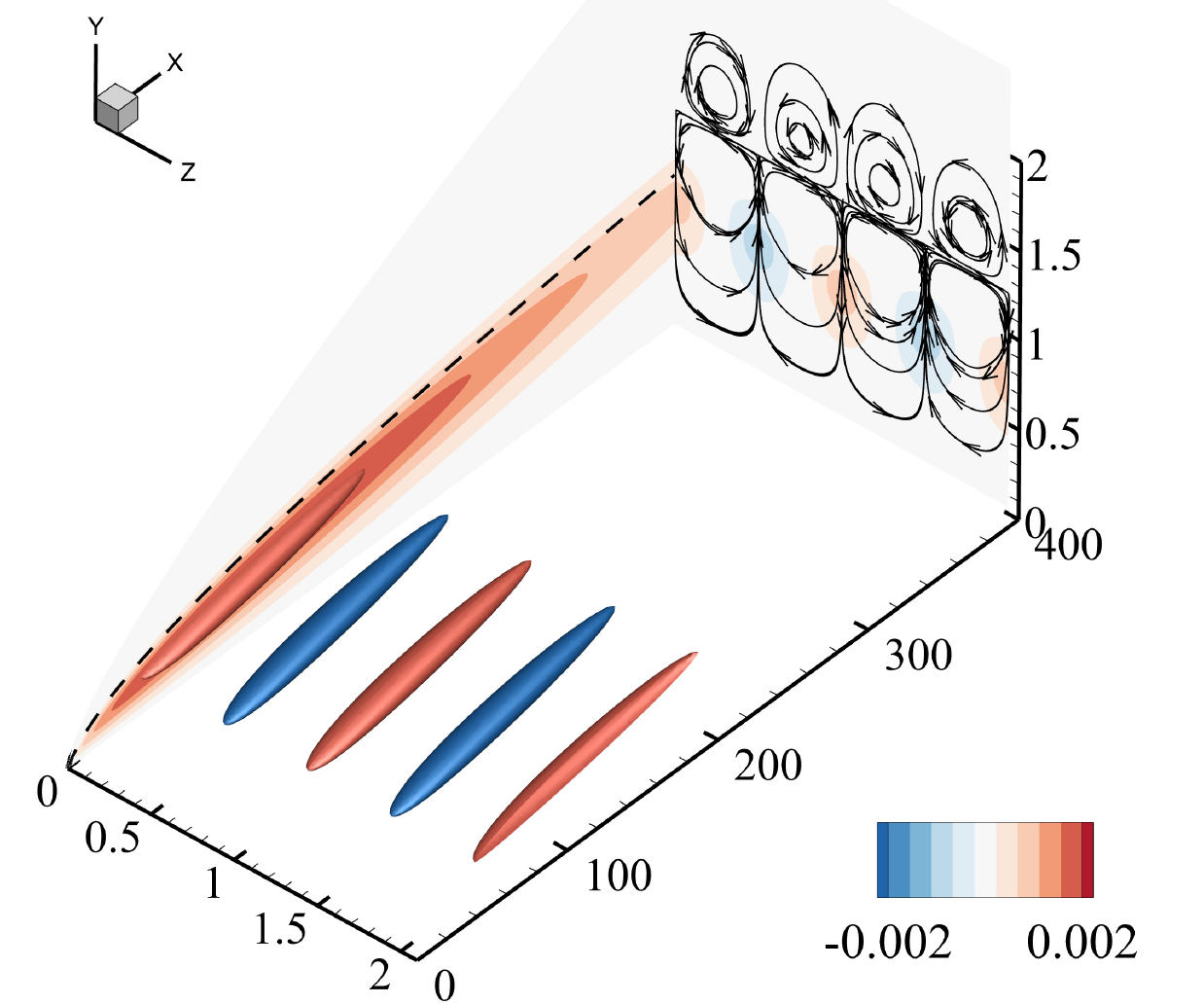}
  \put(-380,150) {$(c)$}
  \put(-205,100) {$\eta $}
  \put(-360,5) {$z/z_l$}
  \put(-265,25) {$\xi$}
  \put(-236,33) {$R^{-1} {\bar u_\xi}$}
  \put(-320,105) {B.L.} 
  \put(-190,150) {$(d)$}
  \put(-15,100) {$\eta $}
  \put(-170,5) {$z/z_l$}
  \put(-75,25) {$\xi$}
  \put(-46,33) {$R^{-1} {\bar u_\xi}$}
  \put(-130,105) {B.L.}
  \caption{LPSE predictions of non-modal perturbations identified by iso-surface of $R^{-1} {\bar u_\xi}=\pm0.002$ for cases RBT1 (a), RBT2 (b), RBT3 (c) and RBT4 (d). Cross-plane streamlines are shown at $\xi=400$, and the dashed lines denote boundary-layer edges. $z_l=2\pi/k_3$. Freestream forcing follows that in figure \ref{fig:com_vw_h_ek3=8_xi=0}.}
  \label{fig:con_streak_T}
  \end{center}
\end{figure}
In figure \ref{fig:con_streak_T},
the streamwise evolution of   non-modal perturbations computed via LPSE is displayed, with the initial perturbations  taken from the composite solutions presented in figure \ref{fig:com_vw_h_ek3=8_xi=0}. {\color{black}As revealed in Part II, the LPSE calculations are insensitive to  $X_0$ as long as $X_0\ll 1$ and the location is downstream of the elliptic zone around the stagnation point. Therefore,  we choose  $X_0=0.2$ for all cases, consistent with Part II.}  Since the downstream non-modal amplification driven by the lift-up mechanism scales with $R$, the perturbation velocity $\tilde u_\xi$ is rescaled by $R^{-1}$. For demonstration, only cases with {$R=1.67\times 10^5$} and various wall temperatures are plotted. Distinct longitudinal streaky structures are identified downstream, with their magnitudes increasing with reducing $T_w$. At $\xi=400$, the cross-plane velocity fields reveal roll-type structures, with the centre of each roll cell situated near the boundary-layer edge. Because the boundary-layer thickness increases with $T_w$, the position of the roll centre shifts outward as $T_w$ rises.

\begin{figure}
  \begin{center}
  \includegraphics[width = 0.49\textwidth]{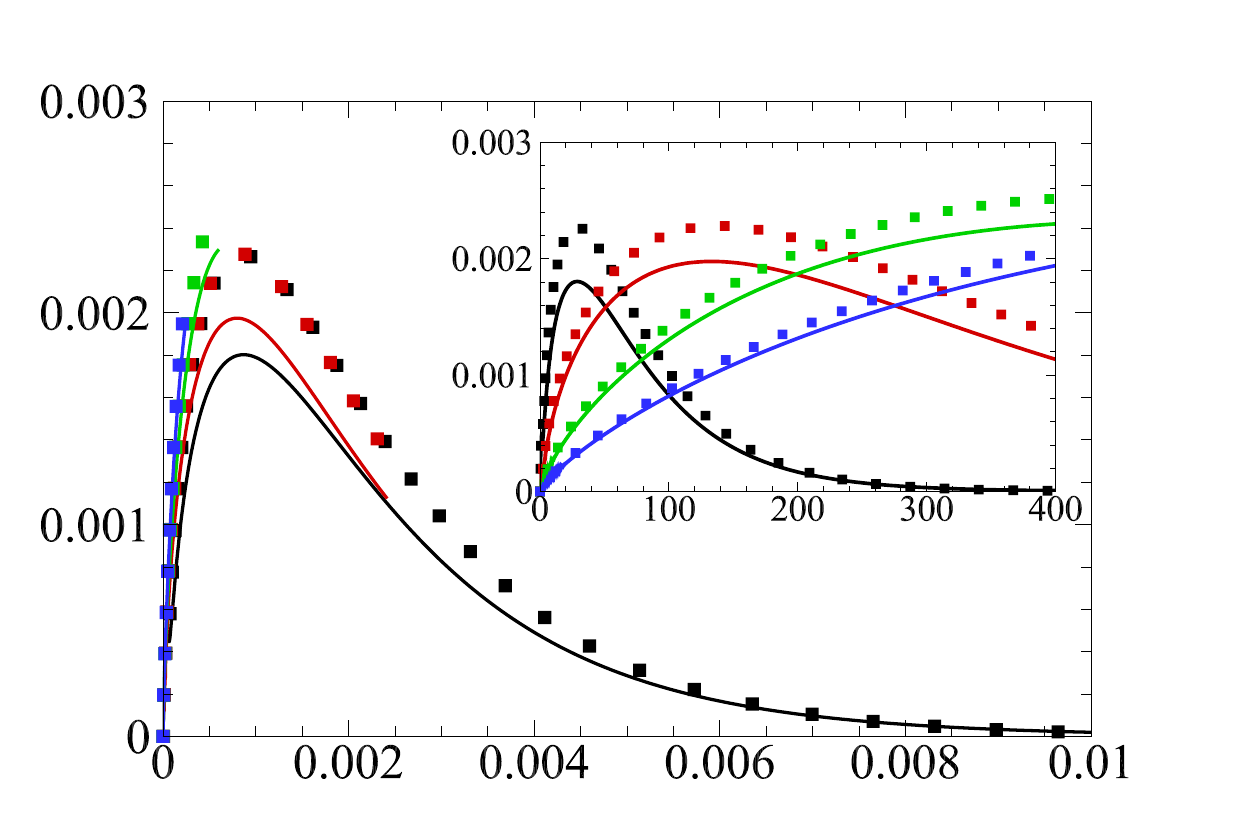}  
  \includegraphics[width = 0.49\textwidth]{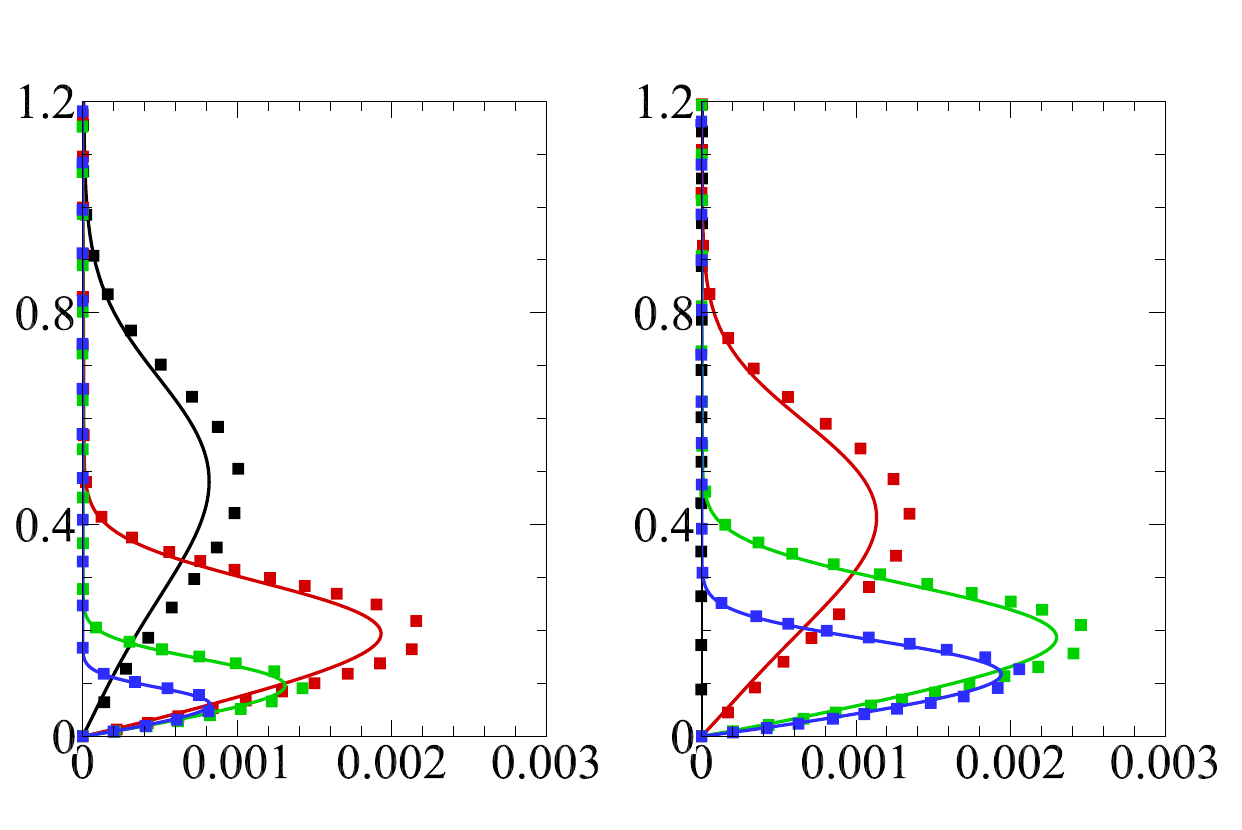}
  \put(-380,120) {$(a)$}
  \put(-195,120) {$(c)$}
  \put(-100,120) {$(e)$}
  \put(-190,60) {$\eta $}
  \put(-385,60) {$R^{-1}A_u$}
  \put(-260,35) {$\xi$}
  \put(-295,-5) {$R^{-1}\xi$}
  \put(-57,102) {\fontsize{6pt}{6pt}\selectfont SF-HLNS}
  \put(-28,102) {\fontsize{6pt}{6pt}\selectfont LPSE}
  \put(-72,95) {\fontsize{6pt}{6pt}\selectfont$RAT2$}
  \put(-72,88) {\fontsize{6pt}{6pt}\selectfont$RBT2$}
  \put(-72,81) {\fontsize{6pt}{6pt}\selectfont$RCT2$}
  \put(-72,74) {\fontsize{6pt}{6pt}\selectfont$RDT2$}
  \put(-50,75){\begin{tikzpicture}
    \draw[black,thick] (0,0) -- (0.4,0);
    \draw[red,thick] (0,-0.25) -- (0.4,-0.25);
    \draw[green,thick] (0,-0.5) -- (0.4,-0.5);
    \draw[blue,thick] (0,-0.75) -- (0.4,-0.75);
    \node[draw,black,rectangle,fill,inner sep=0.8pt] at (1.0,0) {};
    \node[draw,red,rectangle,fill,inner sep=0.8pt] at (1.0,-0.25) {};
    \node[draw,green,rectangle,fill,inner sep=0.8pt] at (1.0,-0.5) {};
    \node[draw,blue,rectangle,fill,inner sep=0.8pt] at (1.0,-0.75) {};
    \end{tikzpicture}}\\
  \includegraphics[width = 0.49\textwidth]{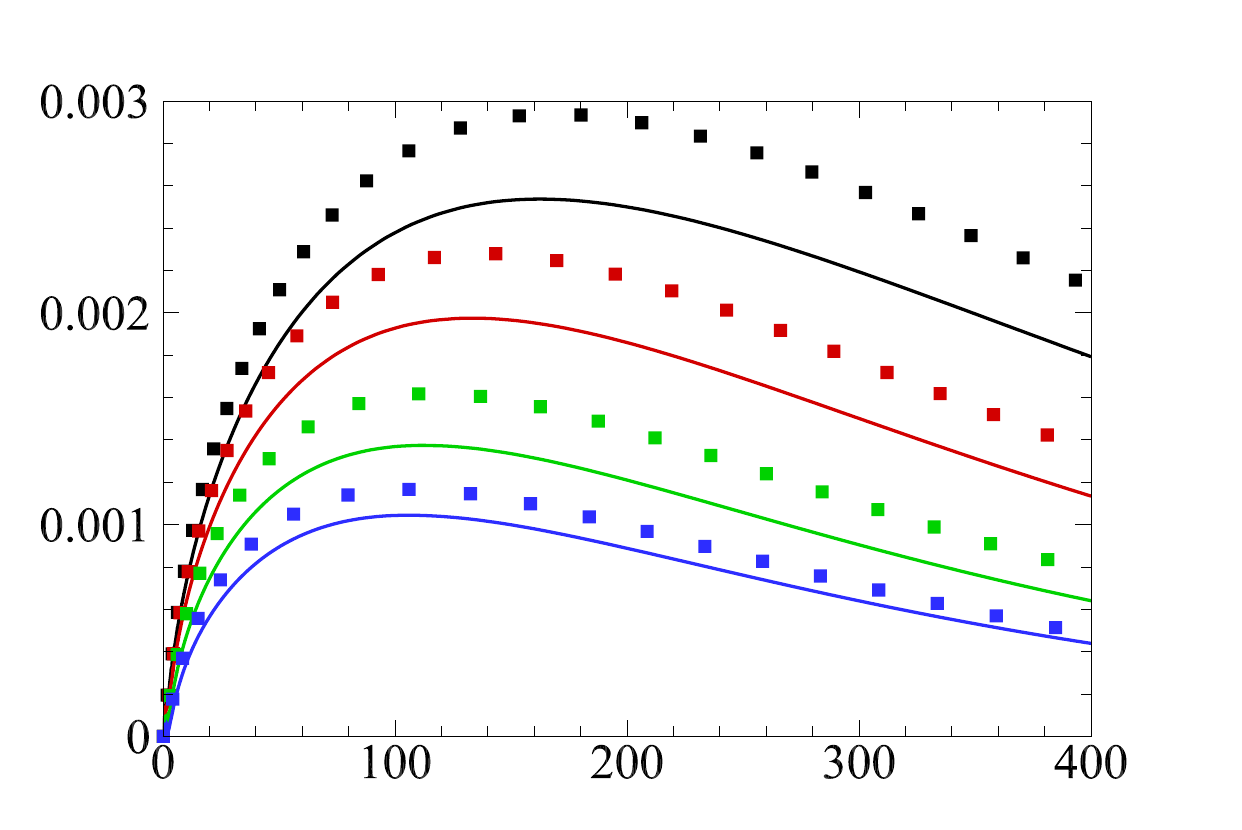}  
  \includegraphics[width = 0.49\textwidth]{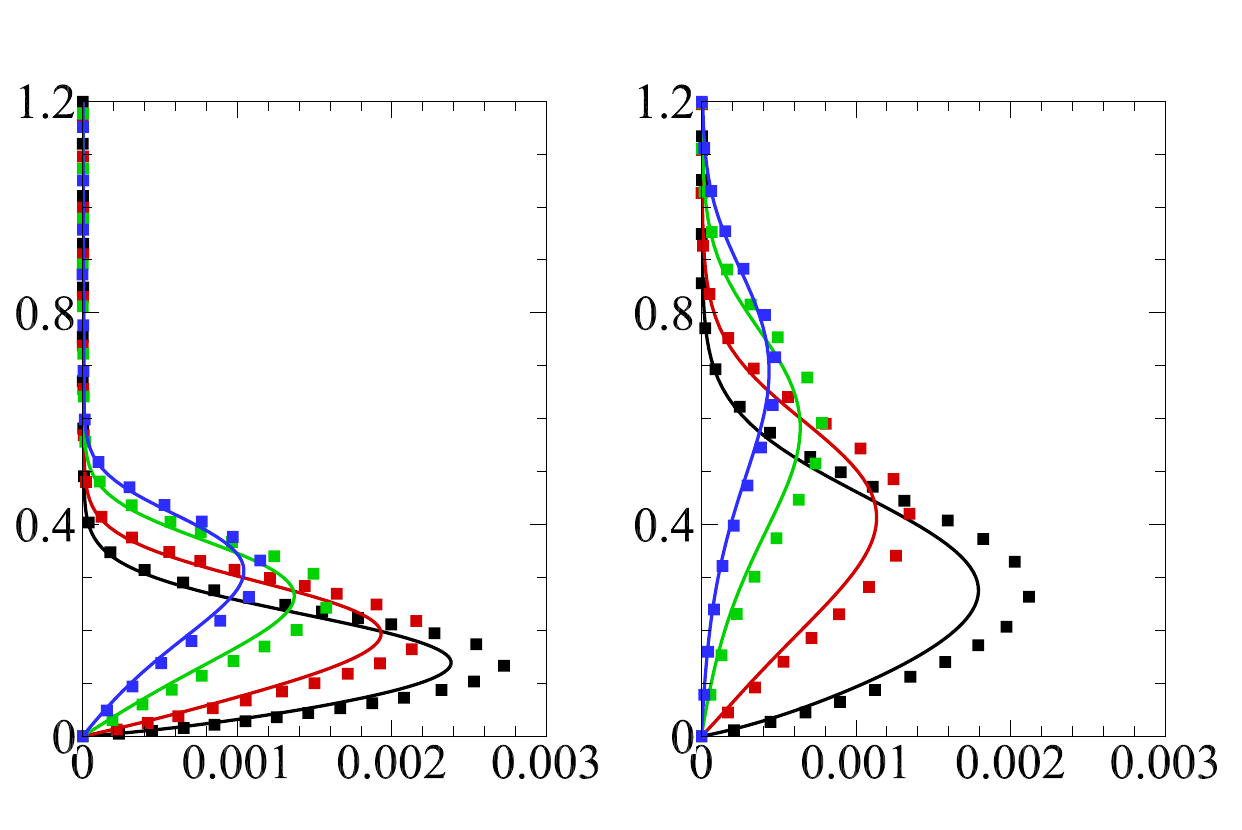}
  \put(-380,120) {$(b)$}
  \put(-195,120) {$(d)$}
  \put(-100,120) {$(f)$}
  \put(-190,60) {$\eta $}
  \put(-385,60) {$R^{-1}A_u$}
  \put(-285,-5) {$\xi$}
  \put(-155,-5) {$R^{-1}|\tilde{u}_{\xi}| $}
  \put(-60,-5) {$R^{-1}|\tilde{u}_{\xi}| $}
  \put(-57,102) {\fontsize{6pt}{6pt}\selectfont SF-HLNS}
  \put(-28,102) {\fontsize{6pt}{6pt}\selectfont LPSE}
  \put(-72,95) {\fontsize{6pt}{6pt}\selectfont$RBT1$}
  \put(-72,88) {\fontsize{6pt}{6pt}\selectfont$RBT2$}
  \put(-72,81) {\fontsize{6pt}{6pt}\selectfont$RBT3$}
  \put(-72,74) {\fontsize{6pt}{6pt}\selectfont$RBT4$}
  \put(-50,75){\begin{tikzpicture}
    \draw[black,thick] (0,0) -- (0.4,0);
    \draw[red,thick] (0,-0.25) -- (0.4,-0.25);
    \draw[green,thick] (0,-0.5) -- (0.4,-0.5);
    \draw[blue,thick] (0,-0.75) -- (0.4,-0.75);
    \node[draw,black,rectangle,fill,inner sep=0.8pt] at (1.0,0) {};
    \node[draw,red,rectangle,fill,inner sep=0.8pt] at (1.0,-0.25) {};
    \node[draw,green,rectangle,fill,inner sep=0.8pt] at (1.0,-0.5) {};
    \node[draw,blue,rectangle,fill,inner sep=0.8pt] at (1.0,-0.75) {};
    \end{tikzpicture}}
  \caption{$(a,b)$ Streamwise evolution of $R^{-1}A_u$;  $(c,d)$ profiles of $R^{-1}|\tilde{u}_{\xi}|$ at $\xi=$ 100;  $(e,f)$  profiles of $R^{-1}|\tilde{u}_{\xi}|$ at $\xi=$400. Top row: cases with $T_w=3.33$ for various  $R$; bottom row: cases with $R=1.67\times 10^5$ for  various $T_w$. The freestream forcing is the same as in figure \ref{fig:com_vw_h_ek3=8_xi=0}.}
  \label{fig:LPSE}
  \end{center}
\end{figure}

To quantify the amplitude of the non-modal perturbations and verify  their accuracy, {figures} \ref{fig:LPSE}-(a) and (b) compare the streamwise evolution of perturbation tangential velocity obtained by {SF-HLNS} and LPSE across all the cases.  Consistent with the established scaling,  the  amplitude is rescaled by $R^{-1}A_u$. {\color{black}The inset in panel (a) illustrates the variation of $R^{-1}A_u$ with  $\xi$ for a fixed wall temperature ($T_w=3.33$) and spanwise wavenumber ($k_3=8$). This reveals that   the transient growth peaks at   $R^{-1}A_u\approx 0.002$ for each case,  with the peaking location shifting downstream as $R$ increases.  Crucially, upon rescaling the streamwise coordinate by   $R^{-1}\xi$}, the curves for different $R$ values almost collapse,   {\color{black}peaking} at $R^{-1}\xi\approx {\color{black}0.0008}$.
With decreasing $T_w$ at a constant $R$, the {\color{black}peak} shifts slightly downstream, while its  amplitude intensifies. Panels (c)-(f) compare the perturbation profiles at representative locations between the SF-HLNS and LPSE calculations. Overall, the LPSE predictions can reproduce the same trends as the {SF-HLNS} calculations, capturing both the streamwise peak locations and the wall-normal profile shapes with high fidelity.   {\color{black}However, a consistent overestimation is observed for all the cases considered. The relative errors, defined as ${\cal E}_r=|A_{u,LPSE}(\xi)-A_{u,SF-HLNS}(\xi)|/A_{u,SF-HLNS}(\xi)$, are tabulated  at $\xi=400$  in Table \ref{tab:error}. These results demonstrate a clear reduction in error as   $R$ increases, and further show that the asymptotic predictions are most accurate for an adiabatic wall. These behaviours are expected and explained as follows. First, our asymptotic model neglects terms of $O(R^{-1/2})$, treating them as higher-order contributions. These neglected terms grow as $R$ decreases, which explains why the discrepancy becomes more pronounced at lower $R$ values.
Second, for a fixed $R$, variations in wall temperature alter the boundary-layer temperature profile. While the asymptotic model accurately captures the zero temperature gradient for an adiabatic wall, decreasing $T_w$ induces increasingly steep temperature gradients. This causes the second-order terms in the asymptotic model to become progressively larger, thereby explaining the greater discrepancy at lower wall temperatures. {\color{red}Notably, the magnitude of the LPSE-prediction error is comparable to the typical discrepancies  reported between high-fidelity numerical simulations and wind-tunnel experiments. Given the low computational cost of LPSE, this level of accuracy indicates that the model holds considerable promise  for practical engineering applications.}

\begin{table}
  \centering
  \begin{tabular}{lccccccc}
  Case  & \textbf{RAT2} & \textbf{RBT2} & \textbf{RCT2} & \textbf{RDT2} & \textbf{RBT1} & \textbf{RBT3} & \textbf{RBT4}  \\
    ${\cal E}_r$($\%$) & 35.9 & 18.7 & 9.51 & 7.00 & 18.4 & 18.0 & 9.07 \\
  \end{tabular}
  \caption { \color{black}{Relative error ${\cal E}_r$ of the LPSE predictions against SF-HLNS calculations at $\xi=400$ for cases presented in figure \ref{fig:com_vw_h_ek3=8_xi=0}.}}
  \label{tab:error}
\end{table}

}

{\color{black}It should be noted that we do not expect exact quantitative agreement between the numerical (SF-HLNS) results  and the asymptotic predictions (composite solution+LPSE). Rather, our objective is to simplify the governing equations to capture the leading-order physics in the high-$R$ limit, which is essential for elucidating the underlying mechanisms. While we accept the presence of numerical errors at finite Reynolds numbers, we have verified that these errors {\color{red}reduce} as $R$ increases and $T_w$ approaches $T_{ad}$.  Therefore, based on the physical mechanisms revealed by the asymptotic model, we will explore a parametric study in the following section, aiming at revealing key scaling relations in the non-modal receptivity processes.}

\section{Parametric studies}
\label{sec:5}
{\color{black}Since all seven cases in Table \ref{tab:1} share the same oncoming Mach number, the post-shock perturbations determined in step (i) remain identical across these cases for a given free-stream forcing. The same invariance holds for the perturbation {\color{red}behaviour} in Region III. Therefore,  the effects of wall temperature and Reynolds number manifest exclusively within the stagnation region (Region IV) and the downstream boundary layer (Region II). The subsequent analyses would be focused on these two regions.
}
\subsection {Impact of Reynolds number (nose radius) on receptivity efficiency}

In this subsection, we study the non-modal receptivity for cases with various nose radii to reveal the Reynolds number effect. As noted in ZD25, the transient growth {\color{black}reaches its peak}  when the boundary-layer thickness $\delta_{BL}$ becomes comparable {\color{black}in magnitude} with the spanwise {wavelength} $2\pi/k_3$. Accordingly, we normalise the spanwise wavenumber as $R^{-1/2}k_3$ for the subsequent analyses. {\color{black}The scaling {\color{red}behaviour} of  $\tilde\Omega_{\xi 0}$ is governed primarily  by the following three factors. (i) For freestream  forcing normalised to a unity energy norm  in (\ref{eq:estimate_FS}), the amplitudes of acoustic and entropy perturbations are insensitive to $k_3$, while the amplitude of vortical forcing is inversely proportional to $k_3$. Therefore, for a fixed  $R^{-1/2}k_3$, the amplitudes of the freestream acoustic/entropy forcing follow an $R^0$ scaling, while the  freestream vortical amplitude varies as $R^{-1/2}$. (ii) In the spanwise momentum equation for region IV, (\ref{eq:perturbation_BL}c),  $\bar w_0$ varies linearly with both $k_3$ and $\hat p_{00}$. Here, $k_3$ scales as $R^{1/2}$ and $\hat p_{00}$ is proportional to the freestream forcing amplitude. It thus follows that $\tilde w\sim\bar w_0\sim R^{1/2}$ for   acoustic/entropy forcing, and $\tilde w\sim\bar w_0\sim R^{0}$  for  vortical forcing. (iii) Within region IV, the streamwise vortical perturbation  $\tilde \Omega_{\xi 0}\sim d\tilde w/d\eta$, whose magnitude scales with  $\bar w$ divided by the stagnation boundary-layer thickness ($O(R^{-1/2})$). As a result, the stagnation vorticity perturbation $\tilde\Omega_{\xi 0}$ scales with $R$ under acoustic/entropy forcing, and with $R^{1/2}$ under vortical forcing.}

\begin{figure}
  \begin{center}
  \includegraphics[width = 0.49\textwidth]{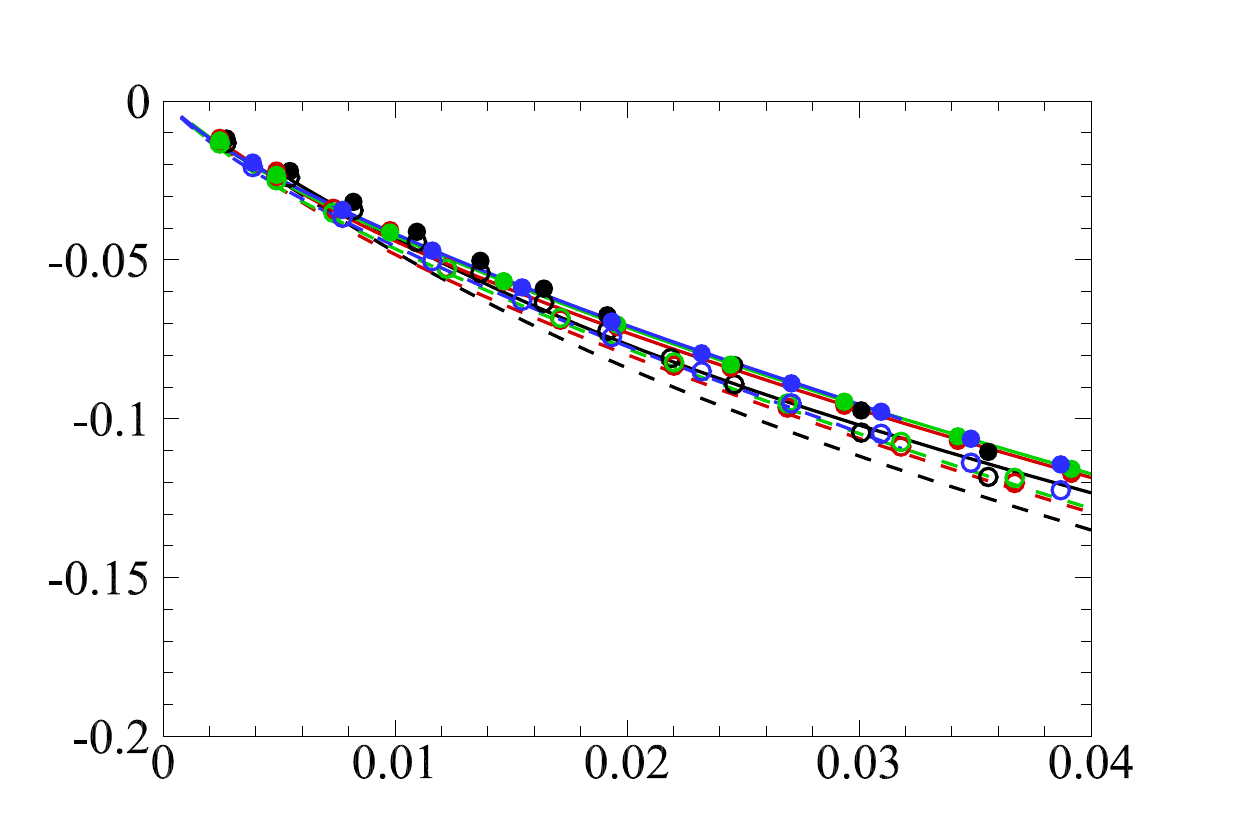} 
  \includegraphics[width = 0.49\textwidth]{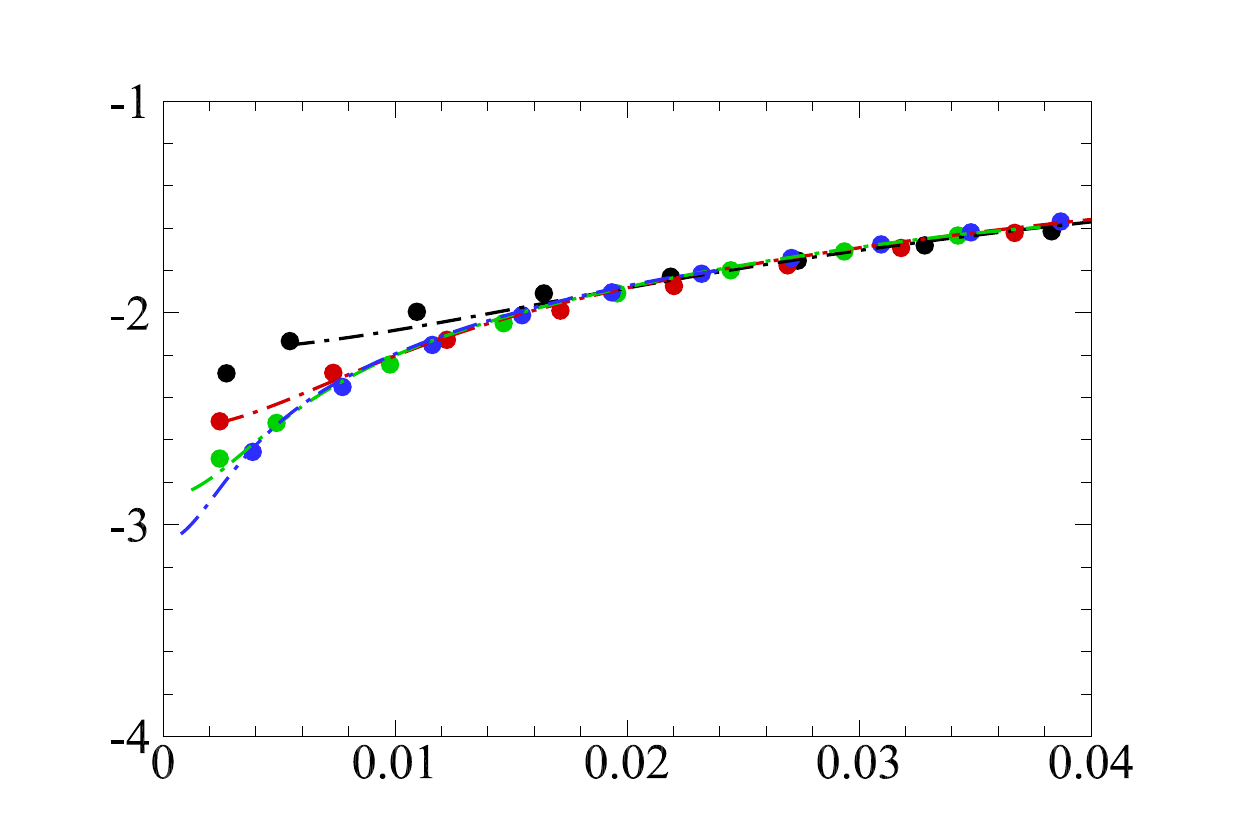} 
  \put(-380,110){$(a) $}
  \put(-190,110){$(b) $}
  \put(-110,-2) { $R^{-1/2}k_3$}
  \put(-300,-2) { $R^{-1/2}k_3$}
  \put(-380,48) {\rotatebox{90}{$R^{-1}\tilde\Omega_{\xi 0}$}}
  \put(-190,48) {\rotatebox{90}{$R^{-1/2}\tilde\Omega_{\xi 0}$}}
  \put(-350,33) {\fontsize{5pt}{5pt}\selectfont$RAT2$}
  \put(-350,28) {\fontsize{5pt}{5pt}\selectfont$RBT2$}
  \put(-350,23) {\fontsize{5pt}{5pt}\selectfont$RCT2$}
  \put(-350,18) {\fontsize{5pt}{5pt}\selectfont$RDT2$}
  \put(-330,19){\begin{tikzpicture}
    \draw[black,thick] (0,0) -- (0.5,0);
    \draw[red,thick] (0,-0.18) -- (0.5,-0.18);
    \draw[green,thick] (0,-0.36) -- (0.5,-0.36);
    \draw[blue,thick] (0,-0.54) -- (0.5,-0.54);
    \node[draw,black,circle,fill,inner sep=0.8pt] at (1.0,0) {};
    \node[draw,red,circle,fill,inner sep=0.8pt] at (1.0,-0.18) {};
    \node[draw,green,circle,fill,inner sep=0.8pt] at (1.0,-0.36) {};
    \node[draw,blue,circle,fill,inner sep=0.8pt] at (1.0,-0.54) {};
    \end{tikzpicture}}
  \put(-330,40) {\fontsize{4pt}{4pt}\selectfont $Asym.$}
  \put(-314,40) {\fontsize{4pt}{4pt}\selectfont $SF-HLNS$}
  \put(-330,45) {\fontsize{5pt}{5pt}\selectfont $Entropy$}
  \put(-285,19){\begin{tikzpicture}
    \draw[black,thick,dashed] (0,0) -- (0.5,0);
    \draw[red,thick,dashed] (0,-0.18) -- (0.5,-0.18);
    \draw[green,thick,dashed] (0,-0.36) -- (0.5,-0.36);
    \draw[blue,thick,dashed] (0,-0.54) -- (0.5,-0.54);
    \node[draw,black,circle,inner sep=0.8pt] at (1,0) {};
    \node[draw,red,circle,inner sep=0.8pt] at (1,-0.18) {};
    \node[draw,green,circle,inner sep=0.8pt] at (1,-0.36) {};
    \node[draw,blue,circle,inner sep=0.8pt] at (1,-0.54) {};
    \end{tikzpicture}}
  \put(-285,40) {\fontsize{4pt}{4pt}\selectfont $Asym.$}
  \put(-269,40) {\fontsize{4pt}{4pt}\selectfont $SF-HLNS$}
  \put(-285,45) {\fontsize{5pt}{5pt}\selectfont $Acoustic$}
  \put(-90,33) {\fontsize{5pt}{5pt}\selectfont$RAT2$}
  \put(-90,28) {\fontsize{5pt}{5pt}\selectfont$RBT2$}
  \put(-90,23) {\fontsize{5pt}{5pt}\selectfont$RCT2$}
  \put(-90,18) {\fontsize{5pt}{5pt}\selectfont$RDT2$}
  \put(-70,19){\begin{tikzpicture}
    \draw[black,thick,dashdotted] (0,0) -- (0.5,0);
    \draw[red,thick,dashdotted] (0,-0.18) -- (0.5,-0.18);
    \draw[green,thick,dashdotted] (0,-0.36) -- (0.5,-0.36);
    \draw[blue,thick,dashdotted] (0,-0.54) -- (0.5,-0.54);
    \node[draw,black,circle,fill,inner sep=0.8pt] at (1,0) {};
    \node[draw,red,circle,fill,inner sep=0.8pt] at (1,-0.18) {};
    \node[draw,green,circle,fill,inner sep=0.8pt] at (1,-0.36) {};
    \node[draw,blue,circle,fill,inner sep=0.8pt] at (1,-0.54) {};
    \end{tikzpicture}}
  \put(-70,40) {\fontsize{4pt}{4pt}\selectfont $Asym.$}
  \put(-54,40) {\fontsize{4pt}{4pt}\selectfont $SF-HLNS$}
  \put(-70,45) {\fontsize{5pt}{5pt}\selectfont $Vortical$}
  \caption{Dependence on $R^{-1/2}k_3$ of rescaled perturbation streamwise vorticity at the stagnation point for cases {RAT2 (black), RBT2 (red), RCT2 (green) and RDT2 (blue) }. (a): Entropy and acoustic forcing with $\omega=0$ and {$\vartheta= 0^{\circ}$} ; (b): vortical forcing with $\omega=0$, {$\vartheta= 0^{\circ}$}  and $\hat \Omega_2=1$. {Lines:} asymptotic prediction; circles: SF-HLNS calculations. }
  \label{fig:omega_R}
  \end{center}
\end{figure}
Figure \ref{fig:omega_R}-(a) shows the stagnation response to freestream acoustic and entropy forcing,   obtained by both asymptotic predictions and SF-HLNS calculations. Notably, fast and slow acoustic forcing yield identical receptivity behaviours and are hence collectively represented by one dashed line per configuration. Following the above scaling argument, all data are presented on the $R^{-1}\tilde \Omega_{\xi 0}$-$R^{-1/2}k_3$ plane. The stagnation vorticity amplitude increases monotonically with $k_3$, and all datasets  with the same $T_w$ nearly collapse into a single curve. Moreover, the agreement  between the asymptotic predictions and SF-HLNS calculations is better at  higher $R$ values.  In  figure \ref{fig:omega_R}-(b), we plot  $R^{-1/2}\tilde \Omega_{\xi 0}$ against $R^{-1/2}k_3$ under freestream vortical forcing. Opposite to the acoustic/entropy counterpart,  the stagnation vorticity perturbation decays as  $k_3$ increases. This trend originates from the inverse-$k_3$ scaling of  unity-energy-norm freestream vortical amplitude, as indicated in (\ref{eq:estimate_FS}). All  curves achieve favourable agreement for $R^{-1/2}k_3>0.02$, which corroborates the scaling laws derived from the asymptotic framework.

\begin{figure}
  \begin{center}
  \includegraphics[width = 0.49\textwidth]{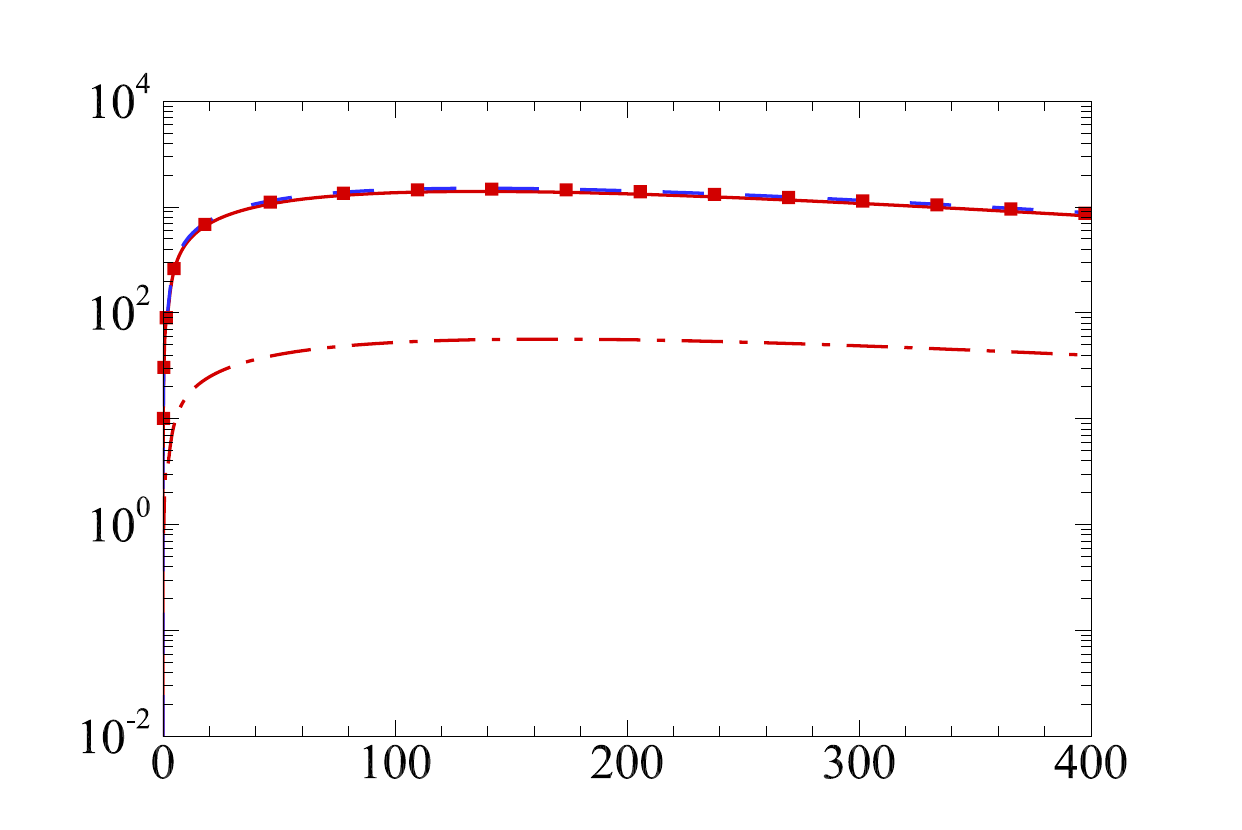}
  \includegraphics[width = 0.49\textwidth]{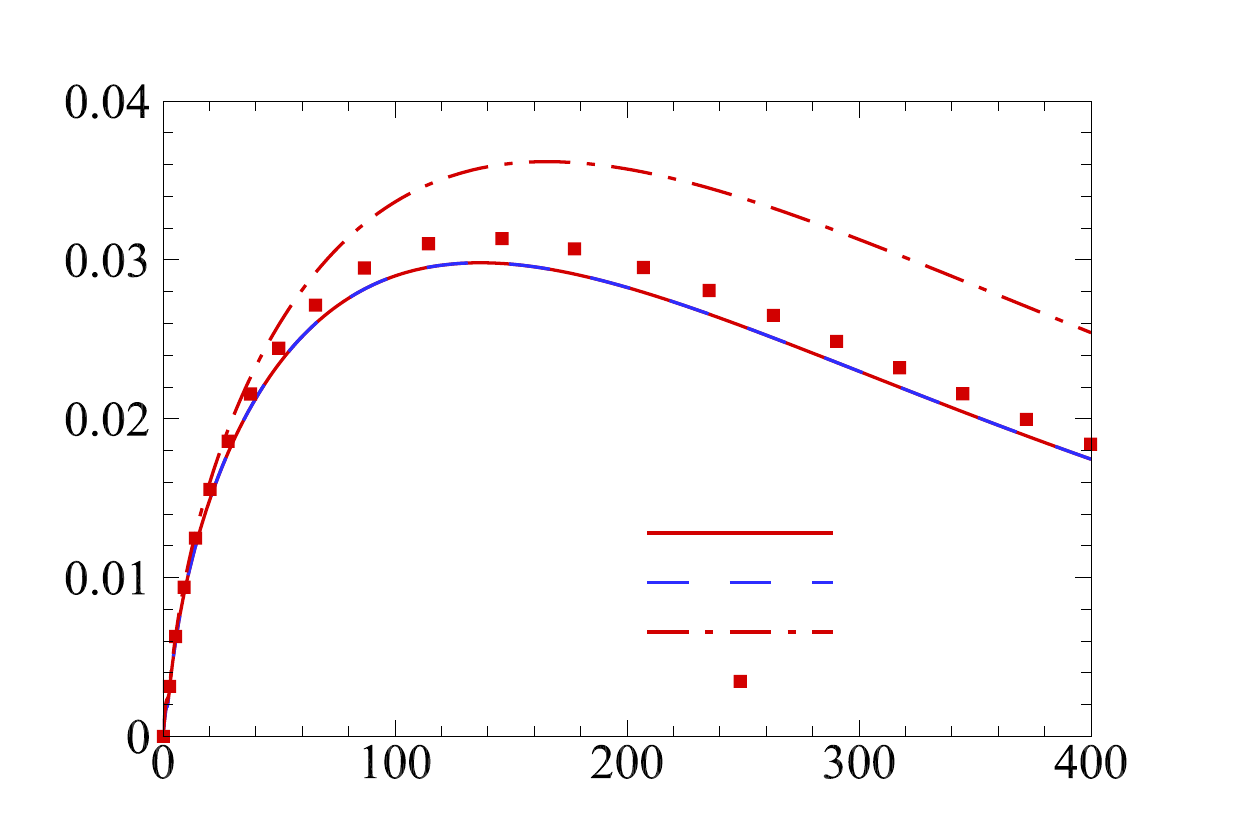}
 \put(-380,120) {$(a)$}
  \put(-190,120) {$(b)$}
  \put(-380,50) {\rotatebox{90}{$A_u$}} 
  \put(-195,50) {\rotatebox{90}{{$\frac{{A_u}}{{|\tilde\Omega_{\xi 0}|}} $}}} 
  \put(-290,-2) {$\xi$}
  \put(-95,-2) {$\xi$}
  \put(-60,44) {\fontsize{5pt}{5pt}\selectfont$Entropy$}
  \put(-60,37) {\fontsize{5pt}{5pt}\selectfont$Acoustic$}
  \put(-60,29) {\fontsize{5pt}{5pt}\selectfont$Vortical$}
  \put(-60,22) {\fontsize{5pt}{5pt}\selectfont$LPSE$}
  \\
  \includegraphics[width = 0.49\textwidth]{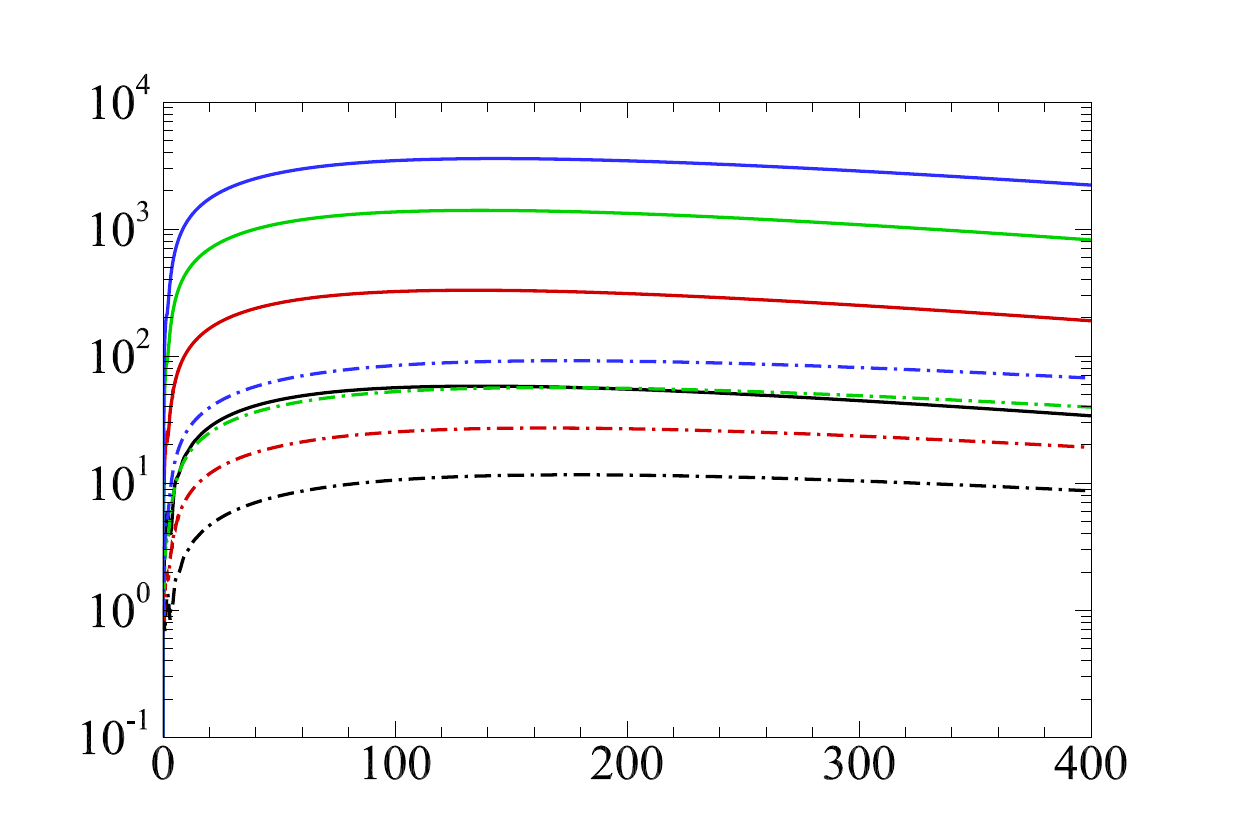}
  \includegraphics[width = 0.49\textwidth]{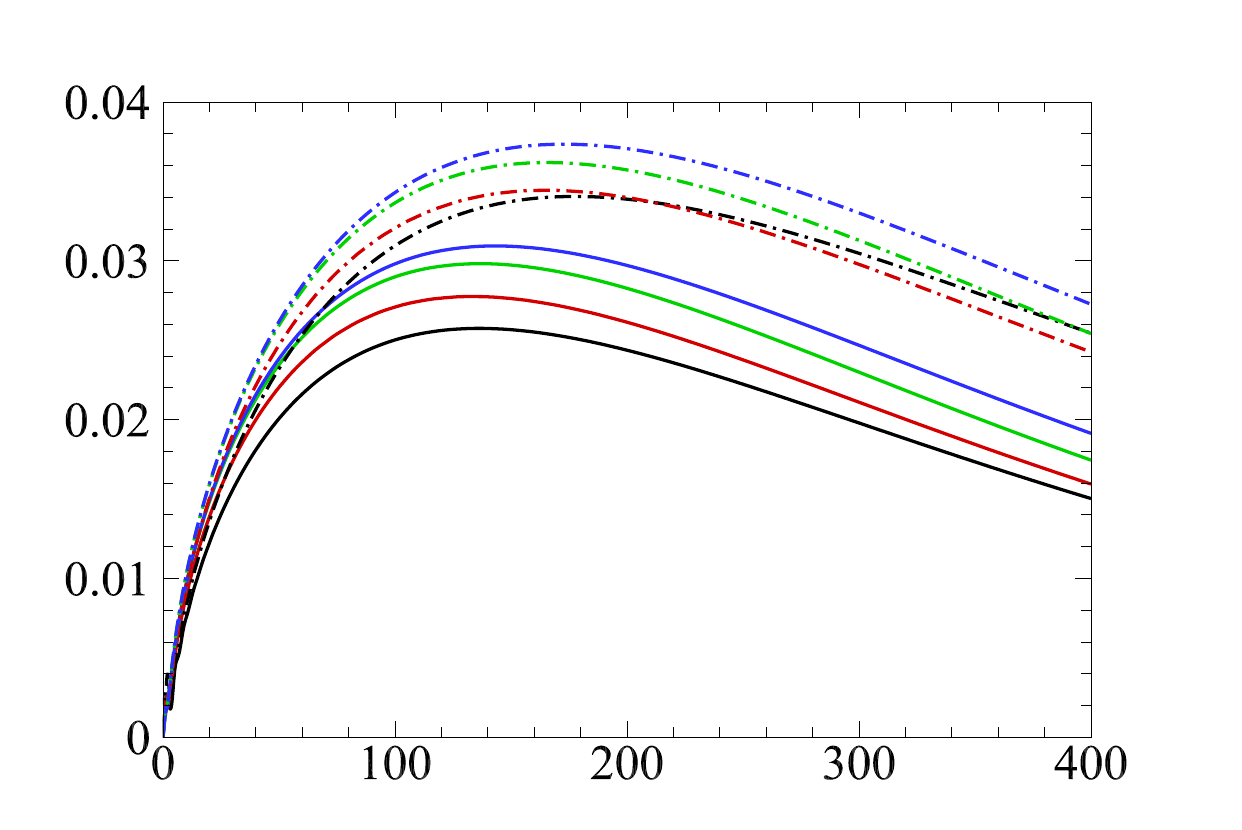}
  \put(-380,120) {$(c)$}
  \put(-190,120) {$(d)$}
  \put(-380,53) {\rotatebox{90}{$A_u$}} 
  \put(-195,50) {\rotatebox{90}{{$\frac{{A_u}}{{|\tilde\Omega_{\xi 0}|}} $}}} 
  \put(-290,-2) {$\xi$}
  \put(-95,-2) {$\xi$}
  \put(-330,33) {\fontsize{5pt}{5pt}\selectfont$RAT2$}
  \put(-330,28) {\fontsize{5pt}{5pt}\selectfont$RBT2$}
  \put(-330,23) {\fontsize{5pt}{5pt}\selectfont$RCT2$}
  \put(-330,18) {\fontsize{5pt}{5pt}\selectfont$RDT2$}
  \put(-310,19){\begin{tikzpicture}
    \draw[black,thick] (0,0) -- (0.5,0);
    \draw[red,thick] (0,-0.18) -- (0.5,-0.18);
    \draw[green,thick] (0,-0.36) -- (0.5,-0.36);
    \draw[blue,thick] (0,-0.54) -- (0.5,-0.54);
    \end{tikzpicture}}
  \put(-317,40) {\fontsize{5pt}{5pt}\selectfont $Entropy$}
  \put(-280,19){\begin{tikzpicture}
    \draw[black,thick,dashdotted] (0,0) -- (0.5,0);
    \draw[red,thick,dashdotted] (0,-0.18) -- (0.5,-0.18);
    \draw[green,thick,dashdotted] (0,-0.36) -- (0.5,-0.36);
    \draw[blue,thick,dashdotted] (0,-0.54) -- (0.5,-0.54);
    \end{tikzpicture}}
  \put(-285,40) {\fontsize{5pt}{5pt}\selectfont $Vortical$}
  \caption{Streamwise  evolution of perturbation amplitude for $R^{-1/2}k_3=$0.02, $\vartheta=0$ and $\omega=0$. (a,b): Response of case RCT2 to various freestream perturbations; (c,d): {\color{black}Comparison among  cases RAT2 to RDT2 for entropy and vortical forcing. Left column: $A_u$; right column: normalised amplitude $A_u/|\tilde\Omega_{\xi 0}|$.}} 
  \label{fig:e_k3Re=0.02_A_u_R}
  \end{center}
\end{figure}

Figure \ref{fig:e_k3Re=0.02_A_u_R}-(a) displays the downstream evolution of the perturbation forced by various freestream perturbations for case RCT2. For comparison, the LPSE result for entropy forcing is also included. In the downstream region, the amplitude of $A_u$ for   acoustic  forcing is slightly higher than that for entropy forcing, and both are significantly stronger than that for vortical forcing. These trends are consistent with the amplitudes observed at the stagnation point  in figure \ref{fig:omega_R}.  When the amplitude is rescaled by the vorticity perturbation amplitude at the stagnation point, {$|\tilde\Omega_{\xi 0}|$}, as shown in panel (b), the curves for acoustic and entropy forcing collapse, while that for vortical forcing  shows the same trend with a comparable magnitude (note  that panel (a) employs a logarithmic scale, whereas panel (b) uses a linear scale). This is a clear demonstration of the transient growth of non-modal perturbations from the leading edge to the downstream region due to the lift-up mechanism. The minor amplitude difference between the acoustic/entropy and vortical forcing is attributed to the difference of the composite solution in region IV.

{\color{black}In figure \ref{fig:e_k3Re=0.02_A_u_R}-(c), we compare the perturbation evolution under entropy and vortical forcing over a range of $R$ values. Results for acoustic forcing are excluded here, given its nearly identical receptivity {\color{red}behaviour} to entropy forcing. While the overall evolution trends remain similar, the downstream   $A_u$ amplitudes span approximately three orders of magnitude, with the entropy-forced response with a higher $R$ being much greater. However, when the downstream amplitude is normalised by the stagnation vorticity perturbation $\tilde \Omega_{\xi 0}$, shown in panel (d), comparable magnitudes are achieved. This suggests that the non-modal perturbation amplification within Region II  induced by the lift-up mechanism is insensitive to freestream forcing types; their downstream amplitudes are directly determined by the perturbation response at the stagnation region. Moreover, the normalised amplitude $A_u/|\tilde\Omega_{\xi 0}|$ exhibits a slight increase with $R$, indicating a weak but systematic strengthening of the lift‑up effect at higher Reynolds numbers. }

\begin{figure}
  \begin{center}
  \includegraphics[width = 0.49\textwidth]{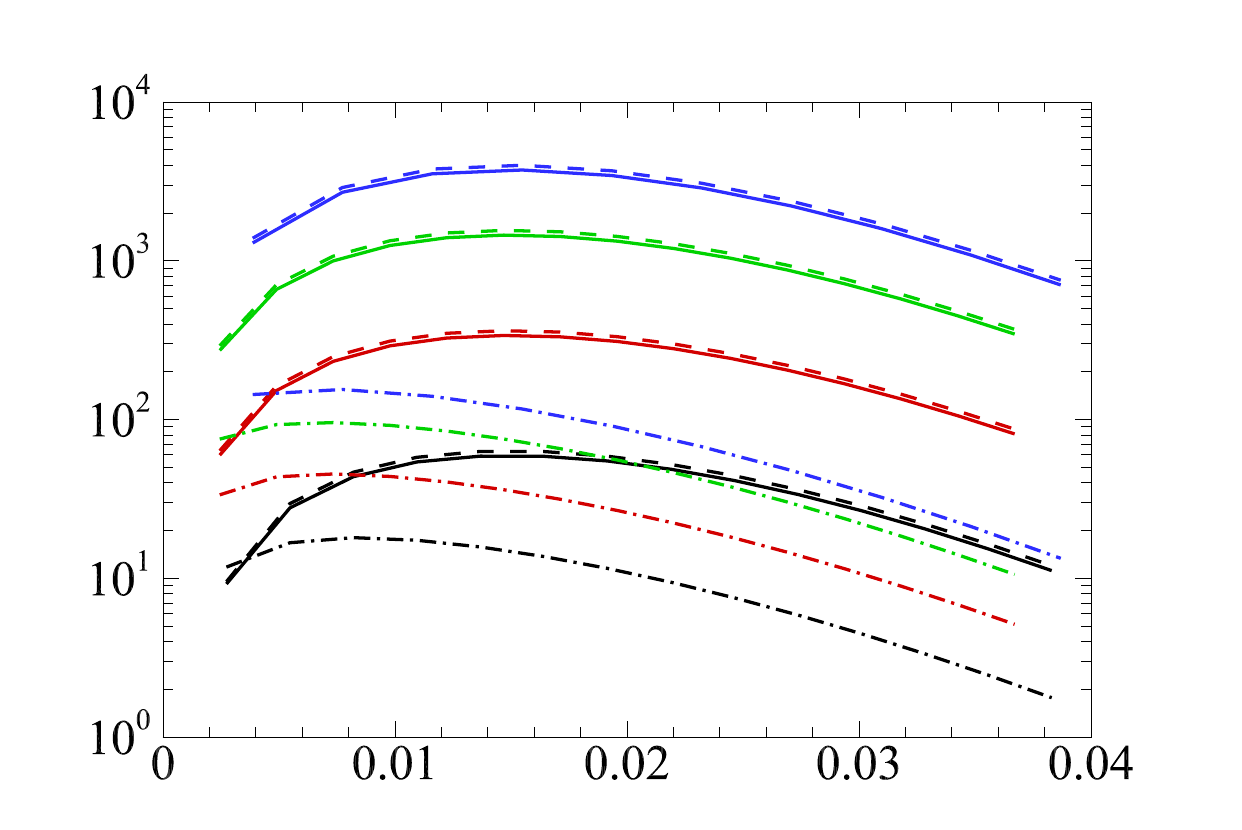}
  \includegraphics[width = 0.49\textwidth]{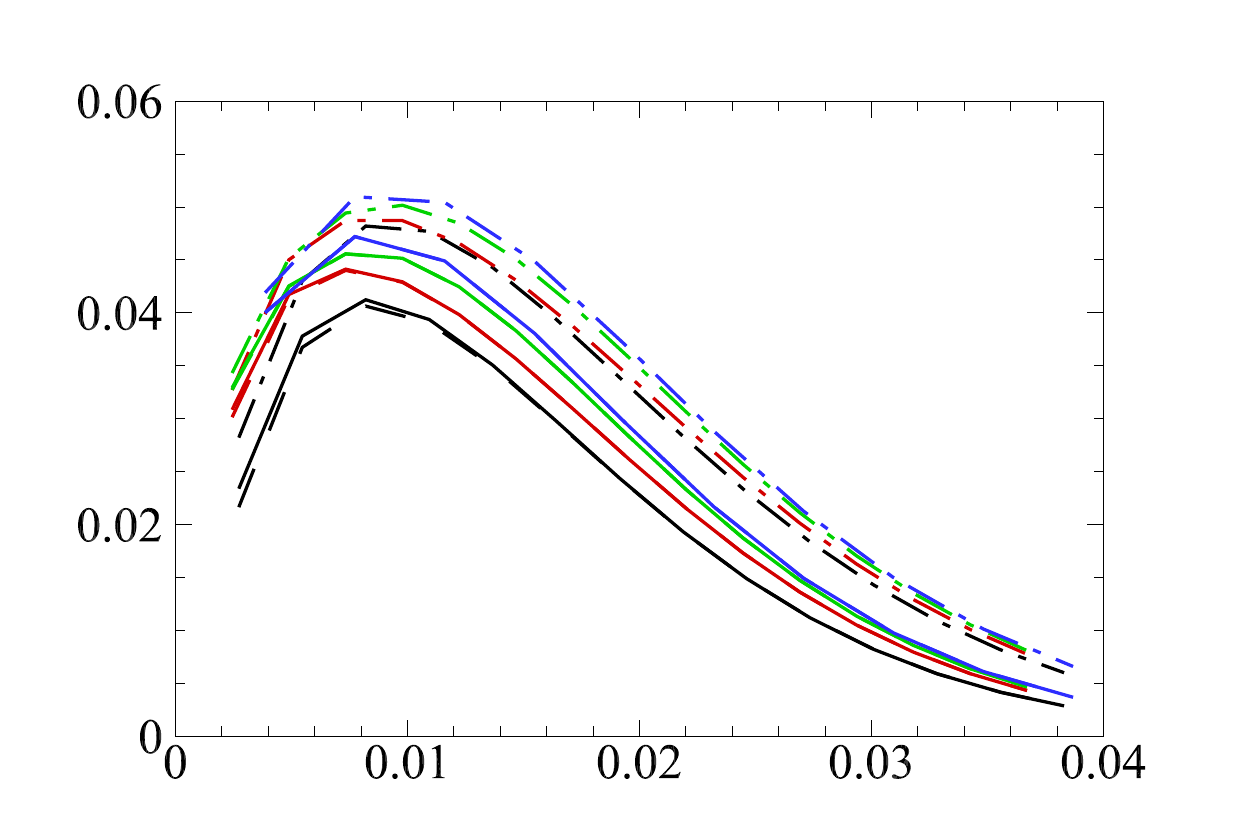}
  \put(-380,120) {$(a)$}
  \put(-190,120) {$(b)$}
  \put(-380,50) {\rotatebox{90}{$A_u(200)$}} 
  \put(-195,50) {\rotatebox{90}{{$\frac{A_u(200)}{|\tilde\Omega_{\xi 0}|}$}}}
  \put(-295,-2) {$R^{-1/2}k_{3}$}
  \put(-105,-2) {$R^{-1/2}k_{3}$}
  \put(-160,33) {\fontsize{5pt}{5pt}\selectfont$RAT2$}
  \put(-160,28) {\fontsize{5pt}{5pt}\selectfont$RBT2$}
  \put(-160,23) {\fontsize{5pt}{5pt}\selectfont$RCT2$}
  \put(-160,18) {\fontsize{5pt}{5pt}\selectfont$RDT2$}
  \put(-140,19){\begin{tikzpicture}
    \draw[black,thick] (0,0) -- (0.4,0);
    \draw[red,thick] (0,-0.18) -- (0.4,-0.18);
    \draw[green,thick] (0,-0.36) -- (0.4,-0.36);
    \draw[blue,thick] (0,-0.54) -- (0.4,-0.54);
  \end{tikzpicture}}
  \put(-118,19){\begin{tikzpicture}
    \draw[black,thick,dashed] (0,0) -- (0.4,0);
    \draw[red,thick,dashed] (0,-0.18) -- (0.4,-0.18);
    \draw[green,thick,dashed] (0,-0.36) -- (0.4,-0.36);
    \draw[blue,thick,dashed] (0,-0.54) -- (0.4,-0.54);
  \end{tikzpicture}}
  \put(-96,19){\begin{tikzpicture}
    \draw[black,thick,dashdotted] (0,0) -- (0.4,0);
    \draw[red,thick,dashdotted] (0,-0.18) -- (0.4,-0.18);
    \draw[green,thick,dashdotted] (0,-0.36) -- (0.4,-0.36);
    \draw[blue,thick,dashdotted] (0,-0.54) -- (0.4,-0.54);
  \end{tikzpicture}}
  \put(-144,38) {\fontsize{4pt}{4pt}\selectfont $Entropy$}
  \put(-122,38) {\fontsize{4pt}{4pt}\selectfont $Acoustic$}
  \put(-100,38) {\fontsize{4pt}{4pt}\selectfont $Vortical$}
  \\
  \includegraphics[width = 0.49\textwidth]{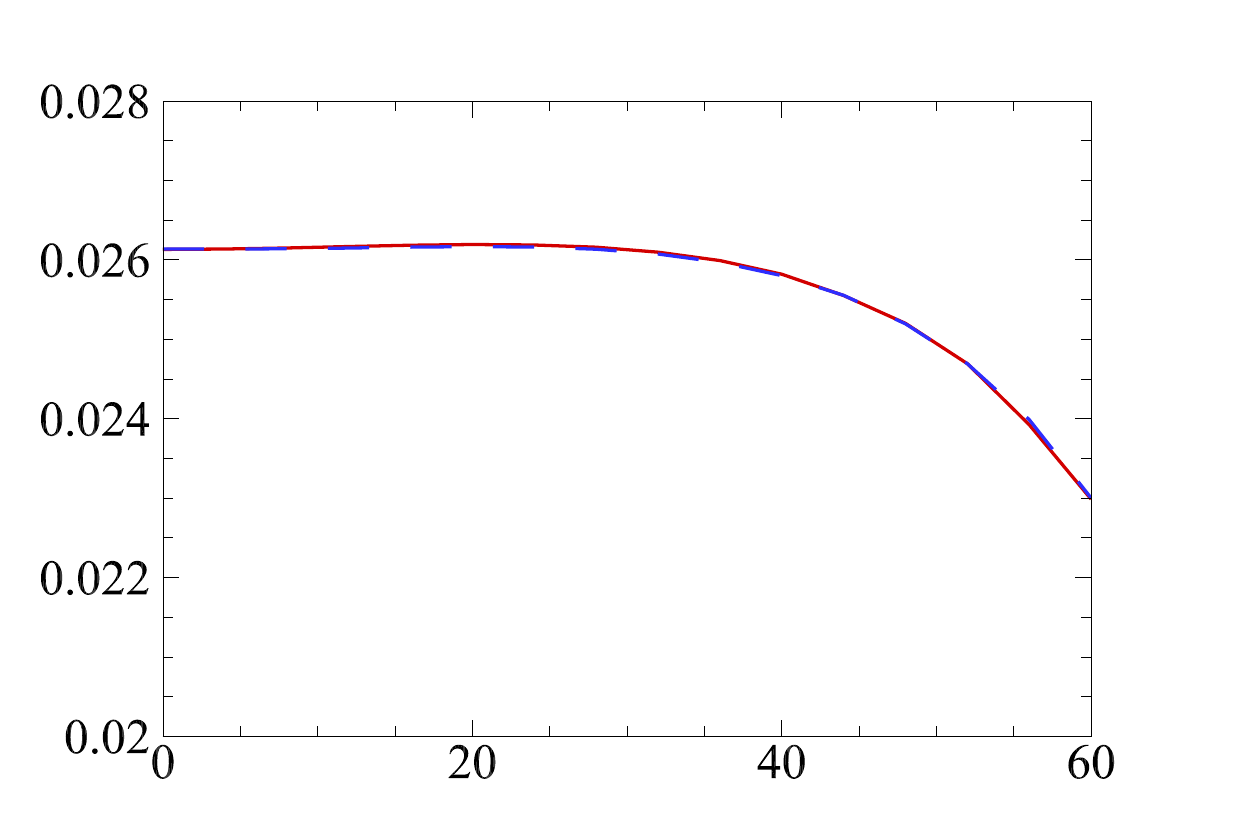}
  \includegraphics[width = 0.49\textwidth]{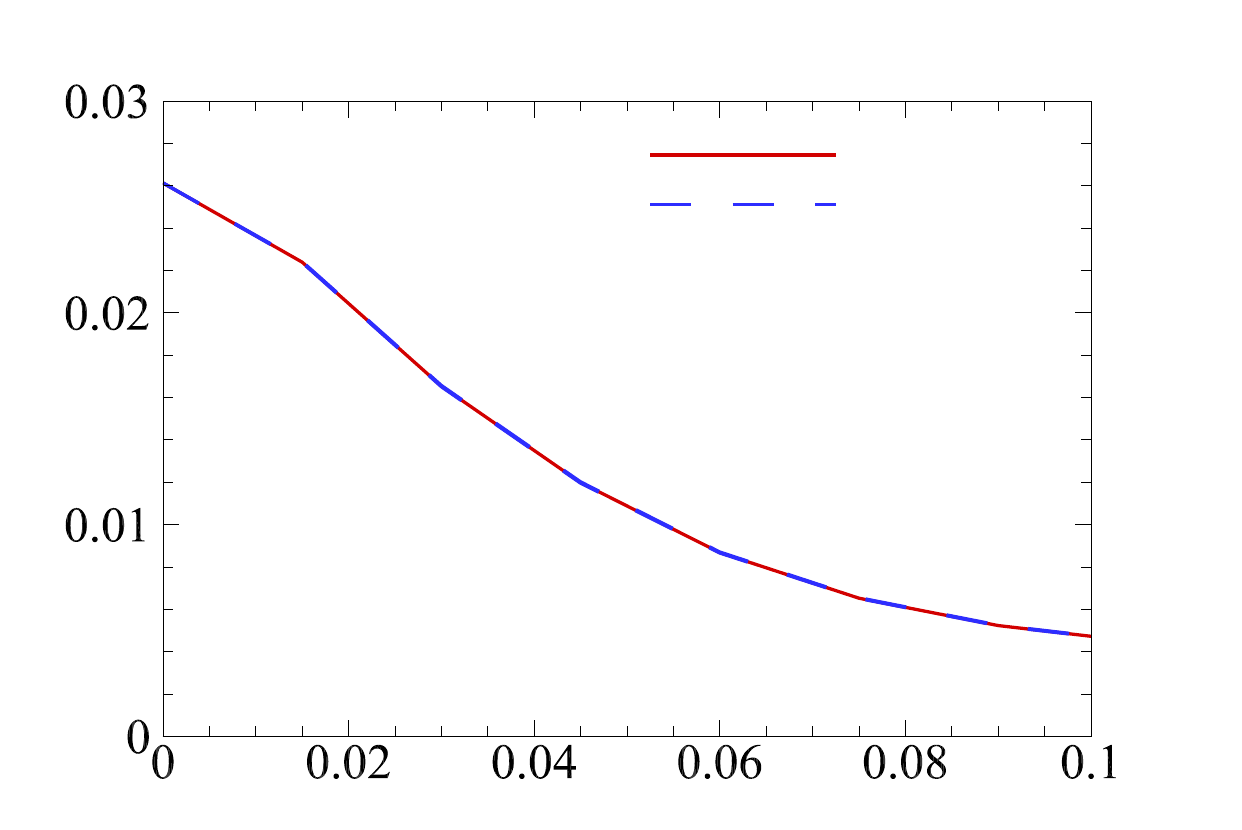}
  \put(-380,120) {$(c)$}
  \put(-190,120) {$(d)$}
  \put(-290,-1) {{$\vartheta$} }
  \put(-95,-1) {$\omega$}
  \put(-390,50) {\rotatebox{90}{{$\frac{A_u(200)}{|\tilde\Omega_{\xi 0}|}$}}} 
  \put(-195,50) {\rotatebox{90}{{$\frac{A_u(200)}{|\tilde\Omega_{\xi 0}|}$}}} 
  \put(-60,101){\fontsize{5pt}{5pt}\selectfont$entropy$}
  \put(-60,93.5) {\fontsize{5pt}{5pt}\selectfont$acoustic$}
  \caption{Dependence of the perturbation amplitude at $\xi=200$ on key parameters. (a,b): $A_u$ (a) and normalised amplitude$A_u(200)/|\tilde\Omega_{\xi 0}|$ (b) as functions of   $R^{-1/2}k_3$ for $\vartheta=0^\circ$ and $\omega=0$;  (c):  $A_u(200)/|\tilde\Omega_{\xi 0}|$ as a function of $\vartheta$ for case RBT2 with $\omega=0$ and  $k_3=$8; (d): $A_u(200)/|\tilde\Omega_{\xi 0}|$ as a function of $\omega$  for case RBT2 with $\vartheta=0^\circ$ and  $k_3=$8. } 
  \label{fig:e_k3Re=0.02_$A_u$_$A_T$_R}
  \end{center}
\end{figure}

Figure \ref{fig:e_k3Re=0.02_$A_u$_$A_T$_R}-(a) summarise the variations of $A_u$ at $\xi=200$ with $R^{-1/2} k_3$ for cases with various $R$,  subject to all types of freestream forcing. For the efficient forcing, acoustic and entropy forcing, the downstream amplitude $A_u(200)$ exhibits a distinct peak near $k_3R^{-1/2}\approx 0.015$, further increasing $k_3R^{-1/2}$ leads to a reduced amplitude. Comparison with the stagnation response in {figure} \ref{fig:omega_R}-(a) suggests that the downstream amplitude results from the combined influence of two factors.  The first  is the stagnation response, which serves as the initial perturbation for  downstream evolution; the second is the  position {\color{black}where the transient growth reaches its maximum}, which is determined by the lift-up mechanism. {\color{black}The peak value of figure \ref{fig:e_k3Re=0.02_$A_u$_$A_T$_R}-(a) at $k_3R^{-1/2}= 0.015$ appears because the transient growth reaches its maximum at around $\xi=200$.}  For higher $R^{-1/2}k_3$ values, the {\color{black}peaking} position occurs upstream of $\xi=200$, and so the amplitude $A_u(200)$ becomes lower even though its initial amplitude is stronger. When the curves are normalised by the stagnation vorticity perturbation $\tilde \Omega_{\xi 0}$, as shown in panel {(b)}, the  responses for each forcing type exhibit identical trends and magnitudes, with only marginal deviations in  amplitude. This indicates that the formation and evolution of the streaky non-modal perturbations in the downstream region are governed by the same lift-up mechanism, and the  differences of the amplitude $A_u$ arise primarily from variations in the initial stagnation-point response. 

{\color{black}Although most of the above calculations predominantly feature $\omega=0$ and $\vartheta=0^\circ$, it is important to emphasise that our studies are not constrained to this specific parameter set.As established in ZD25, the SF-HLNS calculations are  robust and valid for any $\omega$ and $\vartheta$ values. The asymptotic model only assumes $\omega\ll 1$ to target the most effective non-modal receptivity regime, without imposing any {\color{cyan}constraints} on $\vartheta$. For completeness,}
figure \ref{fig:e_k3Re=0.02_$A_u$_$A_T$_R}-(c)  shows the dependence of the normalised amplitude $A_u(200)$ on the declination angle for the most efficient acoustic and entropy forcing types. The resulting curves for the two forcing types are nearly indistinguishable. 
Overall, the receptivity efficiency shows weak dependence on $\vartheta$. Specifically, a broad plateau appears for $\vartheta<40^\circ$, while for larger declination angles, the amplitude decreases gradually as $\vartheta$ increases. 
In figure \ref{fig:e_k3Re=0.02_$A_u$_$A_T$_R}-(d), the normalised amplitude $A_u(200)$ is seen to decrease monotonically with increasing frequency $\omega$, which is consistent with the observations in bypass transition that low-frequency streaks are more amplified. Therefore, the previous calculations based on $\omega=0$ and $\vartheta=0^\circ$ can be regarded as a representative parameter set for quantifying the non-modal receptivity. Although these observations are drawn from case RBT2 with a specific wavenumber $k_3=8$,  the same trends hold for other cases and across different values of  $k_3$.

\begin{figure}
  \begin{center}
  \includegraphics[width = 0.49\textwidth]{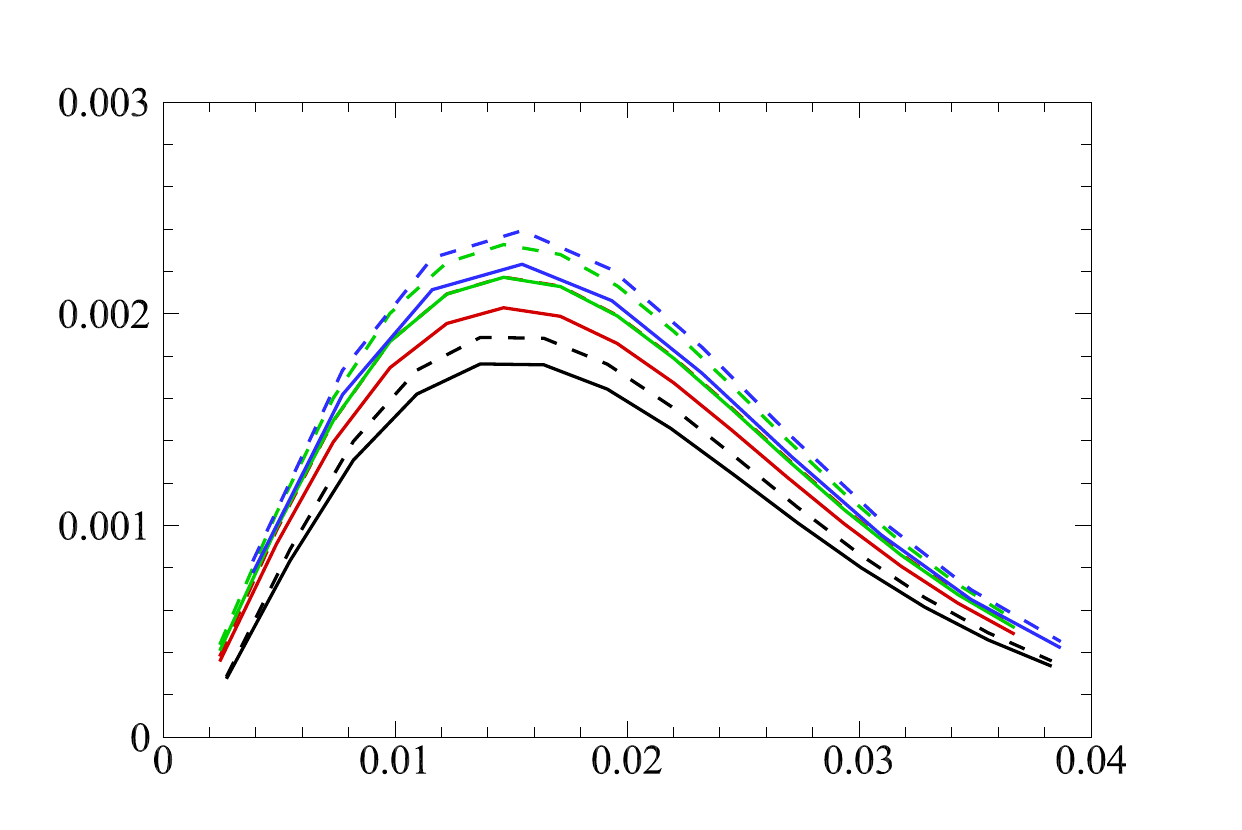} 
  \includegraphics[width = 0.49\textwidth]{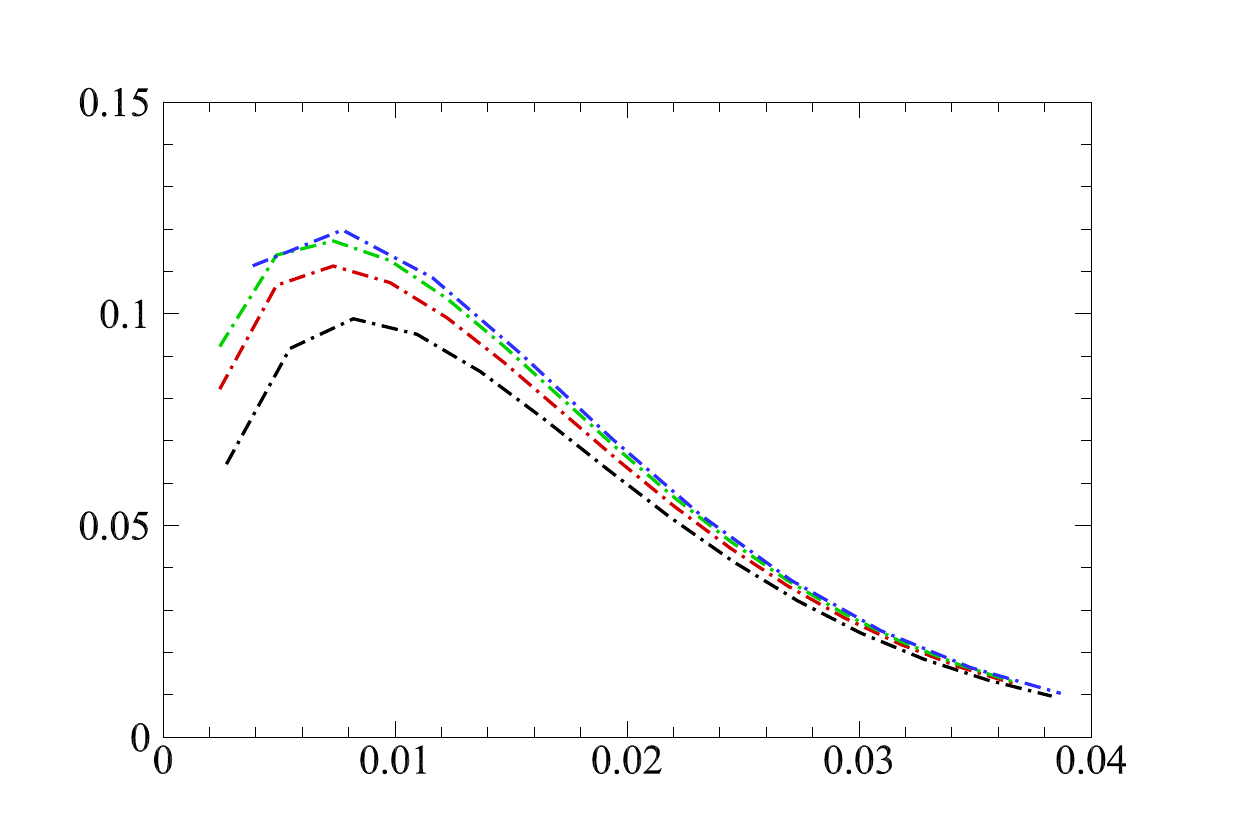} 
  \put(-380,120) {$(a)$}
  \put(-190,120) {$(b)$}
  \put(-110,-2) { $R^{-1/2}k_3$}
  \put(-300,-2) { $R^{-1/2}k_3$}
  \put(-380,48) {\rotatebox{90}{$R^{-1}A_u(200)$}}
  \put(-190,45) {\rotatebox{90}{$R^{-1/2}A_u(200)$}}
  \put(-330,33) {\fontsize{5pt}{5pt}\selectfont$RAT2$}
  \put(-330,28) {\fontsize{5pt}{5pt}\selectfont$RBT2$}
  \put(-330,23) {\fontsize{5pt}{5pt}\selectfont$RCT2$}
  \put(-330,18) {\fontsize{5pt}{5pt}\selectfont$RDT2$}
  \put(-310,19){\begin{tikzpicture}
    \draw[black,thick] (0,0) -- (0.5,0);
    \draw[red,thick] (0,-0.18) -- (0.5,-0.18);
    \draw[green,thick] (0,-0.36) -- (0.5,-0.36);
    \draw[blue,thick] (0,-0.54) -- (0.5,-0.54);
    \end{tikzpicture}}
  \put(-317,40) {\fontsize{5pt}{5pt}\selectfont $Entropy$}
  \put(-285,19){\begin{tikzpicture}
    \draw[black,thick,dashed] (0,0) -- (0.5,0);
    \draw[red,thick,dashed] (0,-0.18) -- (0.5,-0.18);
    \draw[green,thick,dashed] (0,-0.36) -- (0.5,-0.36);
    \draw[blue,thick,dashed] (0,-0.54) -- (0.5,-0.54);
    \end{tikzpicture}}
  \put(-290,40) {\fontsize{5pt}{5pt}\selectfont $Acoustic$}
  \put(-145,33) {\fontsize{5pt}{5pt}\selectfont$RAT2$}
  \put(-145,28) {\fontsize{5pt}{5pt}\selectfont$RBT2$}
  \put(-145,23) {\fontsize{5pt}{5pt}\selectfont$RCT2$}
  \put(-145,18) {\fontsize{5pt}{5pt}\selectfont$RDT2$}
  \put(-120,19){\begin{tikzpicture}
    \draw[black,thick,dashdotted] (0,0) -- (0.5,0);
    \draw[red,thick,dashdotted] (0,-0.18) -- (0.5,-0.18);
    \draw[green,thick,dashdotted] (0,-0.36) -- (0.5,-0.36);
    \draw[blue,thick,dashdotted] (0,-0.54) -- (0.5,-0.54);
    \end{tikzpicture}}
  \put(-125,40) {\fontsize{5pt}{5pt}\selectfont $Vortical$}
  \caption{Dependence of rescaled perturbation amplitude at $\xi=200$ on $R^{-1/2}k_3$ for $\omega=0$ and $\vartheta=0$. (a): $R^{-1}A_u$ under acoustic and entropy forcing {\color{red}with $\omega=0$ and $\vartheta=0$}; (b): $R^{-1/2}A_u$ under vortical forcing {\color{red}with  $\hat \Omega_2=1$.} }
  \label{fig:AuR_xi=200_1}
  \end{center}
\end{figure}
{\color{black}Combining the results from figure \ref{fig:omega_R} and figure \ref{fig:e_k3Re=0.02_$A_u$_$A_T$_R}-(b), the integrated receptivity scaling can be quantified. For fixed $R^{-1/2}k_3$, the stagnation vorticity  perturbation $\tilde\Omega_{\xi 0}$ scales as $R$ for acoustic/entropy forcing and as $R^{1/2}$ for vortical forcing. Meanwhile,  the lift-up mechanism enforces an $R^0$ scaling for the normalised downstream amplitude $A_u/|\tilde\Omega_{\xi 0}|$. Taken together,  for a unit-energy-norm forcing, the downstream non-modal perturbation amplitude obeys  $O(R)$ scaling under  acoustic/entropy forcing, and $O(R^{1/2})$  under  vortical forcing. Figures \ref{fig:AuR_xi=200_1}-(a) and (b) verify these predicted scaled amplitudes by plotting data at   $\xi=200$ against $R^{-1/2}k_3$.  The almost collapse curves confirm the receptivity scaling relations, and the rescaled receptivity efficiency peaks at $R^{-1/2}k_3\approx 0.015$ for acoustic/entropy forcing and at  $R^{-1/2}k_3\approx 0.08$ for vortical forcing.
}

\subsection {Impact of wall temperature on receptivity efficiency }
\begin{figure}
  \begin{center}
  \includegraphics[width = 0.49\textwidth]{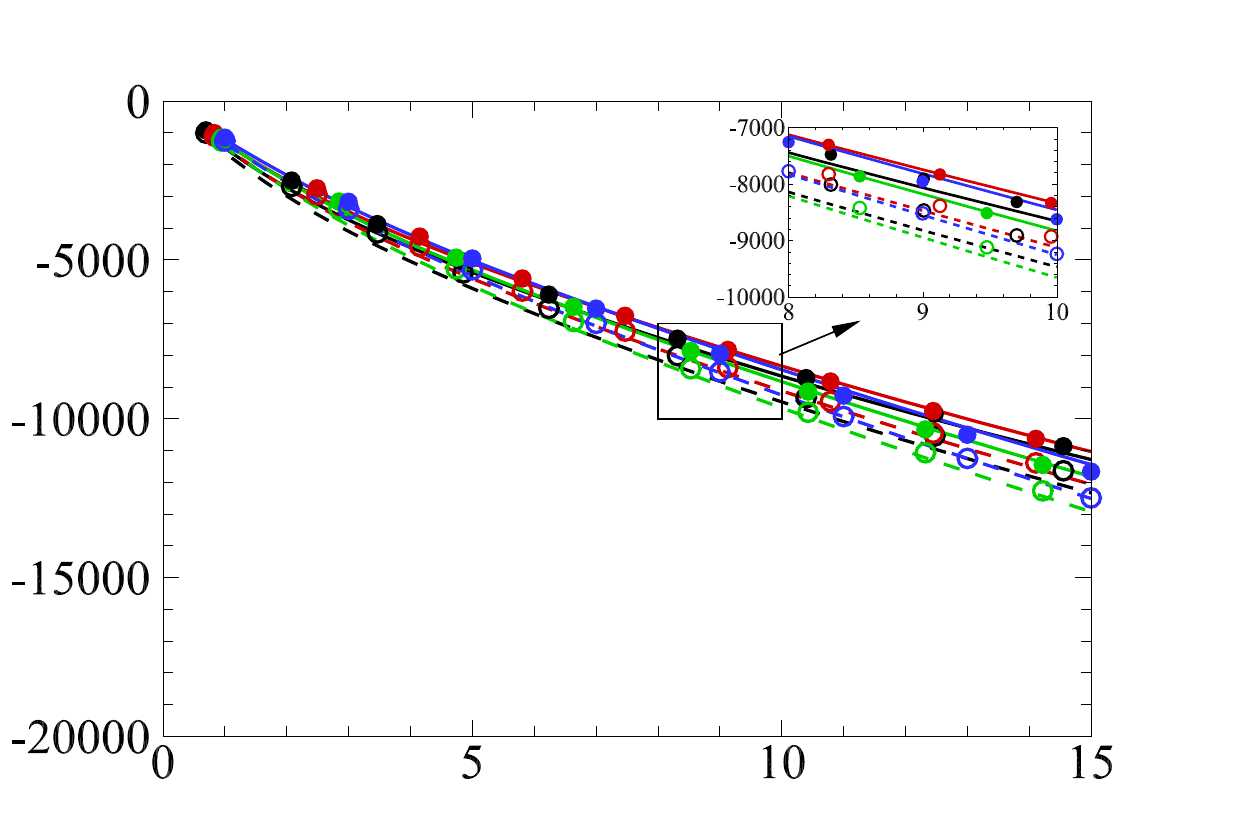} 
  \includegraphics[width = 0.49\textwidth]{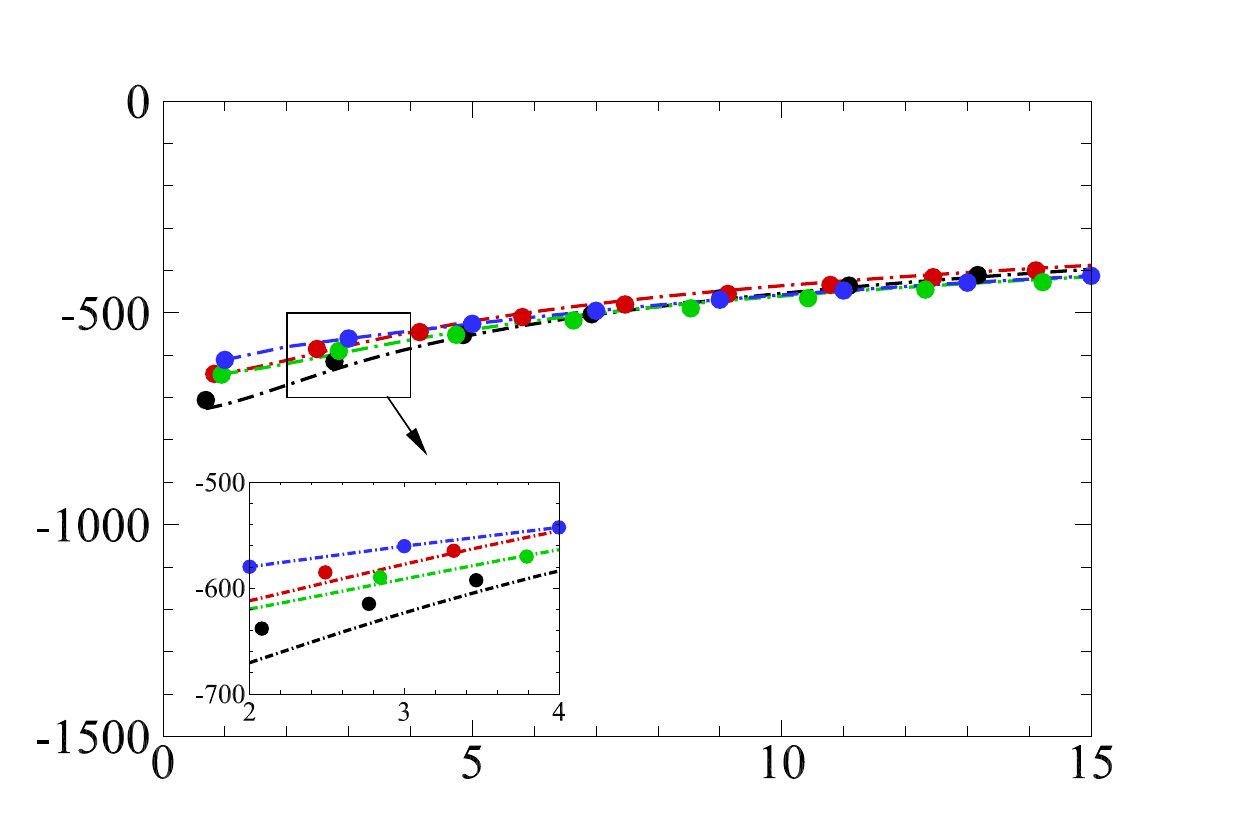} 
  \put(-380,110){$(a) $}
  \put(-190,110){$(b) $}
  \put(-120,-1) { $(T_w/T_{ad})^{1/4}k_3$}
  \put(-310,-1) { $(T_w/T_{ad})^{1/4}k_3$}
  \put(-390,30) {\rotatebox{90}{{$\tilde\Omega_{\xi0}(T_w/T_{ad})^{7/8}$}}}
  \put(-195,30) {\rotatebox{90}{{$\tilde\Omega_{\xi0}(T_w/T_{ad})^{5/8}$}}}
  \put(-350,33) {\fontsize{5pt}{5pt}\selectfont$RBT1$}
  \put(-350,28) {\fontsize{5pt}{5pt}\selectfont$RBT2$}
  \put(-350,23) {\fontsize{5pt}{5pt}\selectfont$RBT3$}
  \put(-350,18) {\fontsize{5pt}{5pt}\selectfont$RBT4$}
  \put(-330,19){\begin{tikzpicture}
    \draw[black,thick] (0,0) -- (0.5,0);
    \draw[red,thick] (0,-0.18) -- (0.5,-0.18);
    \draw[green,thick] (0,-0.36) -- (0.5,-0.36);
    \draw[blue,thick] (0,-0.54) -- (0.5,-0.54);
    \node[draw,black,circle,fill,inner sep=0.8pt] at (1.0,0) {};
    \node[draw,red,circle,fill,inner sep=0.8pt] at (1.0,-0.18) {};
    \node[draw,green,circle,fill,inner sep=0.8pt] at (1.0,-0.36) {};
    \node[draw,blue,circle,fill,inner sep=0.8pt] at (1.0,-0.54) {};
    \end{tikzpicture}}
  \put(-330,40) {\fontsize{4pt}{4pt}\selectfont $Asym.$}
  \put(-314,40) {\fontsize{4pt}{4pt}\selectfont $SF$-$HLNS$}
  \put(-330,45) {\fontsize{5pt}{5pt}\selectfont $Entropy$}
  \put(-285,19){\begin{tikzpicture}
    \draw[black,thick,dashed] (0,0) -- (0.5,0);
    \draw[red,thick,dashed] (0,-0.18) -- (0.5,-0.18);
    \draw[green,thick,dashed] (0,-0.36) -- (0.5,-0.36);
    \draw[blue,thick,dashed] (0,-0.54) -- (0.5,-0.54);
    \node[draw,black,circle,inner sep=0.8pt] at (1,0) {};
    \node[draw,red,circle,inner sep=0.8pt] at (1,-0.18) {};
    \node[draw,green,circle,inner sep=0.8pt] at (1,-0.36) {};
    \node[draw,blue,circle,inner sep=0.8pt] at (1,-0.54) {};
    \end{tikzpicture}}
  \put(-285,40) {\fontsize{4pt}{4pt}\selectfont $Asym.$}
  \put(-269,40) {\fontsize{4pt}{4pt}\selectfont $SF$-$HLNS$}
  \put(-285,45) {\fontsize{5pt}{5pt}\selectfont $Acoustic$}
  \put(-90,33) {\fontsize{5pt}{5pt}\selectfont$RBT1$}
  \put(-90,28) {\fontsize{5pt}{5pt}\selectfont$RBT2$}
  \put(-90,23) {\fontsize{5pt}{5pt}\selectfont$RBT3$}
  \put(-90,18) {\fontsize{5pt}{5pt}\selectfont$RBT4$}
  \put(-70,19){\begin{tikzpicture}
    \draw[black,thick,dashdotted] (0,0) -- (0.5,0);
    \draw[red,thick,dashdotted] (0,-0.18) -- (0.5,-0.18);
    \draw[green,thick,dashdotted] (0,-0.36) -- (0.5,-0.36);
    \draw[blue,thick,dashdotted] (0,-0.54) -- (0.5,-0.54);
    \node[draw,black,circle,fill,inner sep=0.8pt] at (1,0) {};
    \node[draw,red,circle,fill,inner sep=0.8pt] at (1,-0.18) {};
    \node[draw,green,circle,fill,inner sep=0.8pt] at (1,-0.36) {};
    \node[draw,blue,circle,fill,inner sep=0.8pt] at (1,-0.54) {};
    \end{tikzpicture}}
  \put(-70,40) {\fontsize{4pt}{4pt}\selectfont $Asym.$}
  \put(-54,40) {\fontsize{4pt}{4pt}\selectfont $SF$-$HLNS$}
  \put(-70,45) {\fontsize{5pt}{5pt}\selectfont $Vortical$}
  \caption{{Dependence on  $(T_w/T_{ad})^{1/4}k_3$ of rescaled perturbation streamwise vorticity at the stagnation point for cases RBT1 (black), RBT2 (red), RBT3 (green) and RBT4 (blue). (a): Entropy and acoustic forcing with $\omega=0$ and {$\vartheta= 0^{\circ}$}; (b): vortical forcing with $\omega=0$, {$\vartheta= 0^{\circ}$} and $\hat \Omega_2=1$. Lines: asymptotic prediction; circles: SF-HLNS calculations. }}
  \label{fig:omega_Tw}
  \end{center}
\end{figure}

Now we investigate the wall temperature effect on the non-modal receptivity efficiency. {\color{black}Since an explicit analytical dependence on $T_w$ cannot be derived within the present asymptotic framework, we resort to numerical curve fitting. Although the resulting scaling relations are limited to the parameter range considered, they provide  considerable practical utility for  quantifying and visualising  the wall temperature effects at representative parameter spaces.} Following the study in the previous subsection, we first examine the perturbation streamwise vorticity at the stagnation point, which determines the initial state of the downstream formation and evolution of non-modal perturbations. Figure \ref{fig:omega_Tw}-(a) displays the variation of rescaled perturbation $\tilde \Omega_{\xi 0}$ as a function of the spanwise wavenumber for cases with the same $R$ but different $T_w$ under freestream acoustic and entropy forcing. Since the boundary-layer thickness $\delta_{BL}$ scales as $(T_w/T_{ad})^{1/4}$ (see figure \ref{fig:baseflow_shock_bl}-(c)) and the transient growth {\color{black}reaches its maximum} when $\delta_{BL}\sim 2\pi/k_3$, the horizontal axis is rescaled as $(T_w/T_{ad})^{1/4}k_3$. Although the unscaled curves for these cases, illustrated in figure \ref{fig:e_omega_TW}-(b), appear scattered, they collapse much {\color{red}closer} when  $\tilde \Omega_{\xi 0}$ is scaled by $(T_w/T_{ad})^{{-7/8}}$. Again, the stagnation-point response to acoustic forcing is slightly greater than that to entropy forcing. In contrast, the response to vortical   forcing, shown in figure \ref{fig:omega_Tw}-(b), is considerably weaker, and the curves for different wall temperatures almost collapse when $\tilde \Omega_{\xi 0}$ is scaled by $(T_w/T_{ad})^{{-5/8}}$. These {negative} exponents ({$-7/8$ and $-5/8$}) indicate an intensified stagnation response to freestream forcing as wall temperatures reduce, agreeing with the behaviours of the perturbation profiles in figures \ref{fig:Per_Bl}-(c) and (d).

\begin{figure}
  \begin{center}
  \includegraphics[width = 0.49\textwidth]{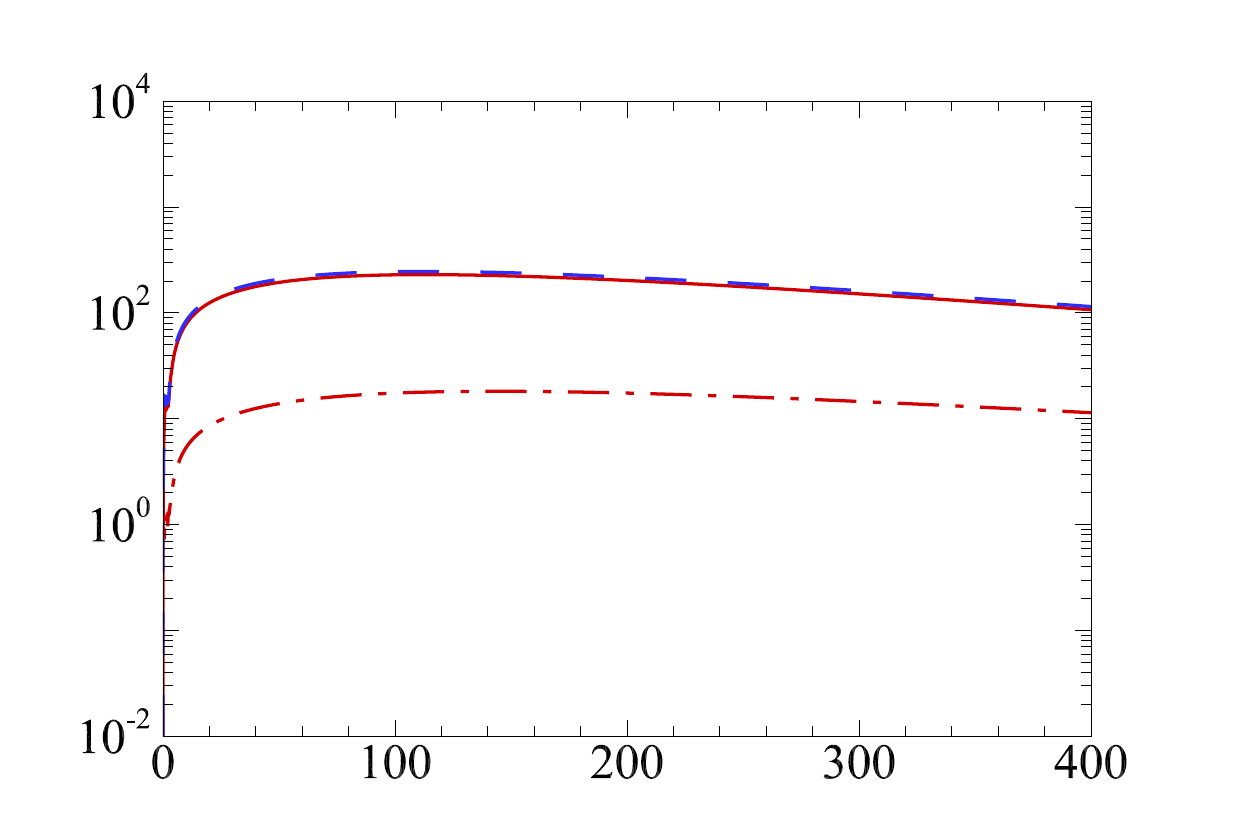}
  \includegraphics[width = 0.49\textwidth]{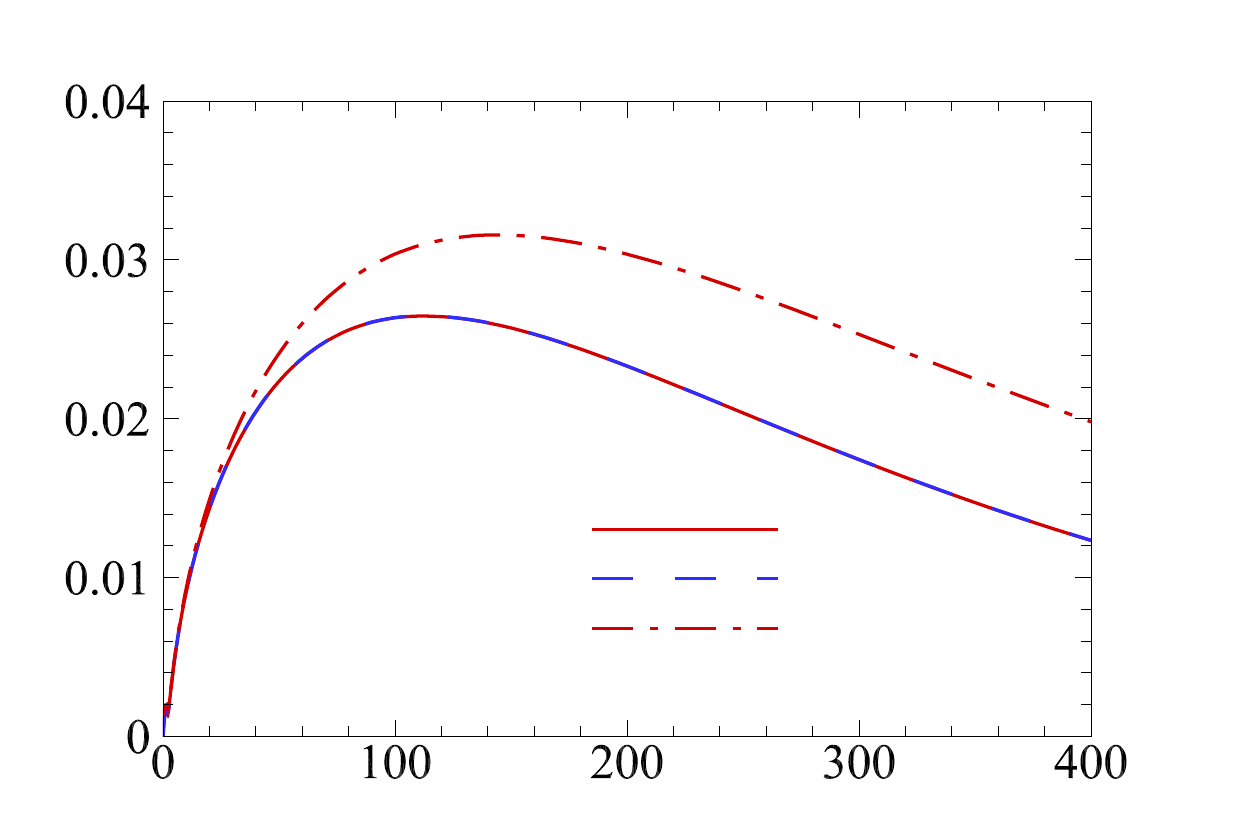}
  \put(-380,120) {$(a)$}
  \put(-190,120) {$(b)$}
  \put(-380,55) {\rotatebox{90}{$A_u$}} 
  \put(-200,50) {{\rotatebox{90}{$\frac{A_u}{|\tilde\Omega_{\xi0}|}$}} }
  \put(-290,-2) {$\xi$}
  \put(-100,-2) {$\xi$}
  \put(-67,45) {\fontsize{5pt}{5pt}\selectfont$Entropy$}
  \put(-67,37) {\fontsize{5pt}{5pt}\selectfont$Acoustic$}
  \put(-67,29) {\fontsize{5pt}{5pt}\selectfont$Vortical$}
  \\
  \includegraphics[width = 0.49\textwidth]{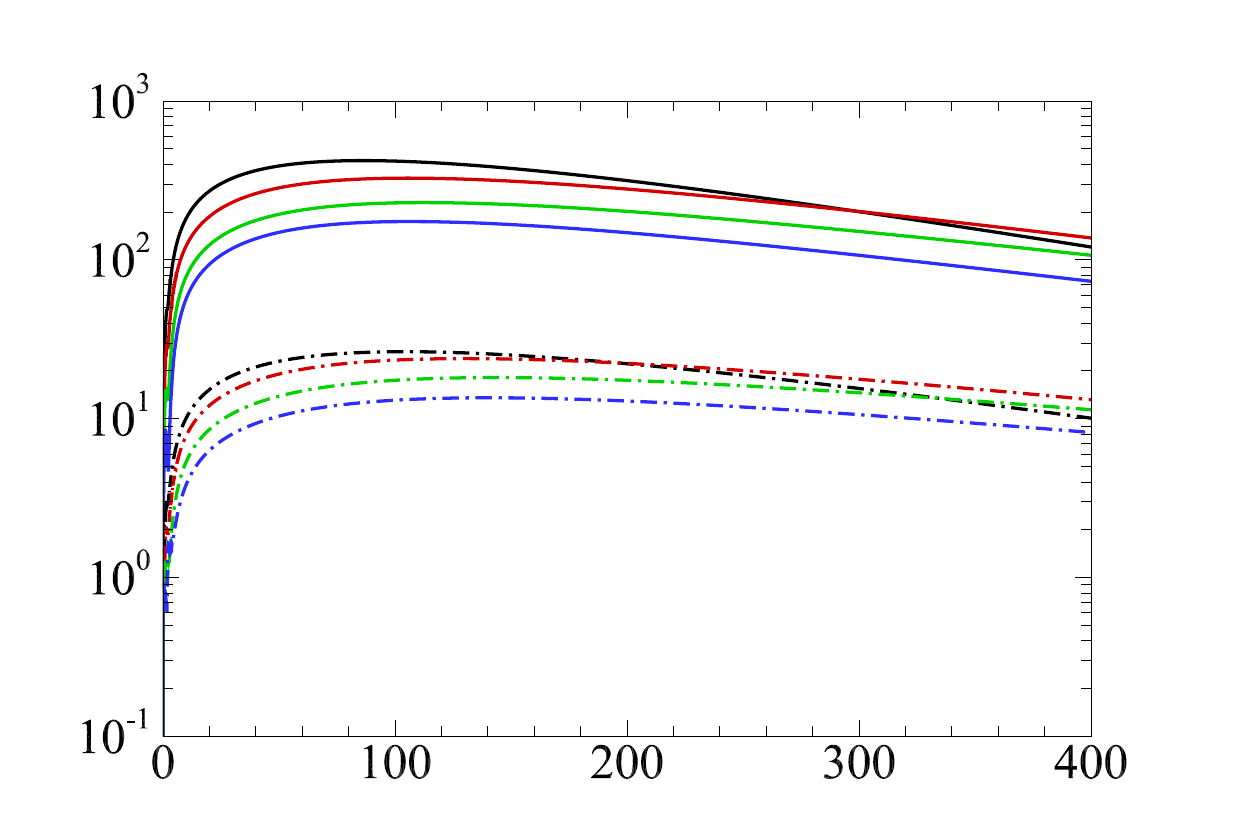}
  \includegraphics[width = 0.49\textwidth]{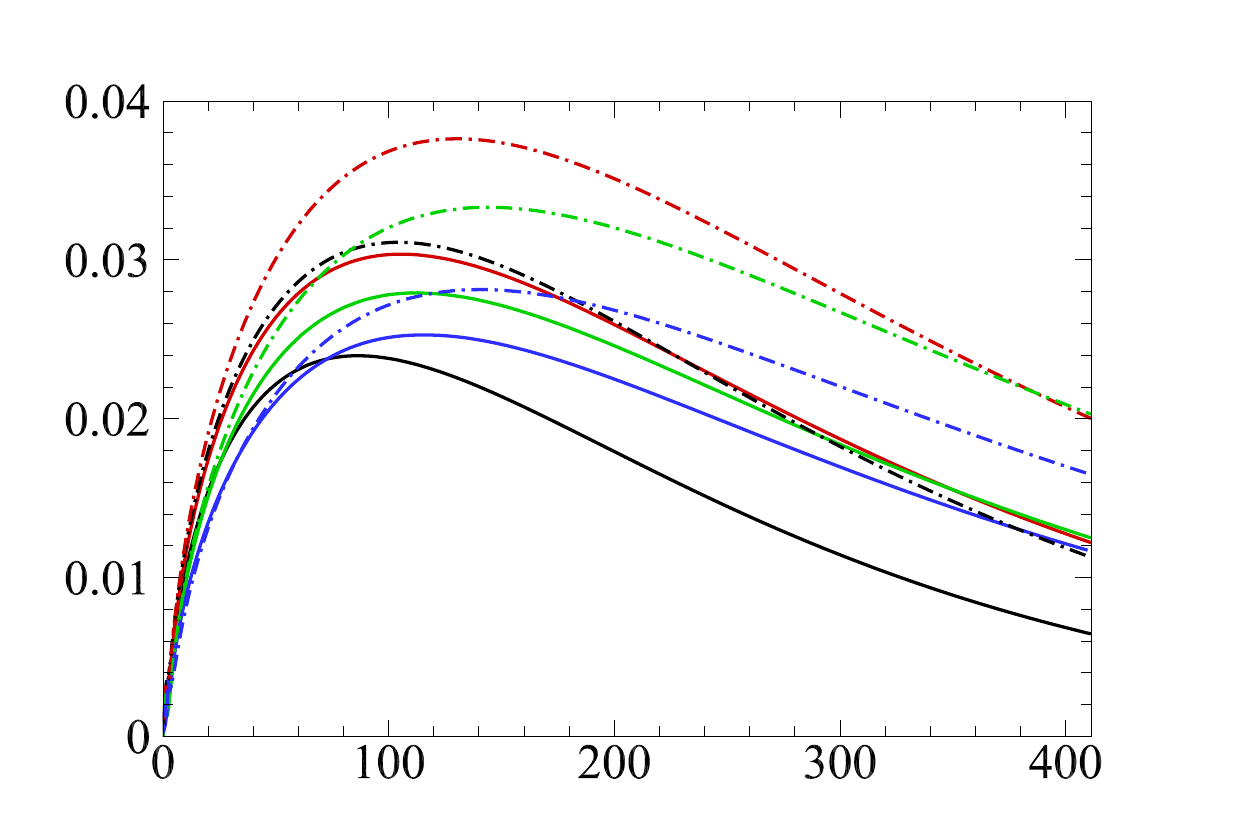}
  \put(-380,120) {$(c)$}
  \put(-190,120) {$(d)$}
  \put(-380,55) {\rotatebox{90}{$A_u$}} 
  \put(-200,50) {{\rotatebox{90}{$\frac{A_u}{|\tilde\Omega_{\xi0}|}$}} }
  \put(-290,-2) {$\xi$}
  \put(-95,-2) {$\xi$}
  \put(-335,33) {\fontsize{5pt}{5pt}\selectfont$RBT1$}
  \put(-335,28) {\fontsize{5pt}{5pt}\selectfont$RBT2$}
  \put(-335,23) {\fontsize{5pt}{5pt}\selectfont$RBT3$}
  \put(-335,18) {\fontsize{5pt}{5pt}\selectfont$RBT4$}
  \put(-315,19){\begin{tikzpicture}
    \draw[black,thick] (0,0) -- (0.5,0);
    \draw[red,thick] (0,-0.18) -- (0.5,-0.18);
    \draw[green,thick] (0,-0.36) -- (0.5,-0.36);
    \draw[blue,thick] (0,-0.54) -- (0.5,-0.54);
    \end{tikzpicture}}
  \put(-322,40) {\fontsize{5pt}{5pt}\selectfont $Entropy$}
  \put(-285,19){\begin{tikzpicture}
    \draw[black,thick,dashdotted] (0,0) -- (0.5,0);
    \draw[red,thick,dashdotted] (0,-0.18) -- (0.5,-0.18);
    \draw[green,thick,dashdotted] (0,-0.36) -- (0.5,-0.36);
    \draw[blue,thick,dashdotted] (0,-0.54) -- (0.5,-0.54);
    \end{tikzpicture}}
  \put(-290,40) {\fontsize{5pt}{5pt}\selectfont $Vortical$}
  \caption{Streamwise evolution of perturbation amplitude for $(T_w/T_{ad})^{1/4}k_3=7.6$, $\vartheta= 0^{\circ}$ and $\omega=0$.
  (a,b) Response of case RBT3 to various freestream perturbations; (c,d): comparison among cases RBT1 to RBT4 for entropy and vortical forcing. Left column:  $A_u$; right column: normalised amplitude {$A_u/|\tilde\Omega_{\xi0}|$}.} 
  \label{fig:e_k3Re=0.02_$A_u$_Tw}
  \end{center}
\end{figure}

\begin{figure}
  \begin{center}
  \includegraphics[width = 0.49\textwidth]{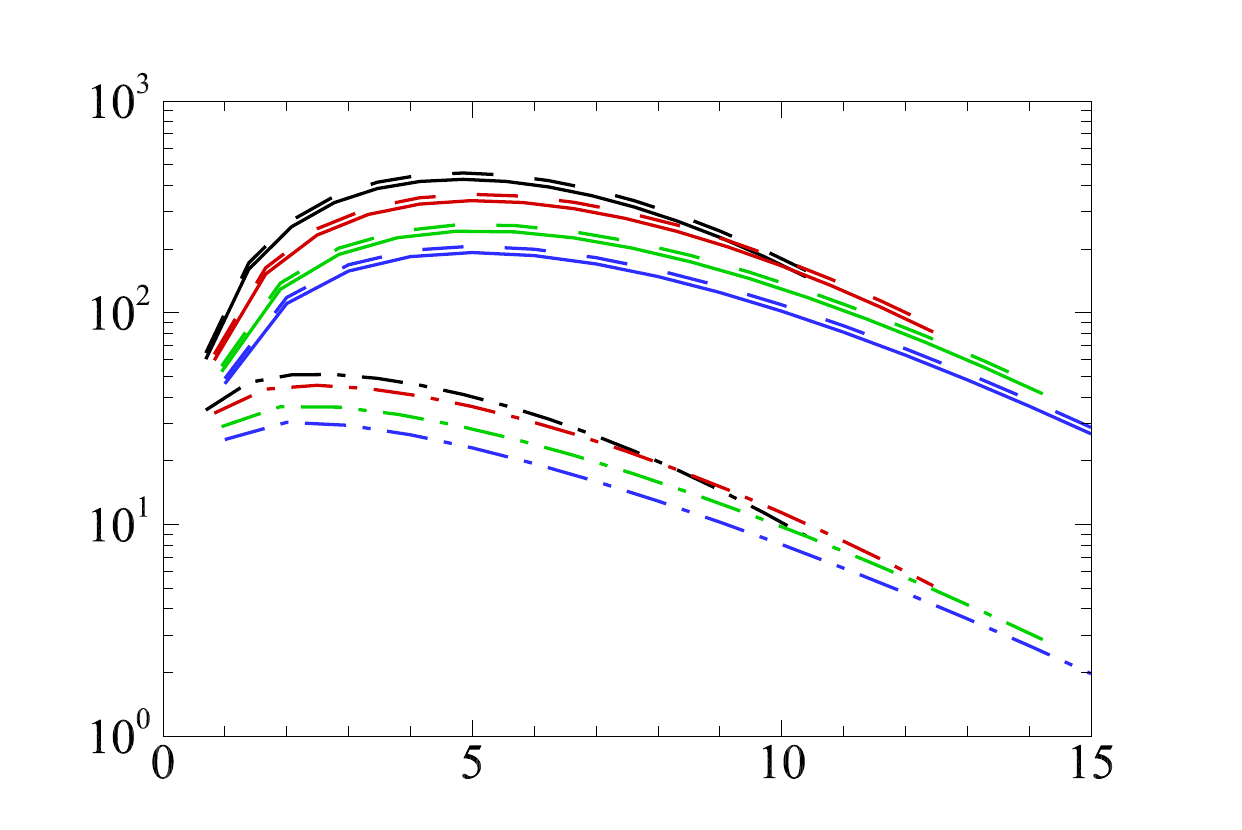}
  \includegraphics[width = 0.49\textwidth]{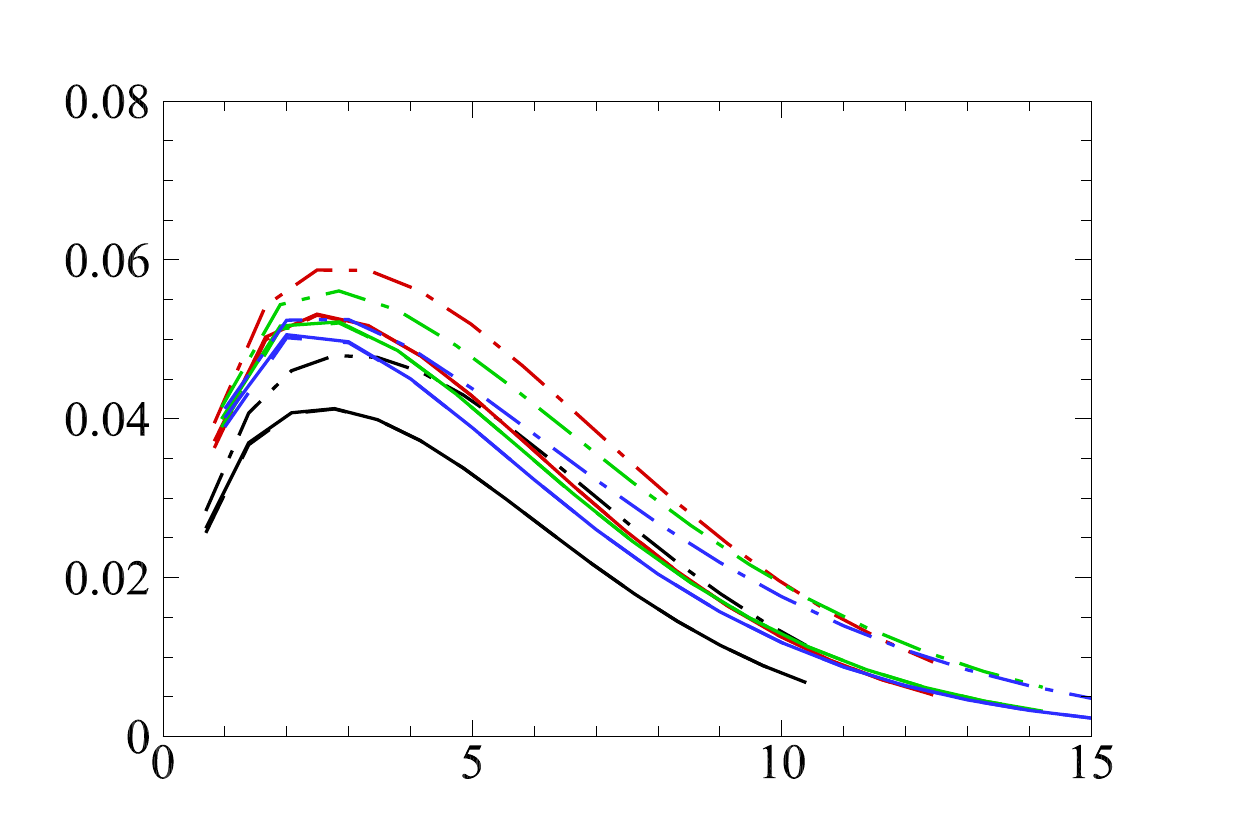}
  \put(-380,120) {$(a)$}
  \put(-190,120) {$(b)$}
  \put(-380,50) {\rotatebox{90}{$A_u(200)$}} 
  \put(-200,30) {{\rotatebox{90}{$\frac{A_u(200)(T_w/T_{ad})^{-1/4}}{|\tilde\Omega_{\xi0}|}$}} }
  \put(-305,-2) {$(T_w/T_{ad})^{1/4}k_{3}$}
  \put(-115,-2) {$(T_w/T_{ad})^{1/4}k_{3}$}
  \put(-109,99) {\fontsize{5pt}{5pt}\selectfont$RBT1$}
  \put(-109,94) {\fontsize{5pt}{5pt}\selectfont$RBT2$}
  \put(-109,89) {\fontsize{5pt}{5pt}\selectfont$RBT3$}
  \put(-109,84) {\fontsize{5pt}{5pt}\selectfont$RBT4$}
  \put(-89,85){\begin{tikzpicture}
    \draw[black,thick] (0,0) -- (0.4,0);
    \draw[red,thick] (0,-0.18) -- (0.4,-0.18);
    \draw[green,thick] (0,-0.36) -- (0.4,-0.36);
    \draw[blue,thick] (0,-0.54) -- (0.4,-0.54);
  \end{tikzpicture}}
  \put(-67,85){\begin{tikzpicture}
    \draw[black,thick,dashed] (0,0) -- (0.4,0);
    \draw[red,thick,dashed] (0,-0.18) -- (0.4,-0.18);
    \draw[green,thick,dashed] (0,-0.36) -- (0.4,-0.36);
    \draw[blue,thick,dashed] (0,-0.54) -- (0.4,-0.54);
  \end{tikzpicture}}
  \put(-45,85){\begin{tikzpicture}
    \draw[black,thick,dashdotted] (0,0) -- (0.4,0);
    \draw[red,thick,dashdotted] (0,-0.18) -- (0.4,-0.18);
    \draw[green,thick,dashdotted] (0,-0.36) -- (0.4,-0.36);
    \draw[blue,thick,dashdotted] (0,-0.54) -- (0.4,-0.54);
  \end{tikzpicture}}
  \put(-93,105) {\fontsize{4pt}{4pt}\selectfont $Entropy$}
  \put(-71,105) {\fontsize{4pt}{4pt}\selectfont $Acoustic$}
  \put(-49,105) {\fontsize{4pt}{4pt}\selectfont $Vortical$}
 \caption{Dependence of perturbation amplitude at $\xi=200$ on $(T_w/T_{ad})^{1/4}k_{3}$ for cases RBT1, RBT2, RBT3 and RBT4. (a): $A_u(200)$; (b): {$(T_w/T_{ad})^{-1/4}A_u(200)/|\tilde\Omega_{\xi0}|$}. } 
  \label{fig:A_u_Tw}
  \end{center}
\end{figure}

\begin{figure}
  \begin{center}
  \includegraphics[width = 0.49\textwidth]{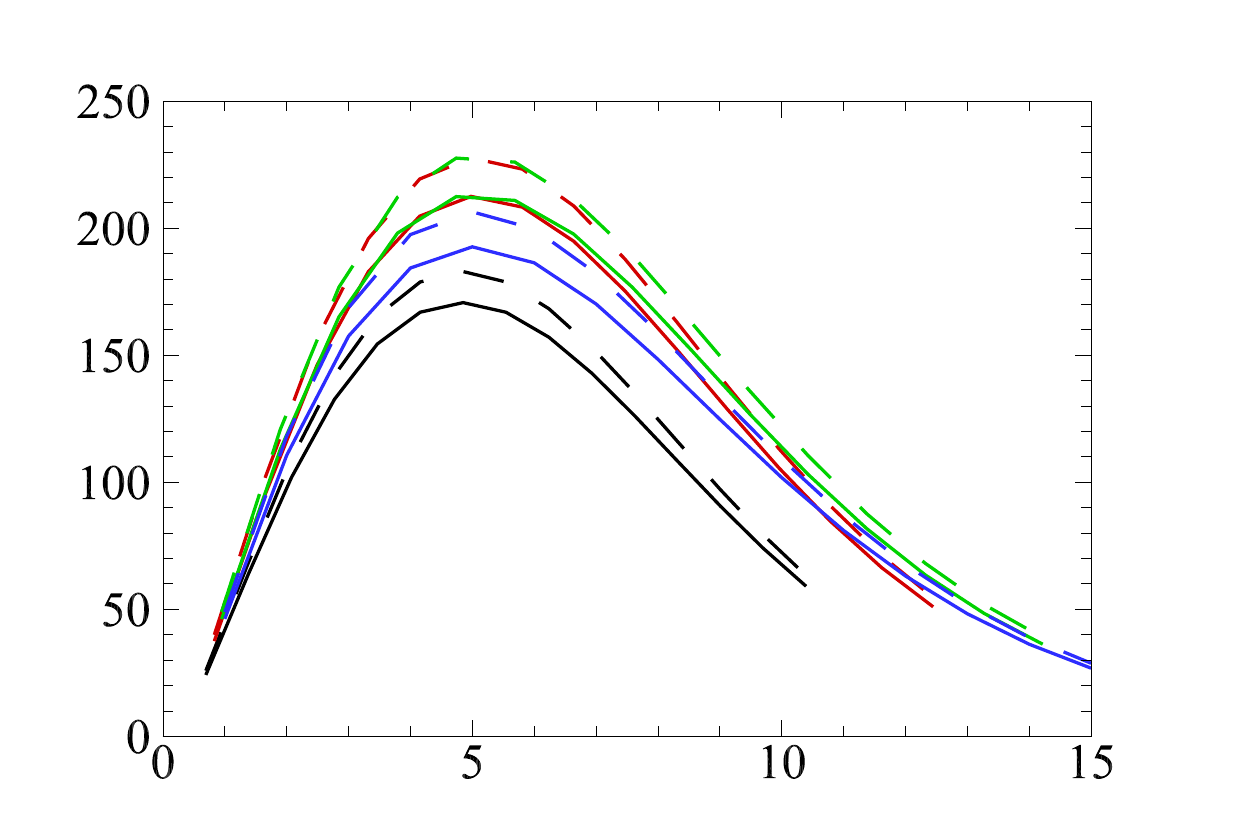} 
  \includegraphics[width = 0.49\textwidth]{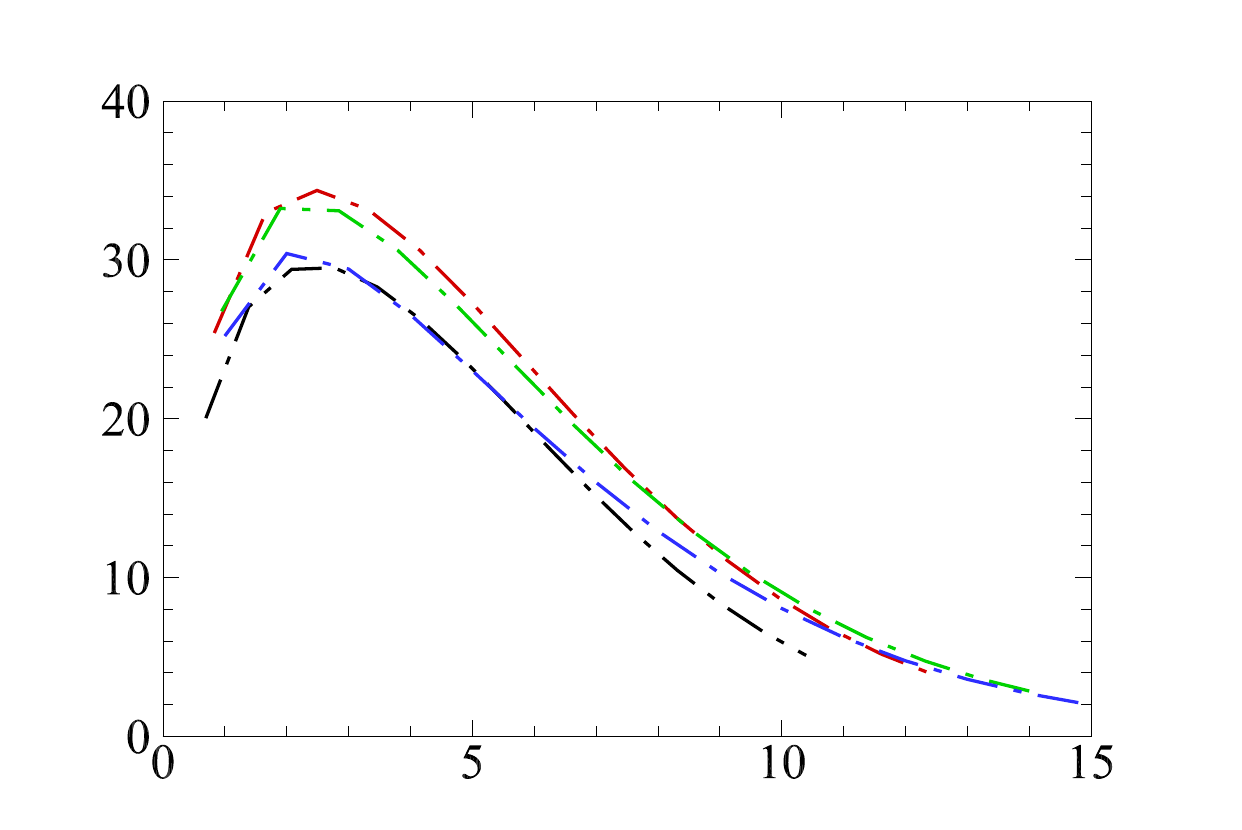} 
  \put(-380,120) {$(a)$}
  \put(-190,120) {$(b)$}
  \put(-305,-2) {$(T_w/T_{ad})^{1/4}k_{3}$}
  \put(-115,-2) {$(T_w/T_{ad})^{1/4}k_{3}$}
  \put(-380,25) {\rotatebox{90}{$(T_w/T_{ad})^{5/8}A_u(200)$}}
  \put(-190,25) {\rotatebox{90}{$(T_w/T_{ad})^{3/8}A_u(200)$}}
  \put(-335,33) {\fontsize{5pt}{5pt}\selectfont$RBT1$}
  \put(-335,28) {\fontsize{5pt}{5pt}\selectfont$RBT2$}
  \put(-335,23) {\fontsize{5pt}{5pt}\selectfont$RBT3$}
  \put(-335,18) {\fontsize{5pt}{5pt}\selectfont$RBT4$}
  \put(-315,19){\begin{tikzpicture}
    \draw[black,thick] (0,0) -- (0.5,0);
    \draw[red,thick] (0,-0.18) -- (0.5,-0.18);
    \draw[green,thick] (0,-0.36) -- (0.5,-0.36);
    \draw[blue,thick] (0,-0.54) -- (0.5,-0.54);
    \end{tikzpicture}}
  \put(-322,40) {\fontsize{5pt}{5pt}\selectfont $Entropy$}
  \put(-285,19){\begin{tikzpicture}
    \draw[black,thick,dashed] (0,0) -- (0.5,0);
    \draw[red,thick,dashed] (0,-0.18) -- (0.5,-0.18);
    \draw[green,thick,dashed] (0,-0.36) -- (0.5,-0.36);
    \draw[blue,thick,dashed] (0,-0.54) -- (0.5,-0.54);
    \end{tikzpicture}}
  \put(-295,40) {\fontsize{5pt}{5pt}\selectfont $Acoustic$}
  \put(-150,33) {\fontsize{5pt}{5pt}\selectfont$RBT1$}
  \put(-150,28) {\fontsize{5pt}{5pt}\selectfont$RBT2$}
  \put(-150,23) {\fontsize{5pt}{5pt}\selectfont$RBT3$}
  \put(-150,18) {\fontsize{5pt}{5pt}\selectfont$RBT4$}
  \put(-125,19){\begin{tikzpicture}
    \draw[black,thick,dashdotted] (0,0) -- (0.5,0);
    \draw[red,thick,dashdotted] (0,-0.18) -- (0.5,-0.18);
    \draw[green,thick,dashdotted] (0,-0.36) -- (0.5,-0.36);
    \draw[blue,thick,dashdotted] (0,-0.54) -- (0.5,-0.54);
    \end{tikzpicture}}
  \put(-130,40) {\fontsize{5pt}{5pt}\selectfont $Vortical$}
  \caption{Dependence of perturbation amplitude at $\xi=200$ on $(T_w/T_{ad})^{1/4}k_{3}$ for cases RBT1, RBT2, RBT3 and RBT4. (a): $(T_w/T_{ad})^{5/8}A_u$; (b): $(T_w/T_{ad})^{3/8}A_u$. (a): Entropy and acoustic forcing with $\omega=0$ and $\vartheta=0$; (b): vortical forcing with $\omega=0$, $\vartheta=0$ and $\hat \Omega_2=1$. }
  \label{fig:AuTw_xi=200}
  \end{center}
\end{figure}

Figure \ref{fig:e_k3Re=0.02_$A_u$_Tw}-(a) displays the streamwise evolution of the perturbation amplitude $A_u$ under various forcing types for case RBT3. Similar to figure \ref{fig:e_k3Re=0.02_A_u_R}-(a), $A_u$ in each case exhibits a transient growth, reaching a peak near $\xi\approx {110}$. Although the magnitude of $A_u$ under vortical forcing is much smaller than that under acoustic and entropy forcing, the responses become comparable in magnitude  when scaled by the perturbation vorticity amplitude  at the stagnation point, as shown in panel (b).  {\color{black}Figure \ref{fig:e_k3Re=0.02_$A_u$_Tw}-(c) further presents the $A_u$ evolution for cases with different wall temperatures under entropy and vortical forcing, revealing variations spanning approximately two orders of magnitude. When $A_u$ is normalised by the stagnation vorticity perturbation $\tilde \Omega_{\xi 0}$, the curves, shown in panel (d), become of the same trend and magnitude. This process highlights the formation of non-modal streaks due to the lift-up mechanism. }

Choosing $\xi=200$ as a representative {\color{red}detection} position, we compare the amplitude $A_u(200)$ across various forcing types  and spanwise wavenumbers  for various wall temperatures,  shown on a log-linear scale in figure \ref{fig:A_u_Tw}-(a). Each curve exhibits a local peak, with amplitudes spanning roughly two orders of magnitude. However, after normalising by the stagnation-point response using the scaling  {$A_u(200)(T_w/T_{ad})^{-1/4}/|\tilde \Omega_{\xi0}|$}, as shown in panel (b), the results for all cases and forcing types  become approximately aligned. Notably, the maximum receptivity efficiency appears at  $(T_w/T_{ad})^{1/4}k_3\approx 2.5$ for all curves. The fact that  for given $(T_w/T_{ad})^{1/4}k_3$ values, the rescaled amplitude $A_u/|\tilde\Omega_{\xi 0}|$ scales with $(T_w/T_{ad})^{1/4}$ indicates  that the downstream transient growth of non‑modal perturbations, from the stagnation point onward, is relatively less efficient for colder walls, which is  independent of the type of freestream forcing and its associated control parameters.

{\color{black}The above observations indicate that decreasing $T_w$ strengthens the stagnation vorticity response while slightly {\color{red}weakening} the lift-up effect. The net effect is an overall enhancement of the downstream amplitude $A_u$, with freestream acoustic and entropy forcing proving more effective than vortical forcing.
We now combine the stagnation‑point vorticity perturbation characteristics presented in figure~\ref{fig:omega_Tw} with the lift‑up mechanism discussed in figure~\ref{fig:A_u_Tw}-(b), and examine the integrated receptivity scaling with respect to $T_w/T_{ad}$, as shown in figure \ref{fig:AuTw_xi=200}. Panel (a) illustrates the scaling relation under entropy and acoustic forcing, showing data collapse onto a common trend in the $(T_w/T_{ad})^{5/8}A_u(200)$-$(T_w/T_{ad})^{1/4}k_3$ plane. The maximum rescaled amplitude appears near $(T_w/T_{ad})^{1/4}k_3\approx 5$. In contrast, under vortical forcing, the curves show a collapse trend when plotted in the $(T_w/T_{ad})^{3/8}A_u(200)$-$(T_w/T_{ad})^{1/4}k_3$ plane, with the maximum response appearing at $(T_w/T_{ad})^{1/4}k_3\approx 2.5$. }

{\color{cyan}We acknowledge that these scaling relations are obtained from calculations confined to a single, albeit representative, Mach number under {\color{red}centreline} symmetry (zero angle of attack), and are based on a limited set of wall temperatures. Nevertheless, the physical insights and the asymptotic framework developed herein provide valuable guidance for understanding receptivity mechanisms in hypersonic blunt boundary layers under more general flow conditions. }

\section{Conclusions and discussions}
\label{sec:6}
This series of studies presents a systematic investigation of the excitation of non‑modal perturbations in hypersonic blunt‑nose boundary layers. Part I (ZD25) introduced an efficient and accurate numerical framework, the SF‑HLNS approach, while Part II employed asymptotic analysis to elucidate the physical mechanism of non‑modal receptivity,
validating the predictions for an adiabatic-wall configuration. In the present work, we specifically examine the receptivity efficiency for a wide range of Reynolds numbers, wall temperatures, and freestream forcing types.
Collectively, these studies provide a comprehensive understanding of the excitation of non‑modal perturbations in blunt‑nose boundary layers. Unlike adiabatic walls, a cold wall introduces a distinct thermal boundary layer near the stagnation point, which determines the compressible {\color{red}behaviour} of the stagnation flow. This thermal layer induces additional temperature perturbations in the stagnation region, posing additional amplification of the non‑modal perturbations in the downstream boundary layer.

Numerical solutions for the seven cases under consideration confirm that the base flow converges to the Blasius similarity solution downstream, with the boundary-layer thickness growing like $R^{-1/2}(T_w/T_{ad})^{1/4}\sqrt{\xi}$.
{\color{black}Since the transient growth of the non-modal perturbation in the downstream region peaks when the local boundary-layer thickness becomes  comparable to its spanwise wavelength, we rescale the spanwise wavenumber $k_3$ as $R^{-1/2}k_3$ when examining the effect of $R$, and as $(T_w/T_{ad})^{1/4}k_3$ when considering the influence of $T_w$.}
Guided by the asymptotic model, the receptivity of the non-modal perturbations is decomposed into the following three steps. 

 Step (i) involves the interaction of the freestream perturbation with the oscillating bow shock in the {\color{red}centreline} region, which serves as the inhomogeneous forcing {\color{red}for} the region-III system. This  interaction is described by the linearised R-H relation. As indicated by (\ref{eq:Perturbation_ps}), the  post-shock perturbation {\color{red}depends only on} the oncoming Mach number and the freestream perturbation quantities. For acoustic and entropy forcing, the magnitudes of the post-shock perturbations are comparable, exceeding that for  vortical forcing by a factor of   $O(M)$ when $k_3=O(1)$. This discrepancy increases notably as $k_3$ grows. {\color{black}Under a unit-energy-norm forcing, the amplitudes of  acoustic and entropy perturbations remain independent of $k_3$, whereas the vortical perturbation amplitude scales as $k_3$. Consequently, for the scaling $k_3\sim R^{1/2}$ in the receptivity study, the acoustic/entropy amplitudes scale as $R^0$, while the vortical amplitudes scale as $R^{-1/2}$.} 

Step (ii) concerns the amplification of the perturbation streamwise vorticity from the post-shock region to the stagnation region, which is referred to as the 'slow-down convection' mechanism. Asymptotic analysis indicates that {\color{black} the spanwise velocity perturbation $\bar w_0$ depends linearly on both $k_3$ (which scales as $R^{1/2}$ in the present receptivity study) and $\hat p_{00}$ (proportional to the  forcing amplitude). 
Moreover, the stagnation streamwise vorticity $\tilde \Omega_{\xi 0}$ experiences an additional  $O(R^{1/2})$ amplification relative to $O(\bar w_0)$.
Therefore, under unity-energy-norm forcing, the stagnation vorticity response scales as $R$ for acoustic/entropy forcing, and as $R^{1/2}$ for vortical forcing. Such scaling relations are confirmed in figure \ref{fig:omega_R}. The effect of wall temperature, however, is not captured by the present scaling argument. Instead, it is obtained through curve fitting, which reveals that the data collapse in the   $(T_w/T_{ad})^{7/8}\tilde\Omega_{\xi 0}$-$(T_w/T_{ad})^{1/4}k_3$ plane under acoustic/entropy forcing, and in the $(T_w/T_{ad})^{5/8}\tilde\Omega_{\xi 0}$-$(T_w/T_{ad})^{1/4}k_3$ plane under vortical forcing. The implication is that the stagnation vorticity perturbation is more pronounced for higher Reynolds numbers and lower wall temperatures. Additionally, in the asymptotic model, the declination angle $\vartheta$ does not appear in the leading-order equations, which explains the weak dependence of the receptivity efficiency  on $\vartheta$ in SF-HLNS calculations. The receptivity efficiency reaches a maximum as the frequency approaches zero.
}

{\color{black}Step (iii) involves the downstream transient growth of the non-modal perturbations due to the lift-up mechanism. This process is parabolic, and initialised by the boundary-layer perturbation around the stagnation region. The effect of this process can be examined by normalising the downstream amplitude by the stagnation {\color{cyan}vorticity} perturbation $\tilde\Omega_{\xi 0}$. Such a measurement excludes the impact of freestream forcing types and their relevant governing parameters. The net amplification within this stage is systematically studied based on numerical results, which is found to be more pronounced for higher $R$ and $T_w$ values. 

Integrating the three steps outlined above, we assess the combined influence of  $R$ and $T_w$ on the downstream perturbation amplitude. Guided by figures \ref{fig:AuR_xi=200_1} and \ref{fig:AuTw_xi=200}, we  derive that, for given $(T_w/T_{ad})^{1/4}R^{-1/2}k_3$, the downstream amplitude $A_u$ scales as 
\begin{equation}
 A_u\sim  \left\{\begin{array}{ll}
     (T_w/T_{ad})^{-5/8}R &\mbox{under acoustic/entropy forcing},  \\
       (T_w/T_{ad})^{-3/8}R^{1/2} &\mbox{under vortical forcing}.
   \end{array}    \right.
\end{equation}

}

It should be noted that although this study provides a comprehensive characterisation of non-modal receptivity across varying Reynolds numbers and wall temperatures, it is important to acknowledge that the analysis is confined to a single, albeit representative, Mach number ($M=5.96$)  {\color{cyan}under {\color{red}centreline} symmetry (zero angle of attack)}. While the Reynolds-number scaling is derived analytically and  can therefore be considered robust, the wall-temperature dependence remains limited to the parameter range considered. Extensions to a broader range of Mach numbers and wall temperatures represent a natural and important direction for future work.

This series of studies provides an important step toward a physics-based transition prediction framework. Traditional transition models often neglect the receptivity process, yet its inclusion is essential, especially when bow-shock and entropy-layer effects are present. Although the SF-HLNS approach offers an efficient and accurate numerical tool, the asymptotic model developed here is computationally even more efficient, making the overall framework well-suited for practical engineering application.
Furthermore, to accurately predict transition onset, the LPSE calculations should be extended to the nonlinear regime. Nonlinear parabolised stability equations can capture both the saturation of streaky non‑modal perturbations and the onset of their secondary instabilities, following the methodology demonstrated in studies such as \cite{song2026prediction}.

 \vspace{.4cm}
 \noindent\textbf{Funding.} {This work is supported by National Natural Science Foundation of China (grant nos. 12372222, 92371104, 12588201),   the CAS Strategic Priority Research Program (no. XDB0620102) and CAS project for Young Scientists in Basic Research (YSBR-087).}

  \vspace{.4cm}
  \noindent\textbf{Declaration of interests.}
  {The authors report no conflict of interest.}

\bibliographystyle{jfm}

\bibliography{jfm}
\end{document}